\documentclass{emulateapj}
\submitted{Accepted to ApJS}

\newcommand{\Ha }{H$\alpha$}
\newcommand{\HI}{H{\sc i}}
\newcommand{\Dpak } {DensePak}
\newcommand{\kms} {km s$^{-1}$}
\newcommand{\siis}{[\mbox{S\,{\sc ii}}]$\lambda$6717}
\newcommand{\siio}{[\mbox{S\,{\sc ii}}]$\lambda$6731}
\newcommand{\nii}{[\mbox{N\,{\sc ii}}]$\lambda$6584}

\shorttitle{LSB Optical Velocity Fields and Rotation Curves}
\shortauthors{Kuzio de Naray et al.}

\begin{document}
\title{High Resolution Optical Velocity Fields of 11 Low Surface
  Brightness Galaxies }
\author{Rachel Kuzio de Naray\altaffilmark{1}, Stacy S. McGaugh\altaffilmark{1}}
\affil{Department of Astronomy, University of Maryland, College Park,
  MD 20742-2421}
\email{kuzio@astro.umd.edu, ssm@astro.umd.edu}
\author{W.J.G. de Blok\altaffilmark{1}}
\affil{Research School of Astronomy and Astrophysics\\ 
  Mount Stromlo Observatory, Cotter Road, Weston Creek ACT 2611, Australia}
\email{edeblok@mso.anu.edu.au}
\altaffiltext{1}{Visiting Astronomer, Kitt Peak
    National Observatory, National Optical Astronomy Observatory,
    which is operated by the Association of Universities for Research
    in Astronomy, Inc. (AURA) under cooperative agreement with the
    National Science Foundation.}
\and
\author{A. Bosma}
\affil{Observatoire de Marseille, 2 Place Le Verrier, 13248 Marseille
  Cedex 4, France}
\email{bosma@oamp.fr}

\begin{abstract}
We present high resolution two-dimensional velocity fields from
integral field spectroscopy along with derived rotation curves for
eleven low surface brightness galaxies.  We fit NFW
and pseudo-isothermal halo models to the new data combined with
previous long-slit and \HI\ data.  In most cases we find the
pseudo-isothermal halo to better represent the data than the NFW halo,
as the NFW concentrations are often lower than expected for a
$\Lambda$CDM cosmology.  We also compare our results to previous studies
and find that including the new two-dimensional optical data does not
significantly alter the halo parameters, but does decrease the 
uncertainties by roughly a factor of 2.

\end{abstract}


\section{Introduction}
Determining the mass-density profiles of galactic dark matter halos
has been an exciting, yet contentious, field for a number of years
now.  Though it is widely agreed that low surface brightness (LSB)
galaxies are dark matter dominated down to small radii \citep{dBM96,dBM97,
Pick97, Pick99, Blais, Borriello, Simon03, Fuchs} and, hence, are ideal test 
subjects for 
studying the dark matter distribution, a consensus on the
interpretation of their rotation curves has been difficult to achieve.

The disagreement arises when comparing the data to numerical
simulations of cold dark matter (CDM).  The most common description of
CDM halo behavior is given by the analytic approximation of
\citet{NFW96,NFW97} and is known as the NFW profile.  The
cosmologically motivated NFW halo is characterized by a mass-density
that rises very steeply toward the center, a property which makes the
halo ``cuspy''.  Cuspy halos that rise more steeply than the NFW halo
have also been suggested \citep[e.g.][]{Moore, Reed, Navarro2004, Diemand}.
Whether or not the dark matter halos of LSB galaxies 
can be described by cuspy NFW-like profiles has been a matter of debate. 

NFW halos can be fit to the observations, but the fits are usually of 
lower quality than fits with pseudo-isothermal halos.  Moreover, the 
implied cosmological parameters are inconsistent 
with the standard $\Lambda$CDM picture.  In particular, the observed 
concentrations of the NFW halos are too low \citep{McGaugh03,
Swaters03}.  Much better fits to LSB observations are found when using the 
pseudo-isothermal halo model (Simon et al. 2005; de Blok, Bosma, \&
McGaugh 2003; de Blok \& Bosma 2002 (hereafter BB02); Marchesini et
al. 2002; Bolatto et al. 2002; de Blok, McGaugh,  \& Rubin 2001 
(hereafter BMR01); de Blok et al. 2001; Blais-Ouellette, Amram, \& 
Carignan 2001; C\^{o}t\'{e}, Carignan, \& Freeman 2000).  These halos have 
a mass-density that remains at an 
approximately constant value toward the center, thus they are referred 
to as ``cored'' halos.  Unlike the NFW profile, however, the 
pseudo-isothermal halo has no cosmological motivation or theoretical 
basis.

\defcitealias{dBB}{BB02}
\defcitealias{dBMR}{BMR01}

In its favor, the CDM model has been successful on large scales in 
explaining structure formation in the early Universe as well as
abundances of galaxy clusters \citep{Tegmark}.  It is more appealing
to have a halo model with explanations rooted in cosmology rather than
a model that is simply a convenient fit to the data.  It is therefore
no surprise that a number of reasons have been given in an attempt to 
salvage the appropriateness of the NFW profile as a description of 
galactic dark matter halos.

\begin{deluxetable*}{lccrrrrrrrc}
\tabletypesize{\scriptsize}
\tablecaption{Properties of Observed Galaxies}
\tablewidth{0pt}
\tablehead{
 &\colhead{R.A.} &\colhead{Dec.} &\colhead{$\mu_{0}$(R)} &\colhead{Dist.} &\colhead{$i$} &\colhead{$V_{hel}$} &\colhead{$R_{max}$} &\colhead{$V_{max}$} &\colhead{PA} &\colhead{Refs.}\\
\colhead{Galaxy} &\colhead{(J2000)} &\colhead{(J2000)} &\colhead{(mag arcsec$^{-2}$)} &\colhead{(Mpc)} &\colhead{(deg)} &\colhead{(\kms)} &\colhead{(kpc)} &\colhead{(\kms)} &\colhead{(deg)}\\
\colhead{(1)} &\colhead{(2)} &\colhead{(3)} &\colhead{(4)} &\colhead{(5)} &\colhead{(6)} &\colhead{(7)} &\colhead{(8)} &\colhead{(9)} &\colhead{(10)} &\colhead{(11)}
}
\startdata
UGC 4325 &08:19:20.5 &+50:00:35 &21.6 &10.1 &41 &514 &2.3 &86 &52 &2\\
F563-V2  &08:53:03.7 &+18:26:09 &21.2$^{c}$ &61 &29 &4316 &6.7 &104 &328 &1\\
F563-1   &08:55:06.9 &+19:44:58 &22.6 &45 &25 &3482 &5.6 &146 &341 &1\\
DDO 64$^{a}$   &09:50:22.4 &+31:29:16 &- &6.1 &60 &520 &2.1 &51 &97 &2\\
F568-3   &10:27:20.3 &+22:14:22 &22.2$^{c}$ &77 &40 &5905 &8.4 &114 &169 &1\\
UGC 5750 &10:35:45.1 &+20:59:24 &22.6 &56 &64 &4160 &8.5 &61 &167 &2\\
NGC 4395$^{b}$ &12:25:48.9 &+33:32:48 &22.2 &3.5 &46 &310 &0.8 &33 &327 &2\\
F583-4   &15:52:12.7 &+18:47:06 &22.9$^{c}$ &49 &55 &3620 &7.5 &75 &115 &1\\
F583-1   &15:57:27.5 &+20:39:58 &23.2$^{c}$ &32 &63 &2256 &4.9 &83 &355 &1\\
UGC 477  &00:46:13.1 &+19:29:24 &- &35 &82 &2635 &10 &112 &347 &3\\
UGC 1281 &01:49:32.0 &+32:35:23 &22.7$^{d}$ &5.5 &85 &145 &1.9 &38 &218 &3\\
\enddata
\tablecomments{Col.(1): Galaxy name. Col.(2): Right Ascension. Col.(3): Declination. Col.(4): Central surface brightness in $R$-band (mag arcsec$^{-2}$). Col.(5): Distance (Mpc). Col.(6): inclination ($\degr$). Col.(7): Heliocentric systemic velocity (\kms). Col.(8): Maximum radius of the DensePak rotation curve (kpc). Col.(9): Maximum velocity of the DensePak rotation curve (\kms). Col.(10): Position angle of major axis ($\degr$); see Sec. 4.1 for details. Col.(11): References for surface brightness, distance and inclination: (1) \citet{dBMR} (2) \citet{dBB} (3) \citet{Tully}.\\
\indent $^{a}$ DDO 64 = UGC 5272.\\
\indent $^{b}$ NGC 4395 = UGC 7524.\\
\indent $^{c}$ Converted from $B$ band assuming $B$ $-$ $R$ = 0.9.\\
\indent $^{d}$ Taken from reference (2).}
\end{deluxetable*}

The earliest observations which indicated cores in LSB galaxies were 
two-dimensional 21 cm \HI\ velocity fields \citep[][hereafter
BMH96]{Moore94, Flores, DMV}.  Low spatial resolution (i.e., beam smearing) was
suggested to be a systematic effect that would erroneously indicate 
cores \citep{vandenB00,Swaters00}.  The question of beam smearing was 
addressed by long-slit \Ha\ observations which had an order of 
magnitude increase in spatial resolution [see for example, McGaugh,
Rubin, \& de Blok 2001 (hereafter MRB01); BMR01]; cusps did not appear 
when the resolution was increased, showing that beam smearing had been 
of only minor importance in the \HI\ observations.  Possible systematic
errors in the long-slit spectroscopy \citep[e.g.][]{Simon03,Rhee04,Spekkens05}
have since become the focus of attention, with slit misplacement 
\citep{Rob} and non-circular motions among the top concerns.  
If the slit misses the dynamical center of the
galaxy, or if there are non-circular motions from, for instance, a
bar, the circular velocity may be underestimated and lead to the false 
inference of a cored halo.   \citet{dBBM} conducted extensive modeling 
in which the rotation curves of both cuspy and cored halos were
subjected to various effects and concluded that no systematic effect
will entirely mask the presence of a cuspy halo for realistic
observing conditions. \citet{Rob} performed a similar exercise with
similar results, but argued that it might still be possible to retain cuspy 
halos.

\defcitealias{DMV}{BMH96}
\defcitealias{MRdB}{MRB01}

Clearly there are a number of issues which new observations must 
simultaneously address.  The data must be both high resolution and 
two-dimensional in nature.  Observations must have resolution
$\lesssim$ 1 kpc as that is the critical length scale at which the 
distinction between cusps and cores can be determined
\citep{deBlok03}.  Any non-circular motions should be readily 
identifiable in a two-dimensional velocity field. Additionally, slit 
placement is not a concern of two-dimensional velocity fields.  This
observational approach has also been applied by such groups as 
Simon et al. (2003, 2005).

In this paper we present the rotation curves derived from high
resolution two-dimensional velocity fields of a sample of LSB
galaxies.  In $\S$ 2 we discuss the sample and observations; data 
reduction is discussed in $\S$ 3.  The results for the individual 
galaxies  are presented in $\S$ 4.  In $\S$ 5 we discuss halo fits to 
the minimum disk case of the new data combined with previous long-slit 
and \HI\ rotation curves and compare our results to previous studies.  
Non-circular motions are also briefly discussed.  Our conclusions 
and goals for future work are stated in $\S$ 6.

\section{Sample and Observations}
Our primary targets were the galaxies in the ``clean'' sample of
\citet{dBBM}.  In brief, galaxies 
in the ``clean'' sample have inclinations between 30$\degr$ and
85$\degr$, are likely to meet the minimum disk assumption,  have 
long-slit rotation curves which are well resolved in the inner 1 kpc, 
have small errorbars and lack large asymmetries.  The minimum disk 
assumption is considered applicable to those galaxies which require
substantial amounts of dark matter at small radii even in the maximum disk
model.  We then expanded our sample to include other LSB galaxies that
nearly made the ``clean'' cut.  We also searched for low mass dwarf 
galaxies to fill out the RAs available at the telescope, giving preference
to those targets with diffuse \Ha\ emission detected by long-slit 
observations.

\begin{deluxetable*}{lllll}
\tabletypesize{\footnotesize}
\tablecaption{Galaxies without Velocity Fields}
\tablewidth{0pt}
\tablehead{
}
\startdata
CamB & &UGC 2684  & &UGC 9211/DDO 189\\
F750-4 & &UGC 4543 & &UGC 11583\\
KK98 251 & &UGC 4787 & &UGC 12344\\
UGC 891 & &UGC 5414/NGC 3104 & &UGC 12713\\
UGC 1501/NGC 784 & &UGC 7603/NGC 4455 & &UGC 12791\\
UGC 2455/NGC 1156 & &UGC 8837/DDO 185 & &\\
\enddata
\tablecomments{Galaxies which were observed with \Dpak\ but whose detections were not good enough to construct meaningful velocity fields.}
\end{deluxetable*}

We observed 8 ``clean'' galaxies: UGC 4325,
DDO 64, NGC 4395, F583-4, F583-1, DDO 185, DDO 189, and NGC 4455.  These
galaxies  were selected from the ``clean'' sample because the 
well-resolved long-slit \Ha\ observations \citepalias{dBB,dBMR}  
show there is diffuse \Ha\ emission for detectability in 
two-dimensional velocity fields and that the galaxies 
lack indicators of significant non-circular motions (e.g., 
strong bars or gross asymmetries).  While not a criterion considered in 
the selection process, it turns out that the long-slit observations of 
these galaxies imply that the galaxies have either unreasonably low NFW 
concentrations or do not have cusps at all.  These kinds of galaxies pose
the biggest problem for CDM and as such, provide important test cases.

We observed 4 galaxies from \citetalias{MRdB} (F563-V2, F563-1, 
F568-3, UGC 5750) that almost made the ``clean'' sample.  They show 
diffuse \Ha\ emission, but either did not have the required number of 
independent points in the inner 1 kpc of the long-slit rotation curve 
or had an inclination outside the ``clean'' range.

Lastly, we observed 16 galaxies from the Nearby Galaxies Catalogue 
\citep{Tully}.  Selection criteria for these galaxies included positions 
satisfying 18h$\lesssim$ $\alpha$ $\lesssim$08h and 
+10$\degr$$\lesssim$ $\delta$ $\lesssim$+50$\degr$, inclinations between 
30$\degr$ 
and 85$\degr$, heliocentric velocities $\lesssim$ 2500 \kms, and an 
estimated $V_{flat}$ (approximated by $V_{flat}$ $\sim$ 0.5$W_{20}$$sin^{-1}i$)
  between roughly 50 \kms\ and 100 \kms.

Our sample of 28 observed galaxies is both weather and signal limited. 
Poor weather prevented us from observing more of the ``clean'' sample.
Our sample is signal limited in that not all of the galaxies have 
enough \Ha\ emission to construct useful velocity fields.  While 
pre-existing long-slit observations can be used as a guide, there is no
way of knowing how much \Ha\ emission will be detected by the IFU until
the experiment is done.  Our sample is intentionally focused on the most 
dark matter dominated objects.  These tend to be very low surface brightness 
dwarfs that are hard to observe.

The galaxies were observed during the nights of 2004 April 12-15, 2004 
November 14-19 and 2005 September 1-7.  Observations were made using
the \Dpak\ Integrated Field Unit on the 3.5-m WIYN\footnote{Based on 
observations obtained at the WIYN Observatory.  The WIYN Observatory is a 
joint facility of the University of Wisconsin-Madison, Indiana University,
Yale University, and the National Optical Astronomy Observatory.} telescope
at the Kitt Peak National Observatory.  \Dpak\ is comprised of 3\arcsec\ 
diameter fibers arranged in a 43\arcsec\ $\times$ 28\arcsec\ rectangle.  We 
measured the fiber separation to be 3.84\arcsec.  The separation was 
determined by centering a bright star in a fiber and repeatedly shifting 
between fibers and across the array.  There are 85 working fibers in
this arrangement; an additional four sky fibers are arranged outside
the main bundle.  We used the 860 line mm$^{-1}$ grating in second
order, centered near \Ha\ giving a 58 \kms\ velocity resolution. The 
distances to the galaxies in the sample are such that a 3\arcsec\
fiber provides sub-kpc resolution.

Because the galaxies were too faint to be visible on the guide 
camera, we centered the \Dpak\ array on a nearby star and then 
offset to the optical center of the galaxy.  Subsequent pointings on
the galaxy were made by shifting the array by small amounts from its 
current position. For a number of galaxies, these moves were the fine 
shifts required to observe the spaces between the fibers.  These 
interstitial pointings effectively increased the resolution 
to $\sim$ 2\arcsec.  The fiber bundle orientation on the sky and 
the total number of pointings per galaxy were tailored to each galaxy 
so that the critical central regions were covered by the \Dpak\
fibers.  Each exposure was 1800 sec, and two exposures were taken at 
each pointing.  A CuAr lamp was observed before and after each
pointing to provide wavelength calibration.

\begin{figure*}
\plottwo{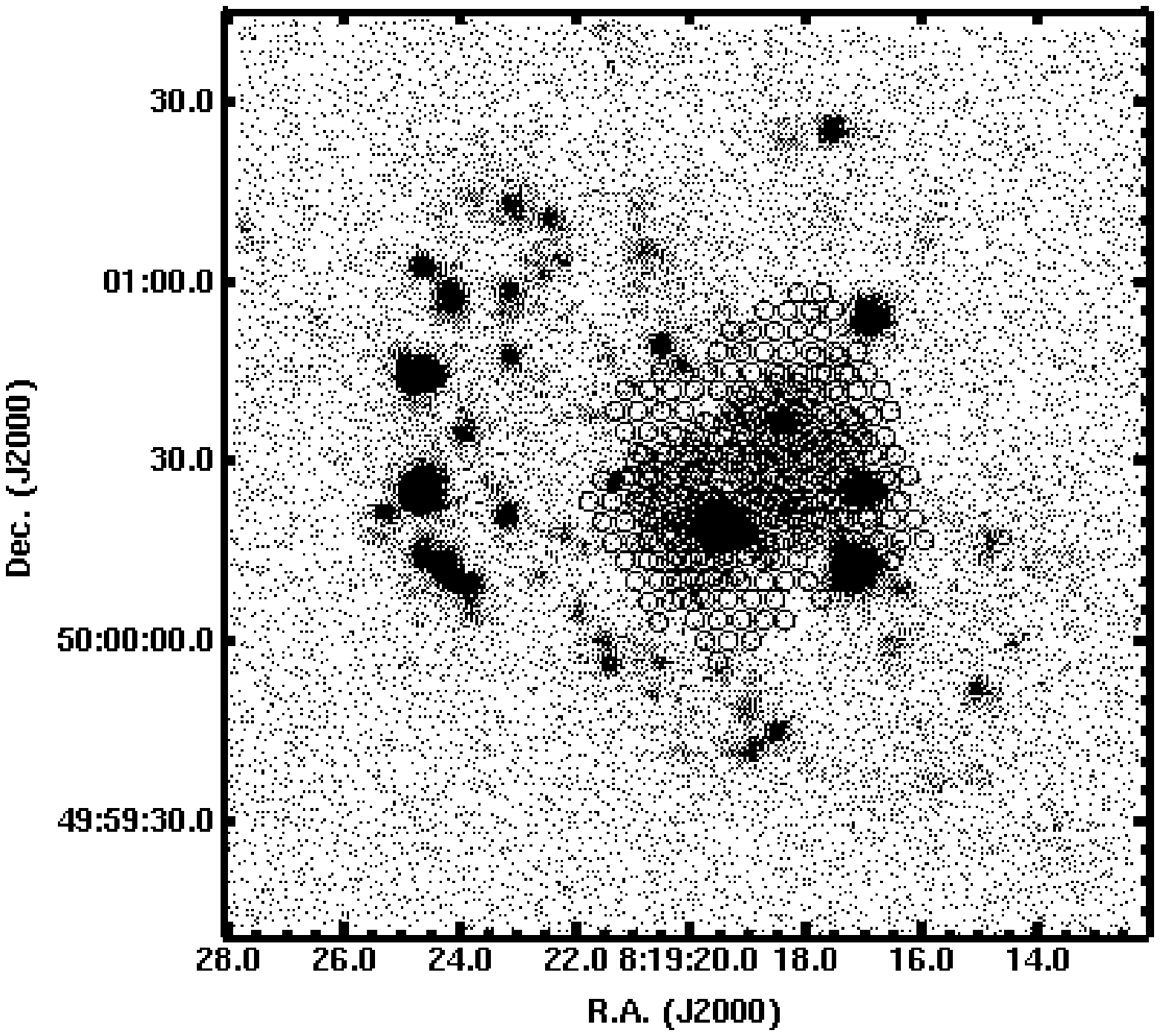}{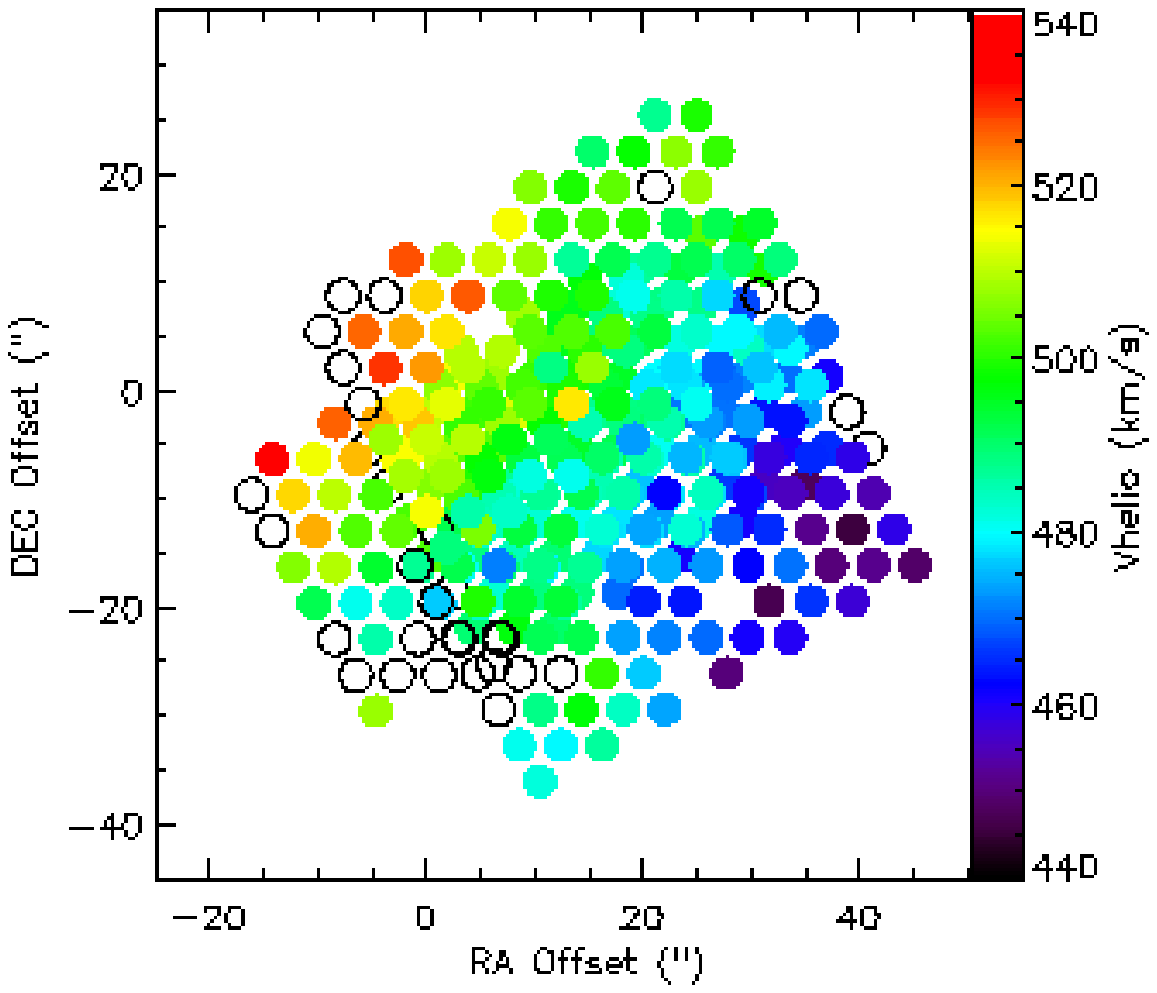}\\
\plottwo{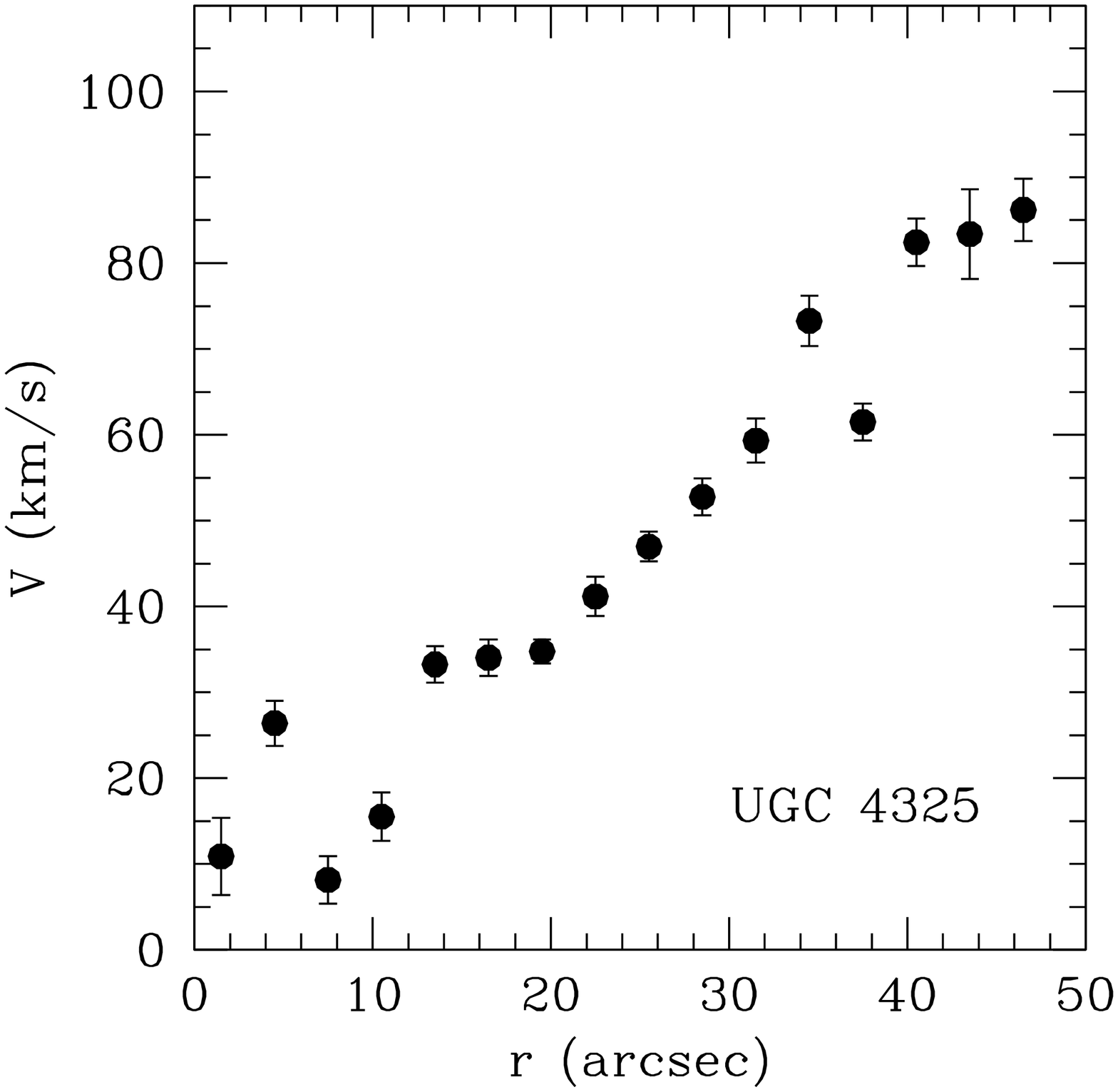}{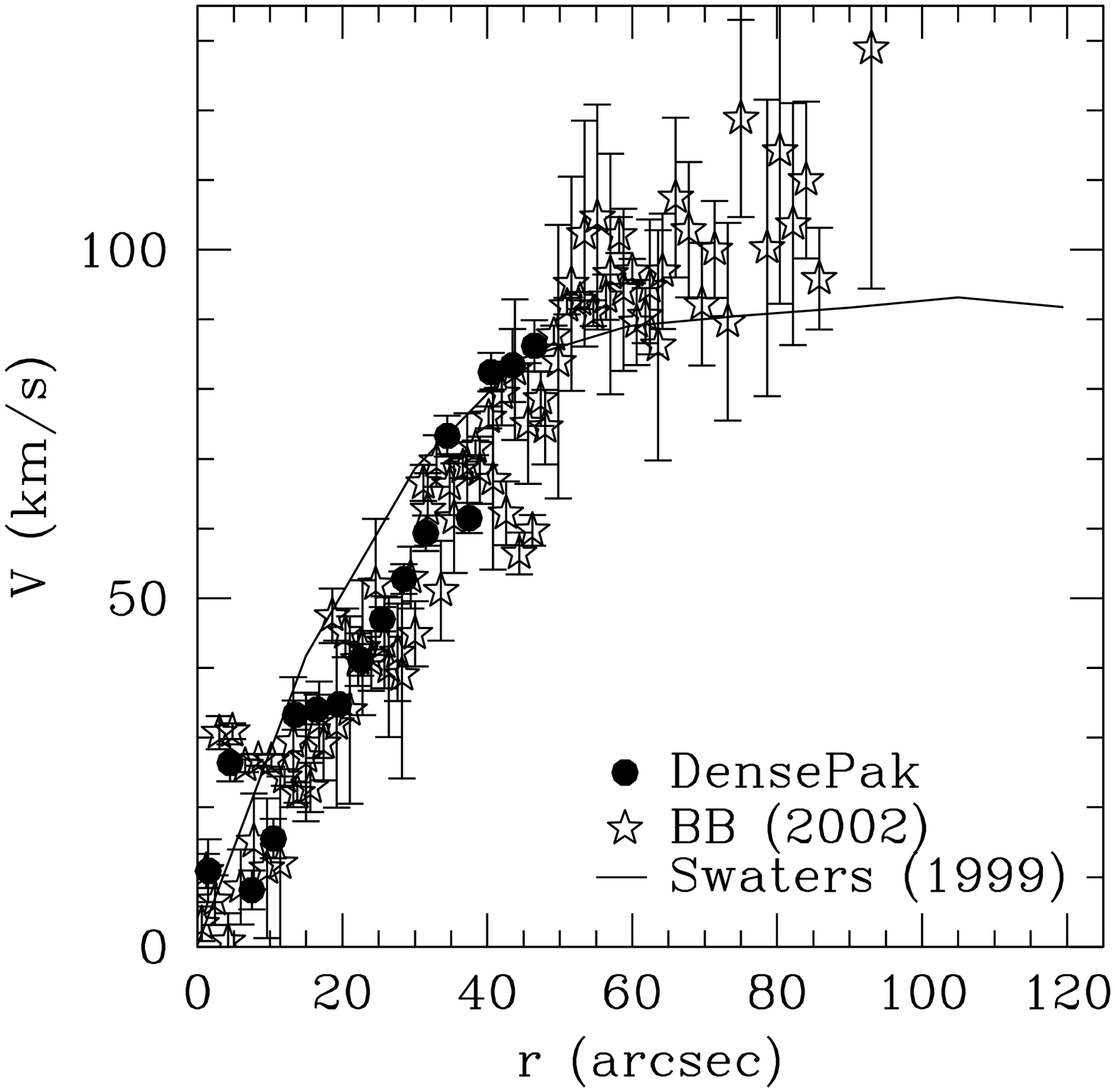}
\caption{Results for UGC 4325: {\it(Upper left)} Position of DensePak
  array on the \Ha\ image of the galaxy.  This is an example of where
  interstitial pointings have been made to fill in the gaps between
  the fibers.  {\it(Upper right)} Observed \Dpak\ velocity field. Empty
  fibers are those without detections. {\it(Lower left)} DensePak rotation 
  curve. {\it(Lower right)} 
  DensePak rotation curve plotted with the raw long-slit \Ha\ rotation 
  curve of \citet{dBB} and the \HI\ rotation curve of \citet{Swatersthesis}.
  The \HI\ data was excluded from the halo fits.  
  Figure appears in color on-line.}
\end{figure*}
\begin{figure*}
\plottwo{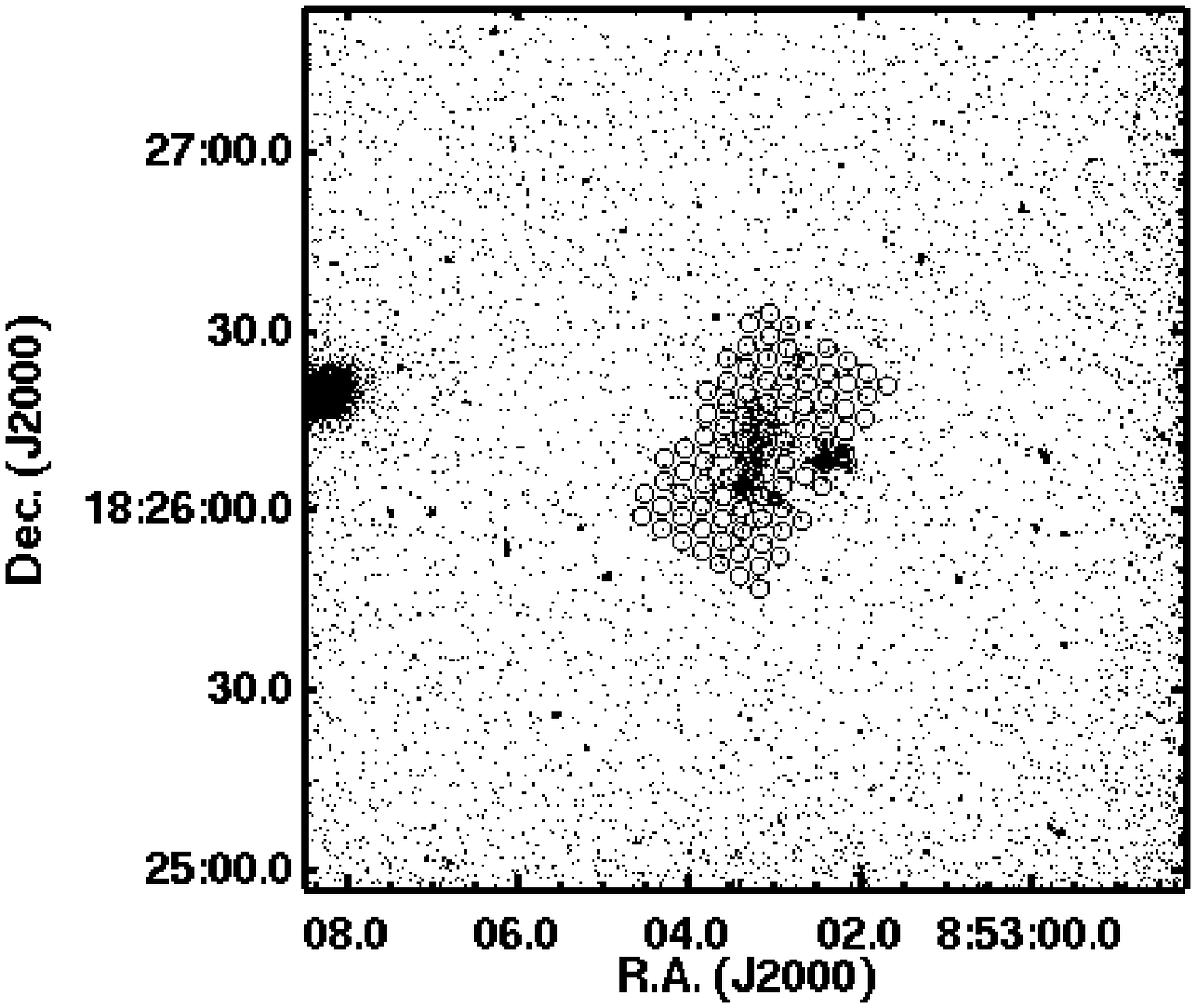}{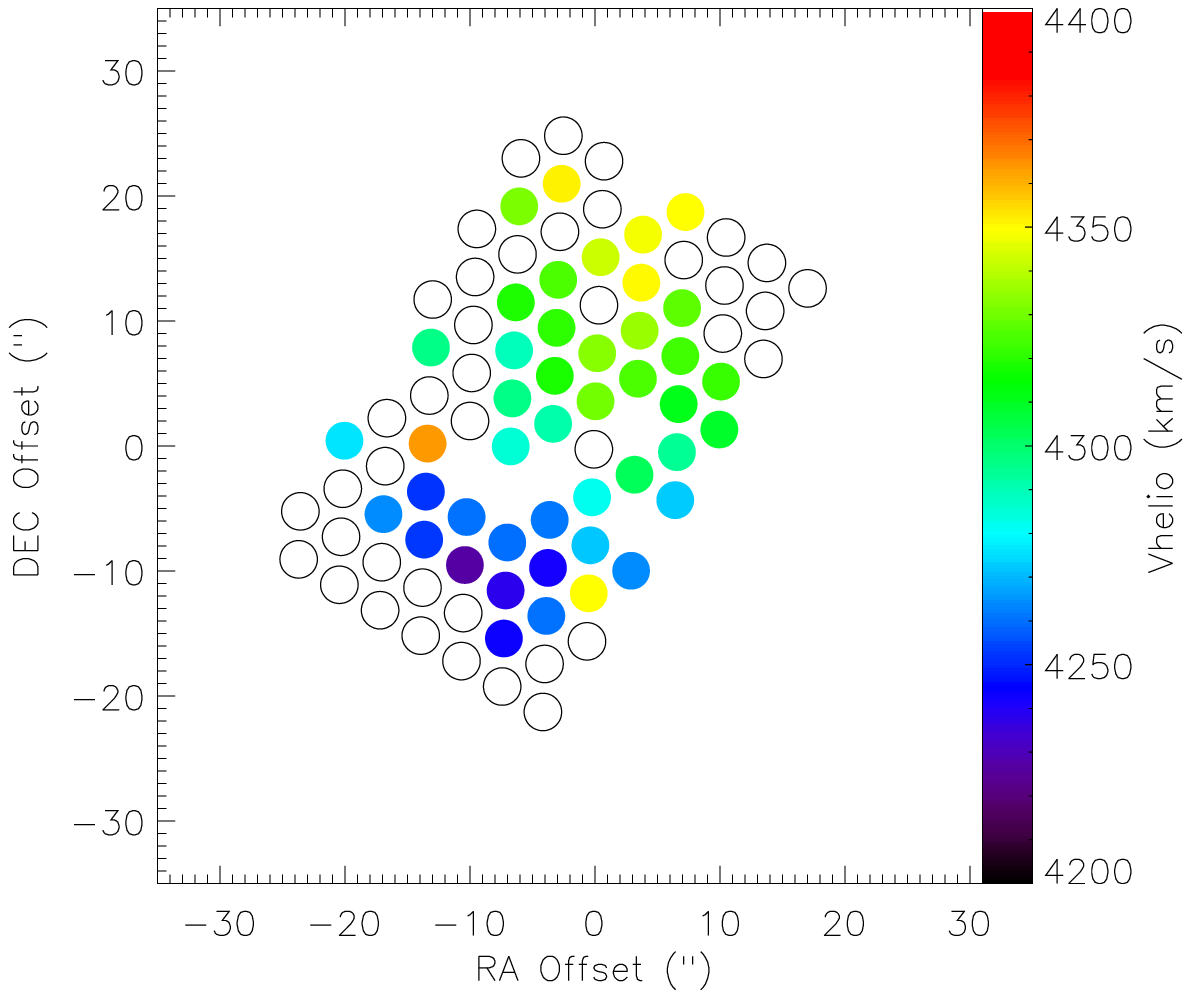}\\
\plottwo{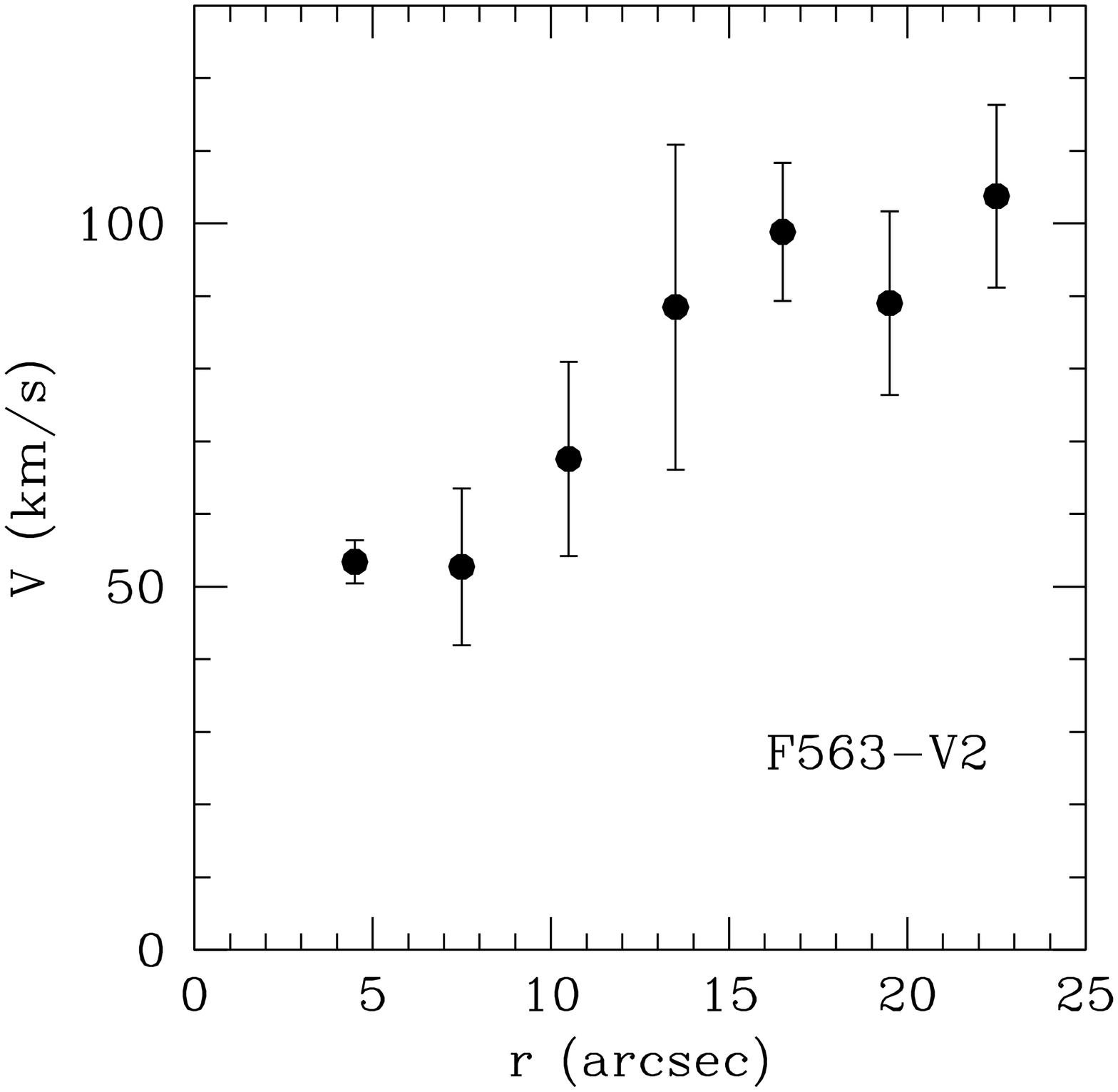}{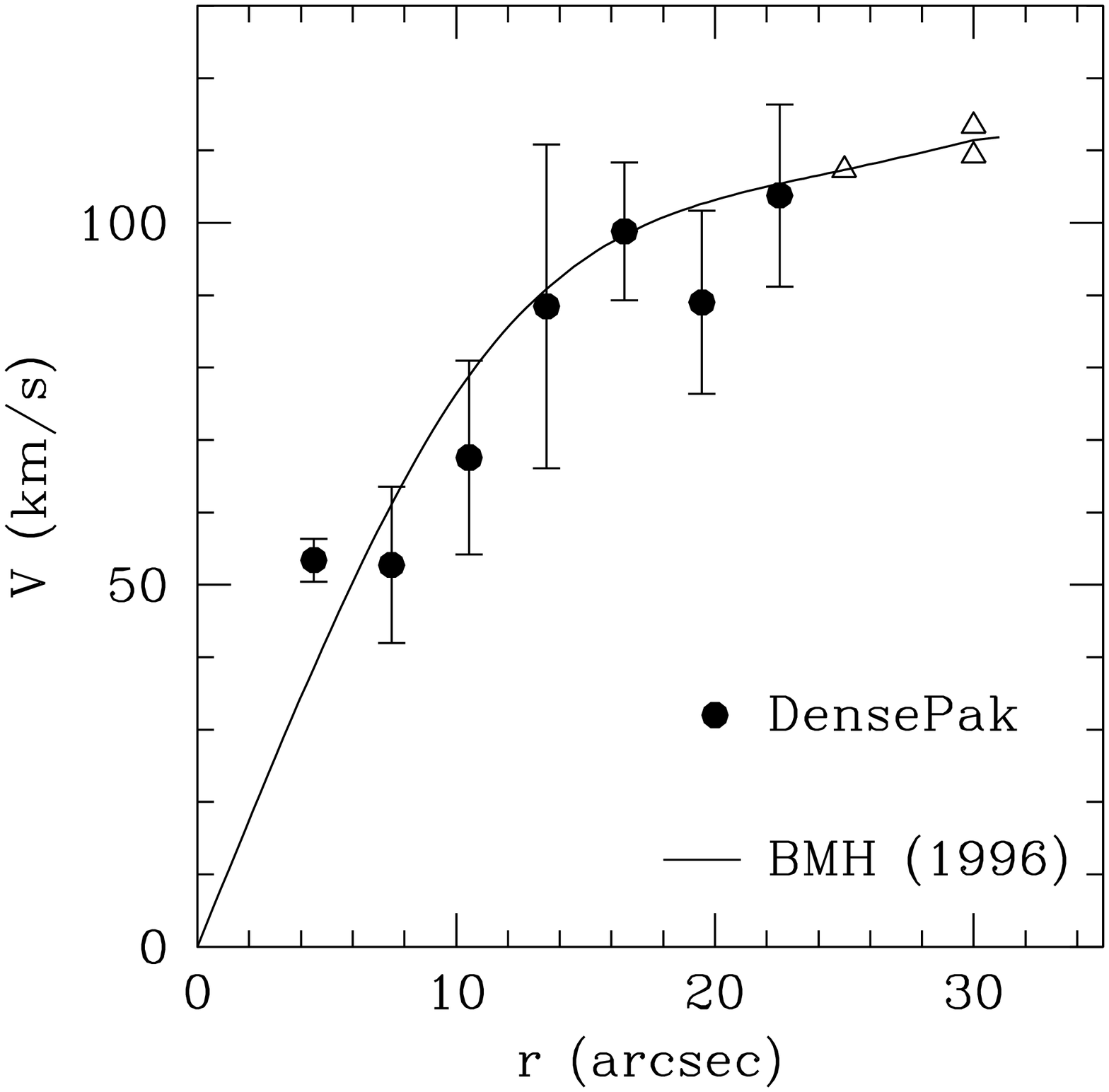}
\caption{Results for F563-V2: {\it(Upper left)} Position of DensePak
  array on an \Ha\ image of the galaxy. {\it(Upper right)} Observed \Dpak\
  velocity field.  Empty fibers are those without detections.  
  {\it(Lower left)} DensePak rotation curve. 
  {\it(Lower right)} DensePak rotation curve plotted with the \HI\ 
  rotation curve of \citet{DMV}.  The triangles represent the \HI\ points
  used in the halo fits.  Figure appears in color on-line.}
\end{figure*}
\begin{figure*}
\plottwo{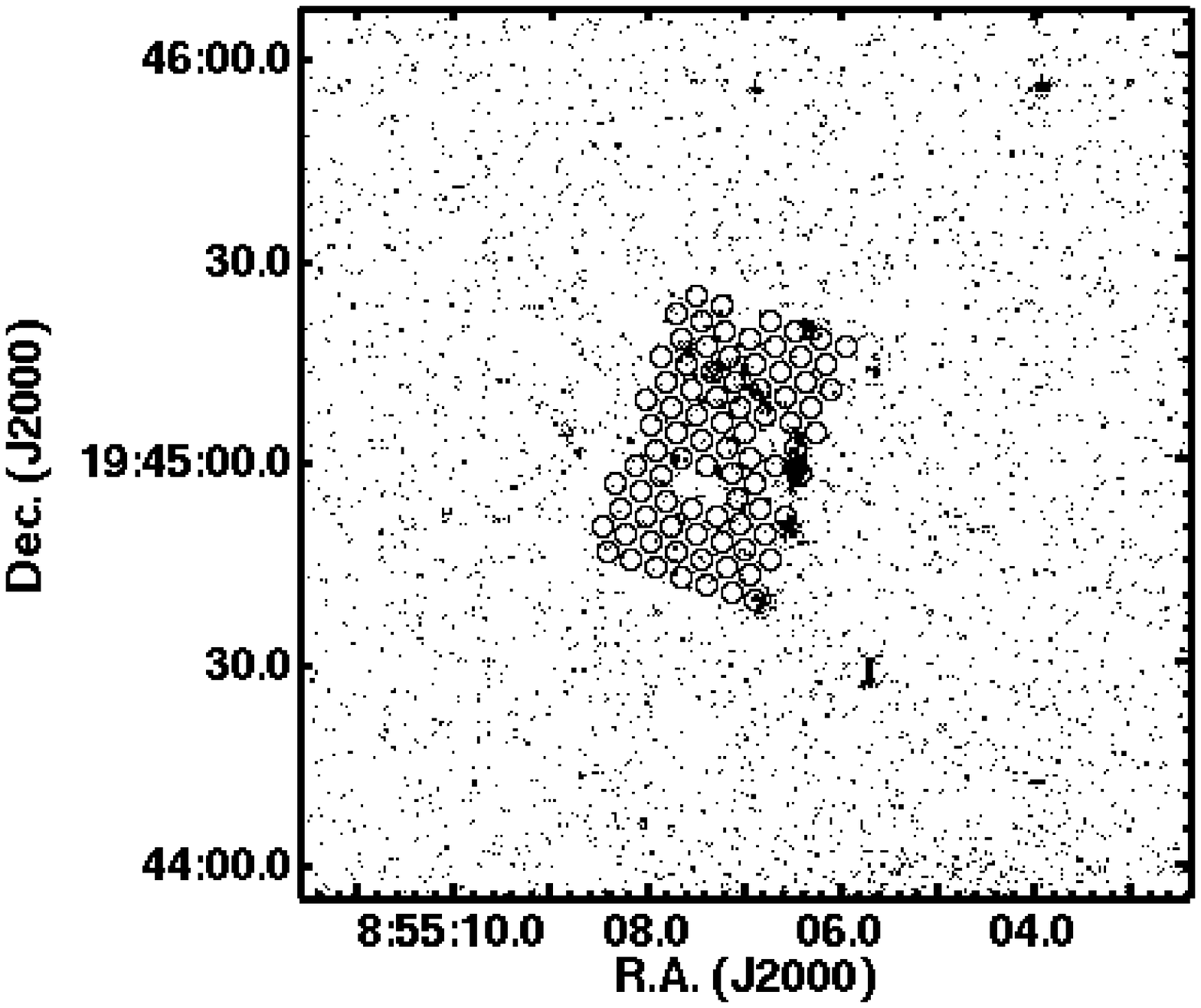}{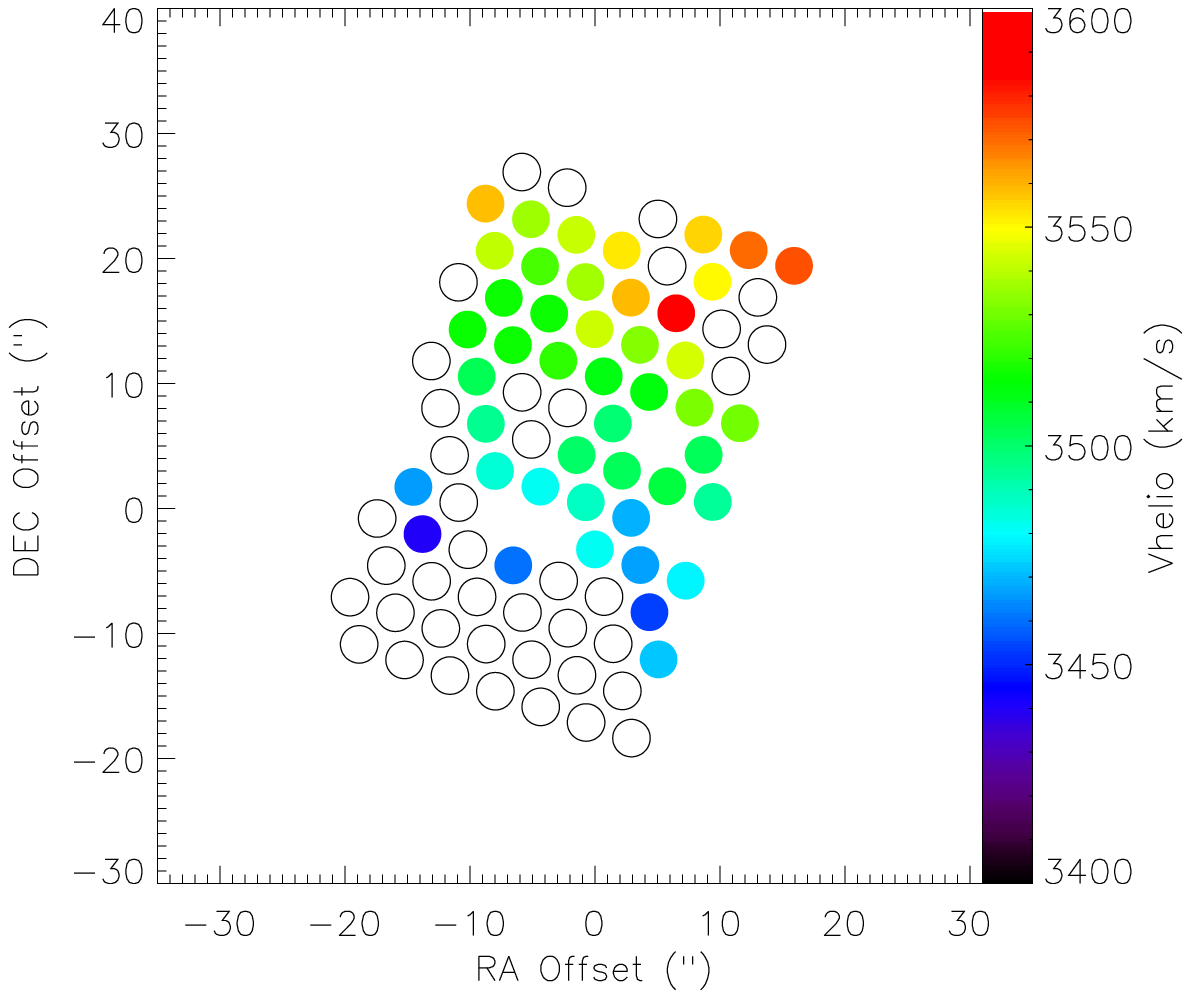}\\
\plottwo{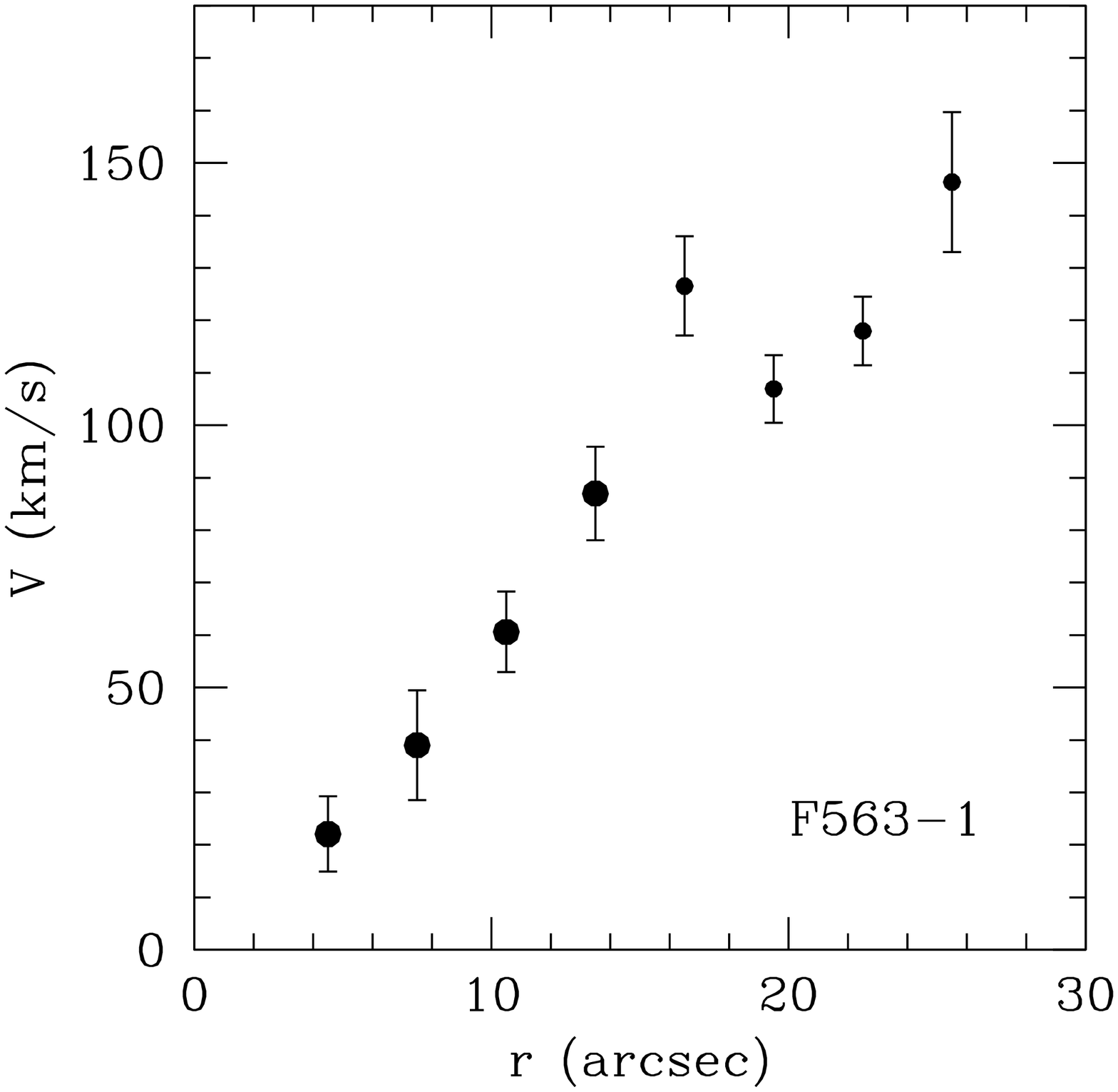}{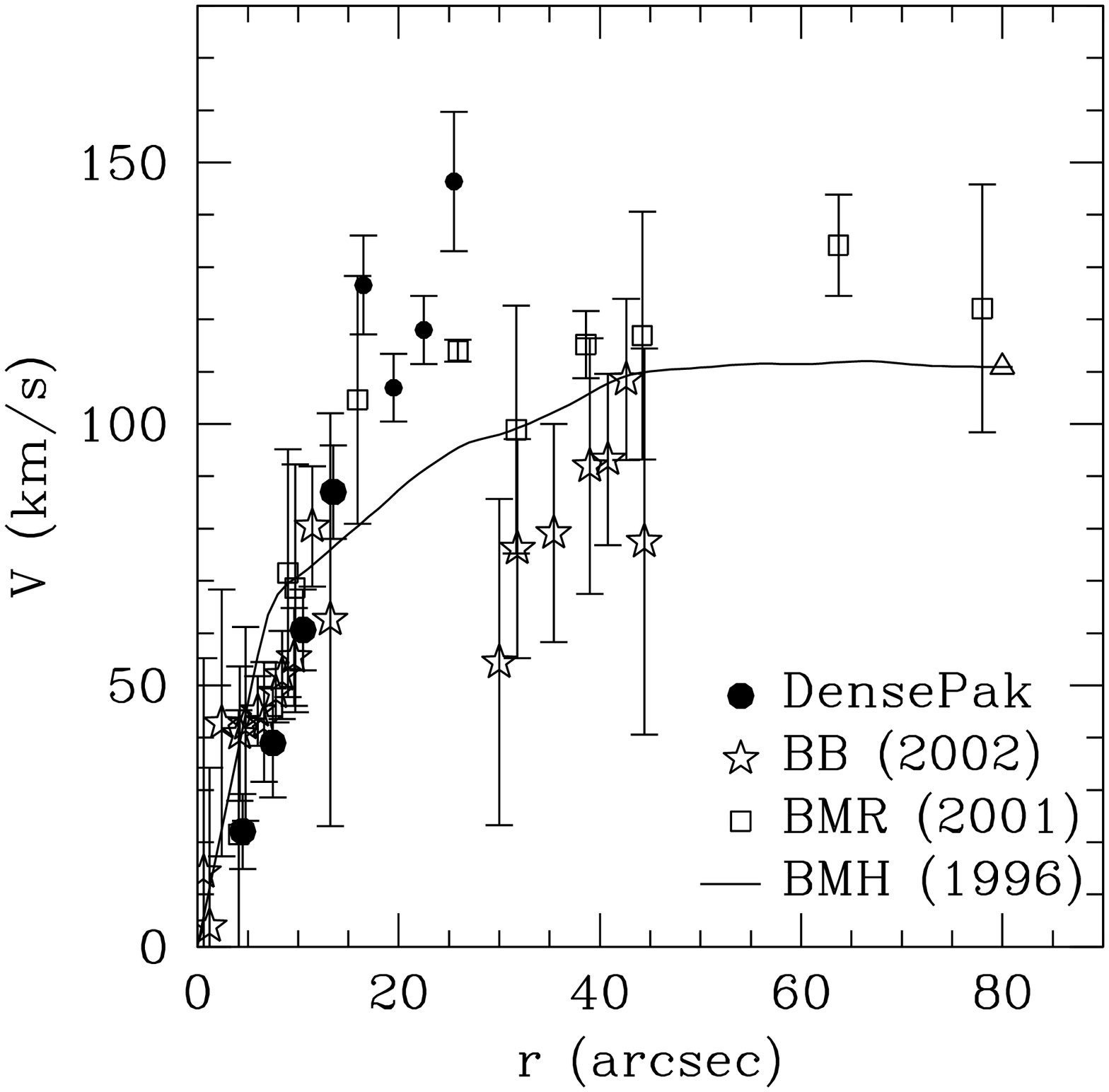}
\caption{Results for F563-1: {\it(Upper left)} Position of DensePak
  array on an \Ha\ image of the galaxy. {\it(Upper right)} Observed 
  \Dpak\ velocity field.  Empty fibers are those without detections. 
  {\it(Lower left)} DensePak rotation curve.  The last four points are
  omitted from the halo fits.
  {\it(Lower right)} DensePak rotation curve plotted with the raw 
  long-slit \Ha\ rotation curves of \citet{dBB} and \citet{dBMR} and 
  the \HI\ rotation curve of \citet{DMV}.  The triangle represents the
  \HI\ point used in the halo fits.  Figure appears in color on-line.}
\end{figure*}
\begin{figure*}
\plottwo{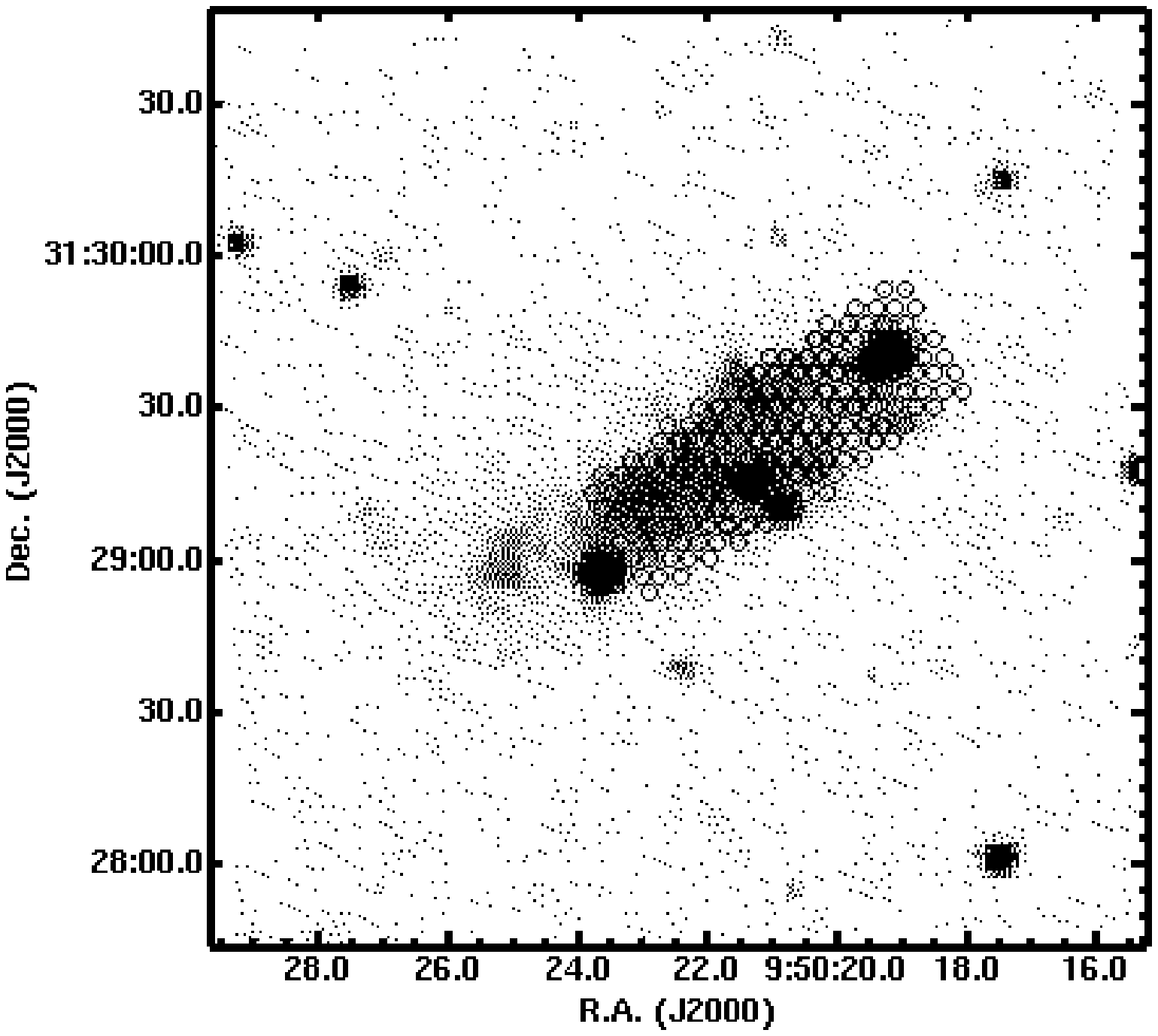}{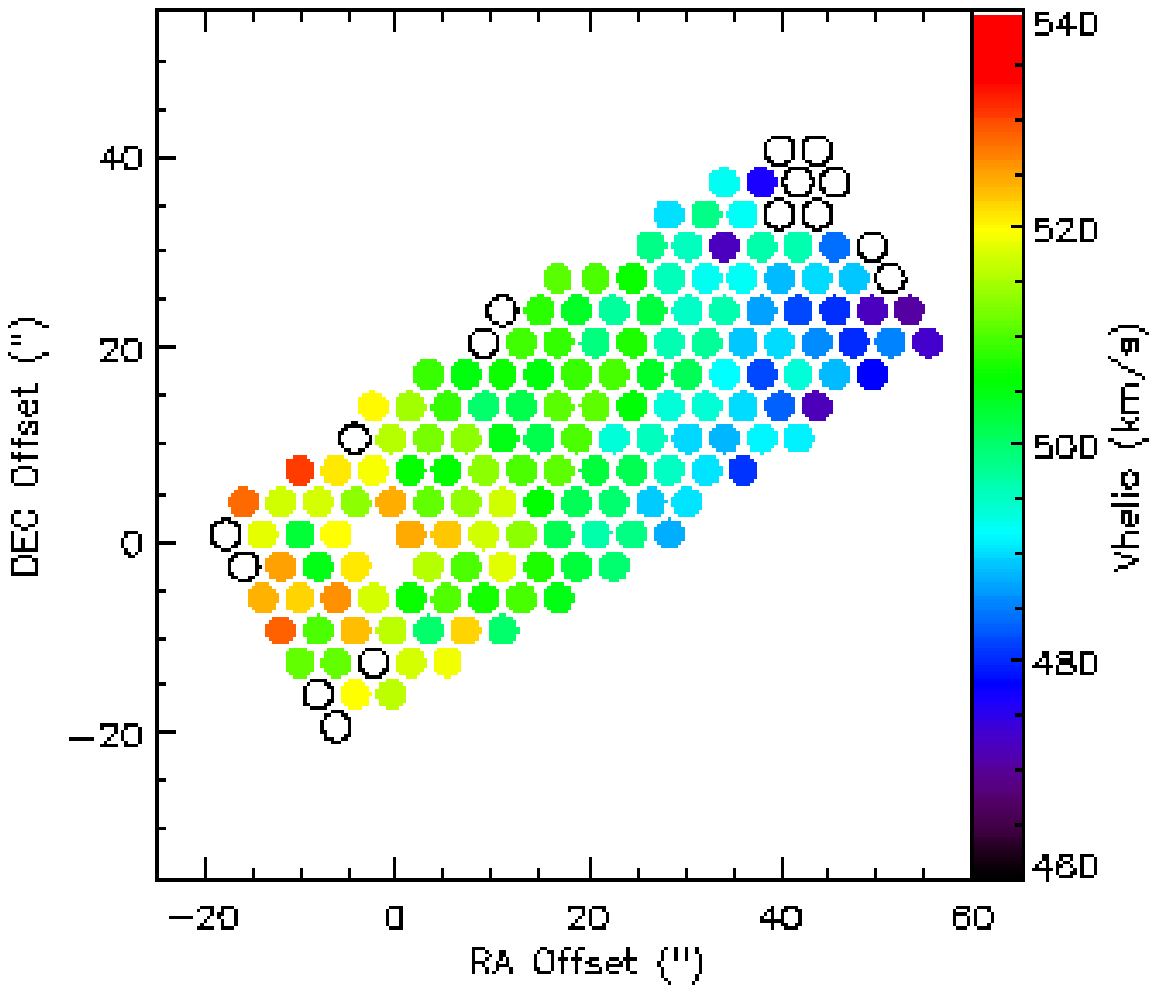}\\
\plottwo{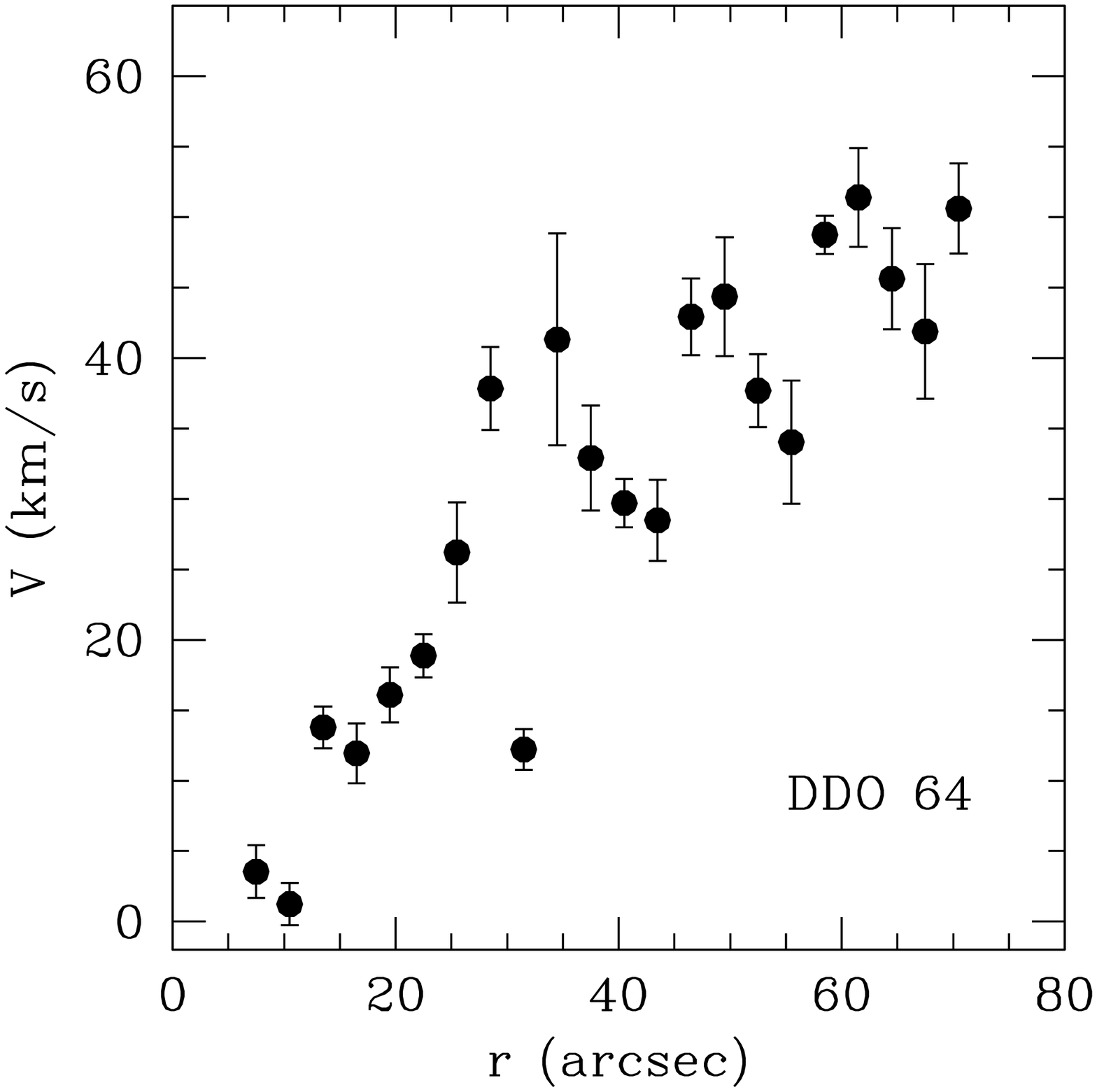}{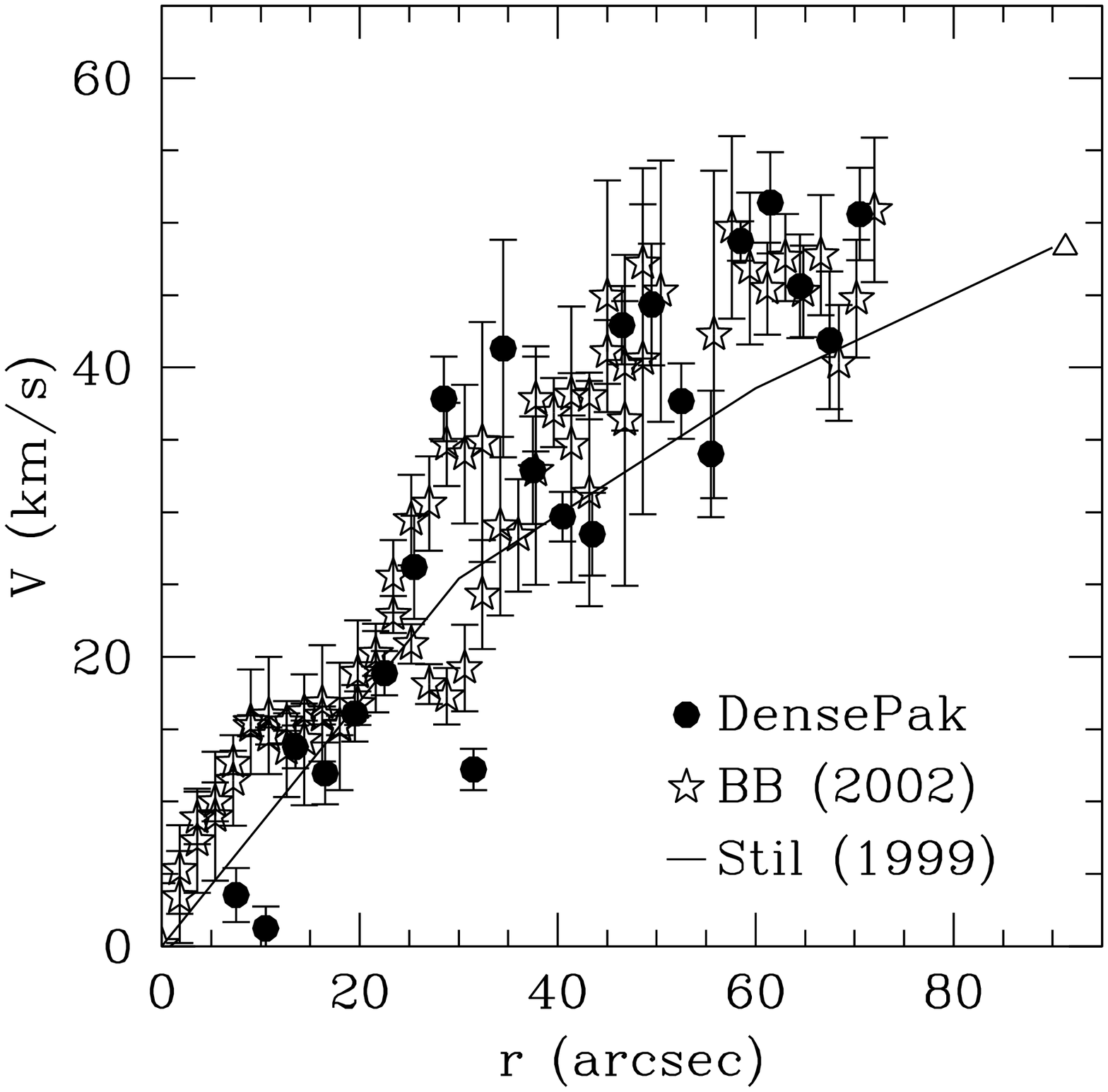}
\caption{Results for DDO 64: {\it(Upper left)} Position of DensePak
  array on a DSS image of the galaxy. {\it(Upper right)} Observed
  \Dpak\ velocity field.  The empty fibers are those without detections.
  {\it(Lower left)} DensePak rotation curve. 
  {\it(Lower right)} DensePak rotation curve plotted with the raw 
  long-slit \Ha\ rotation curve of \citet{dBB} and the \HI\ rotation 
  curve of \citet{Stil}.  The triangle represents the \HI\ point used in 
  the halo fits.  Figure appears in color on-line.}
\end{figure*}
\begin{figure*}
\plottwo{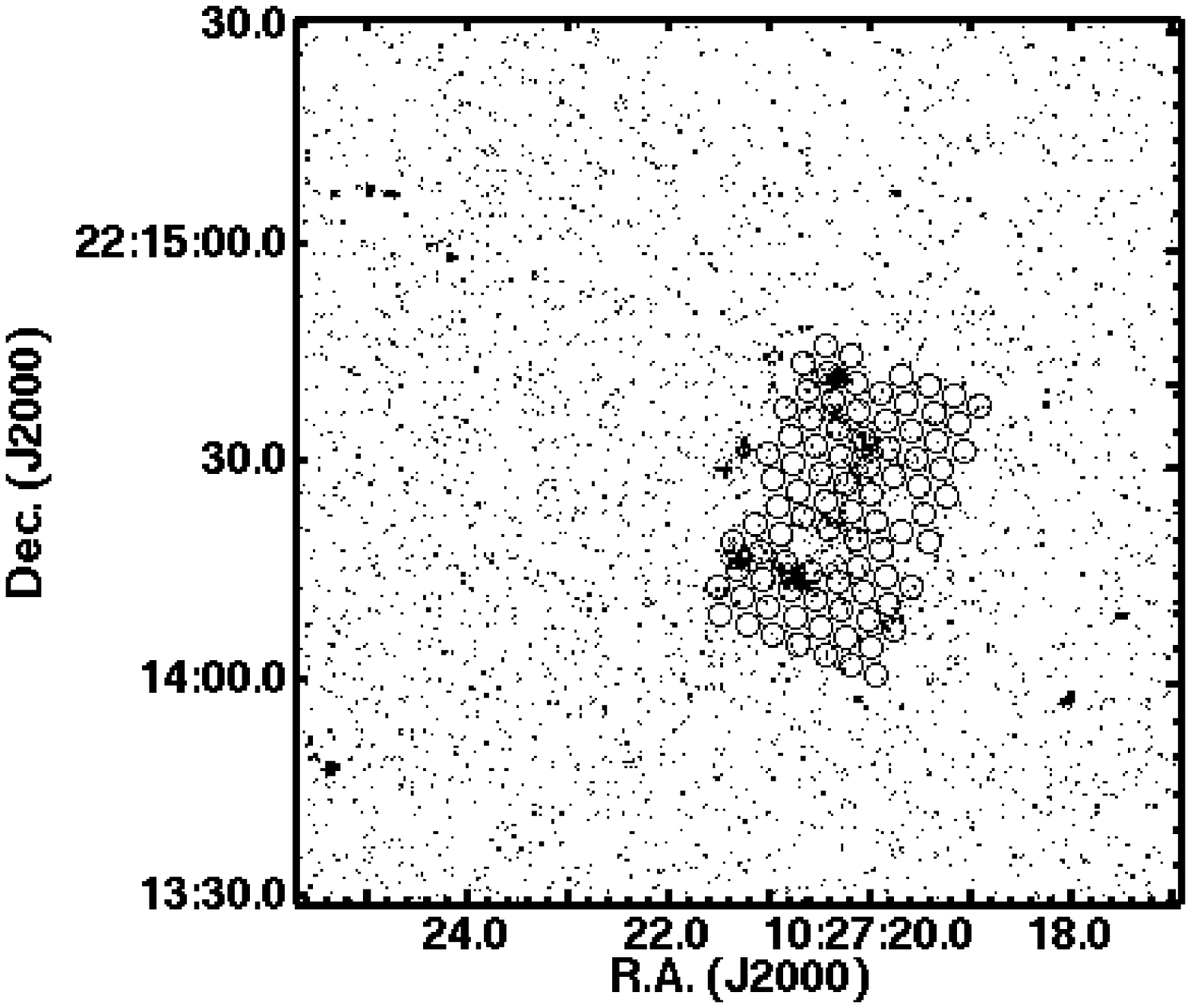}{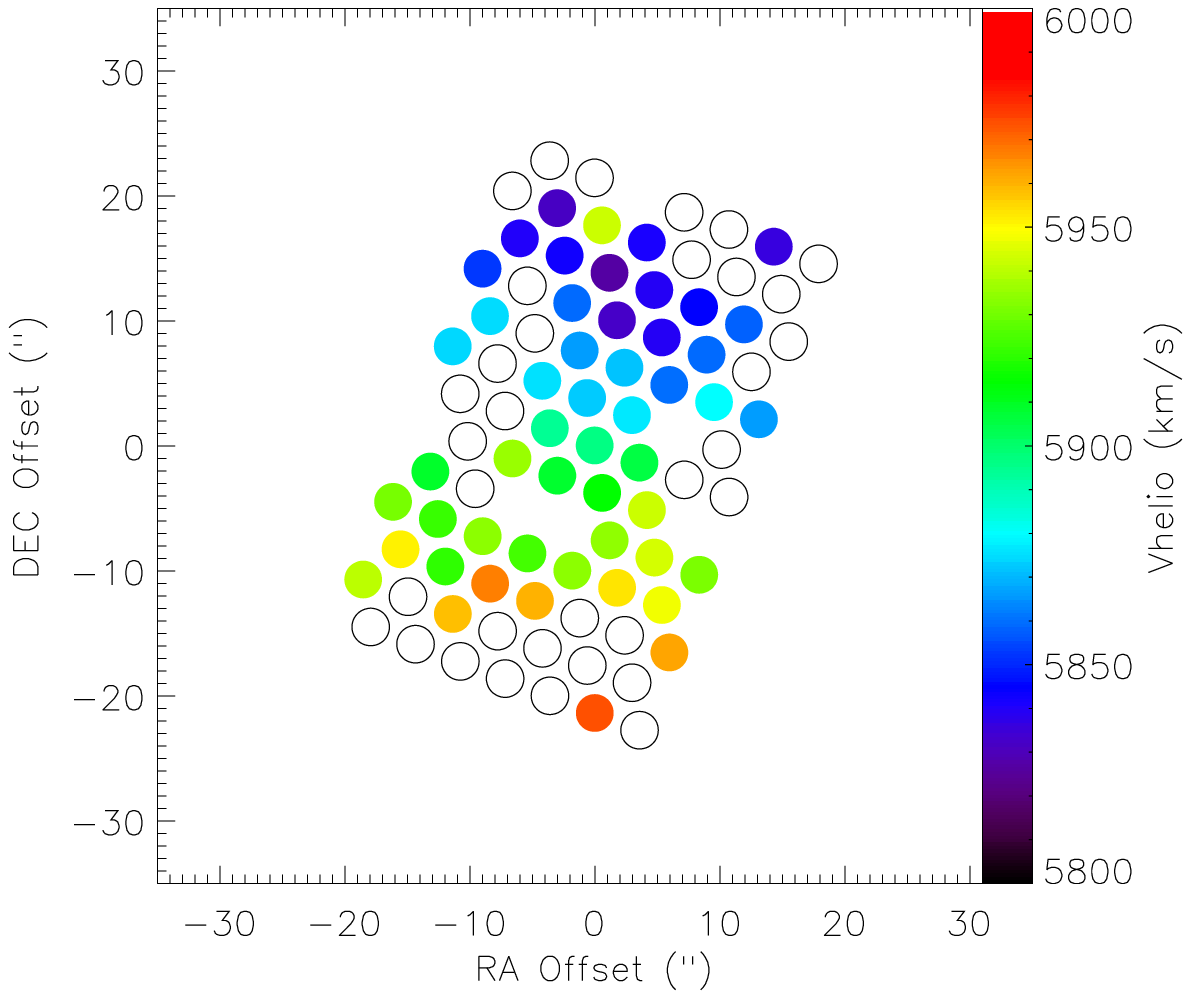}\\
\plottwo{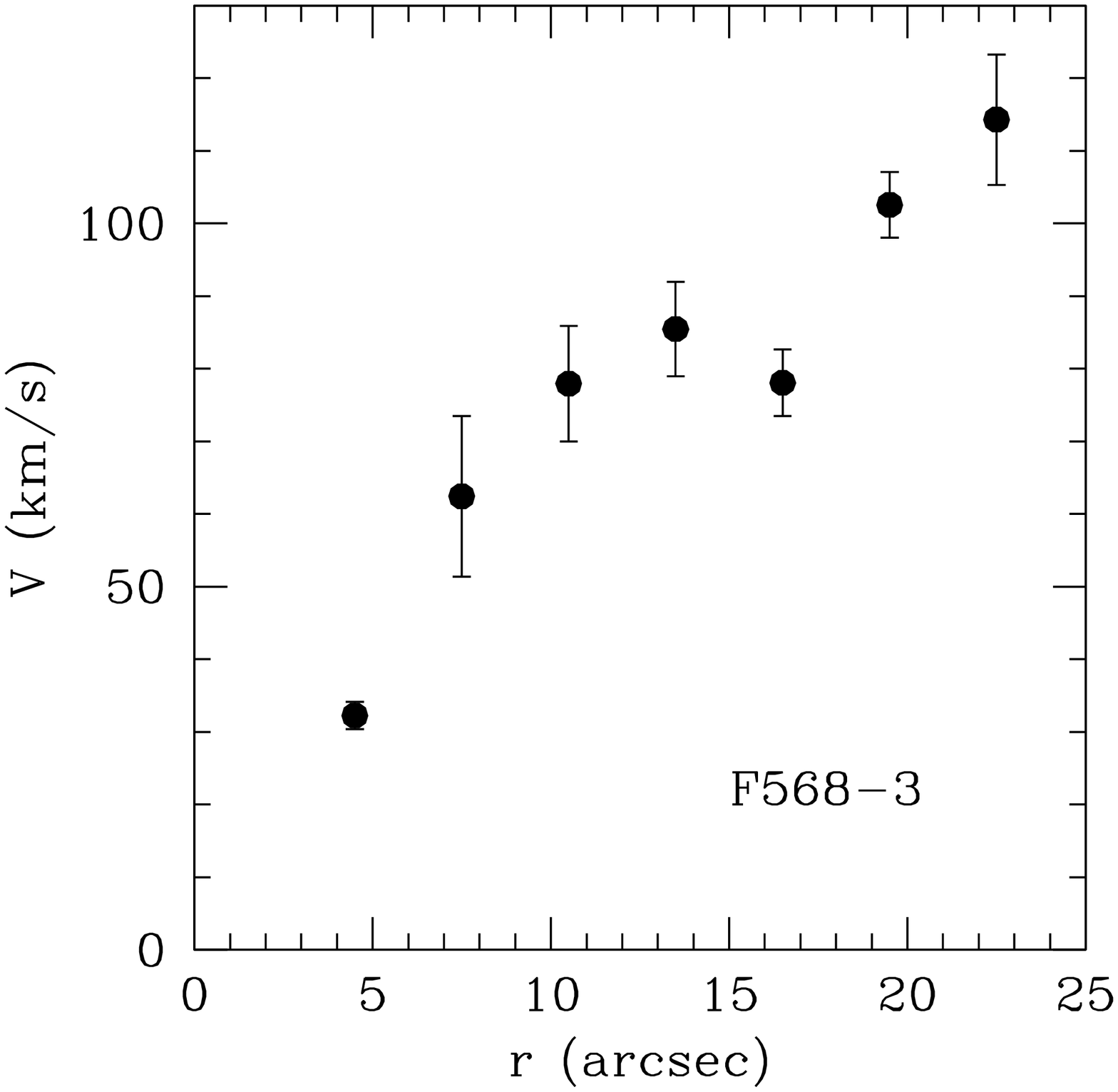}{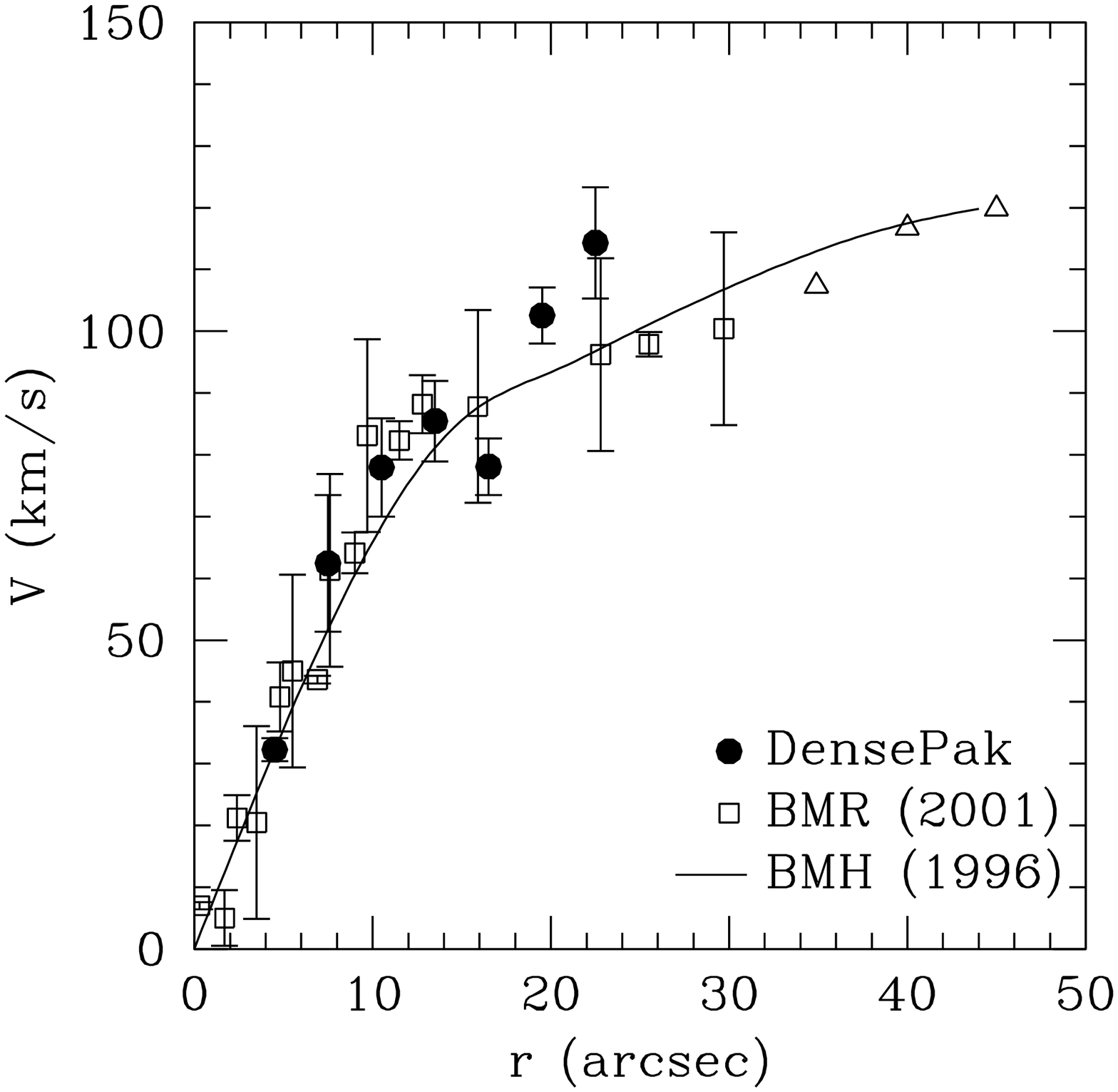}
\caption{Results for F568-3: {\it(Upper left)} Position of DensePak 
  array on an \Ha\ image of the galaxy. {\it(Upper right)} Observed
  \Dpak\ velocity field.  The empty fibers are those without detections.
  {\it(Lower left)} DensePak rotation curve. 
  {\it(Lower right)} DensePak rotation curve plotted with the raw 
  long-slit \Ha\ rotation curve of \citet{dBMR} and the \HI\ rotation 
  curve of \citet{DMV}.  The triangles represent the \HI\ points used in
  the halo fits.  Figure appears in color on-line.}
\end{figure*}
\begin{figure*}
\plottwo{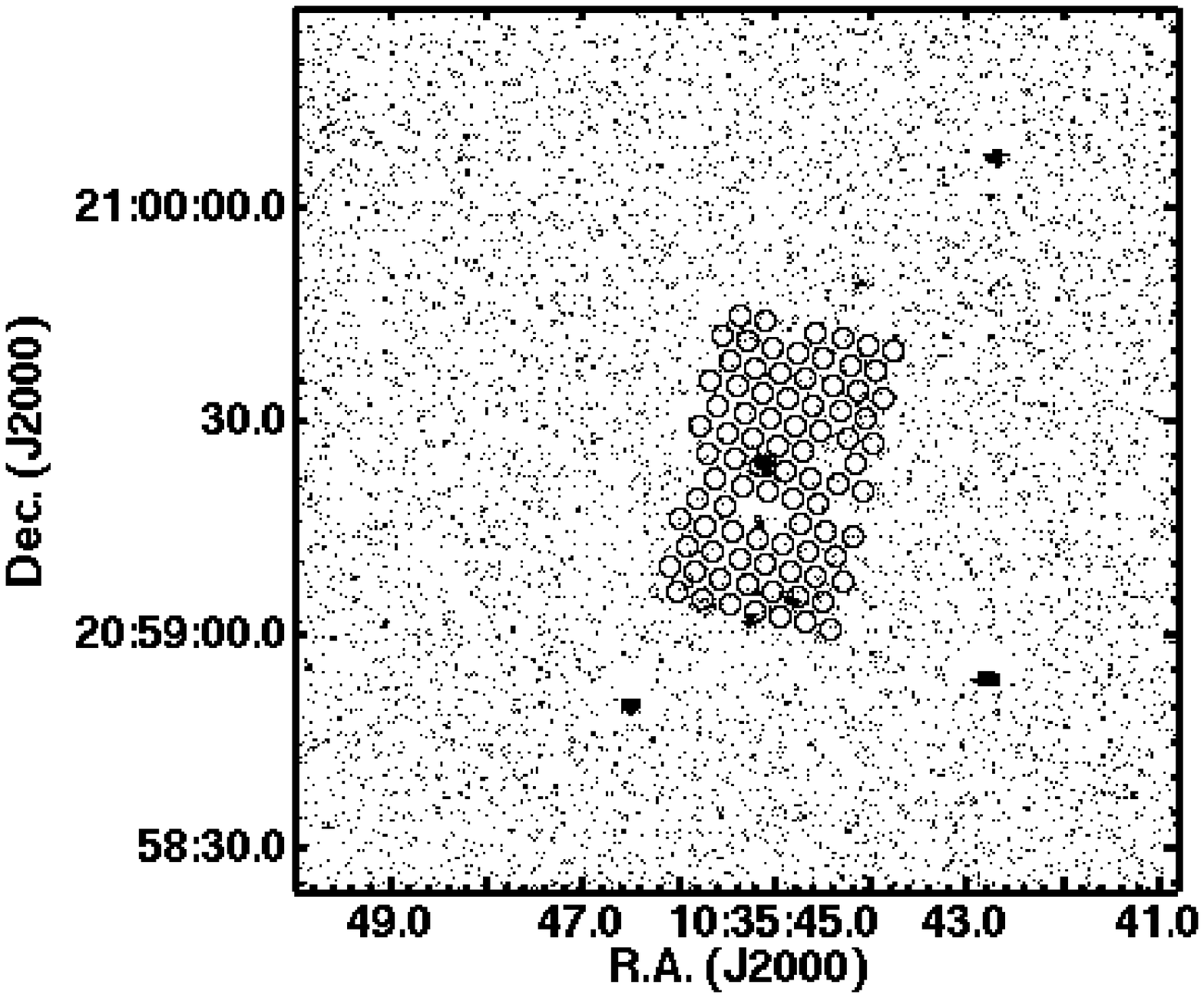}{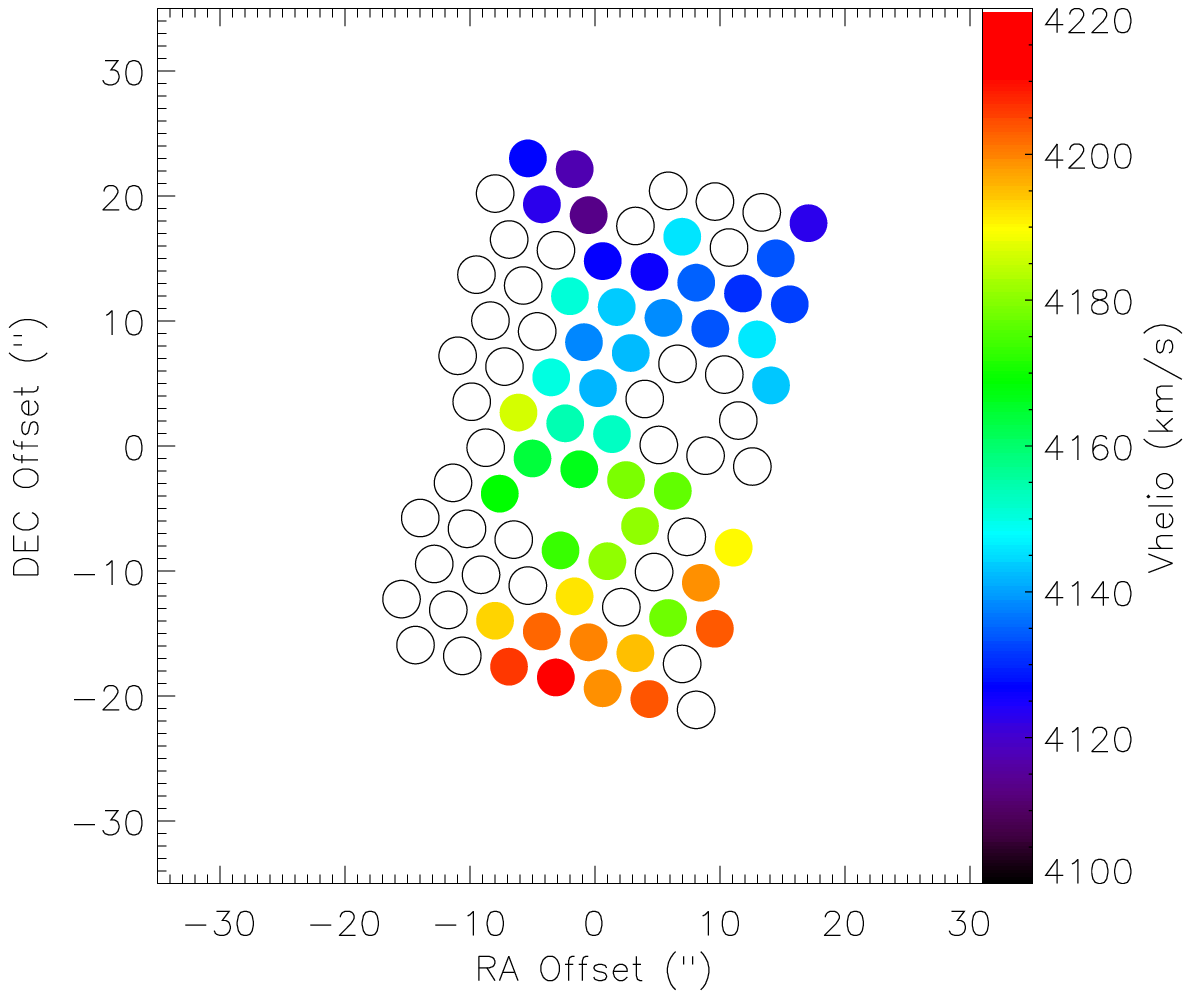}\\
\plottwo{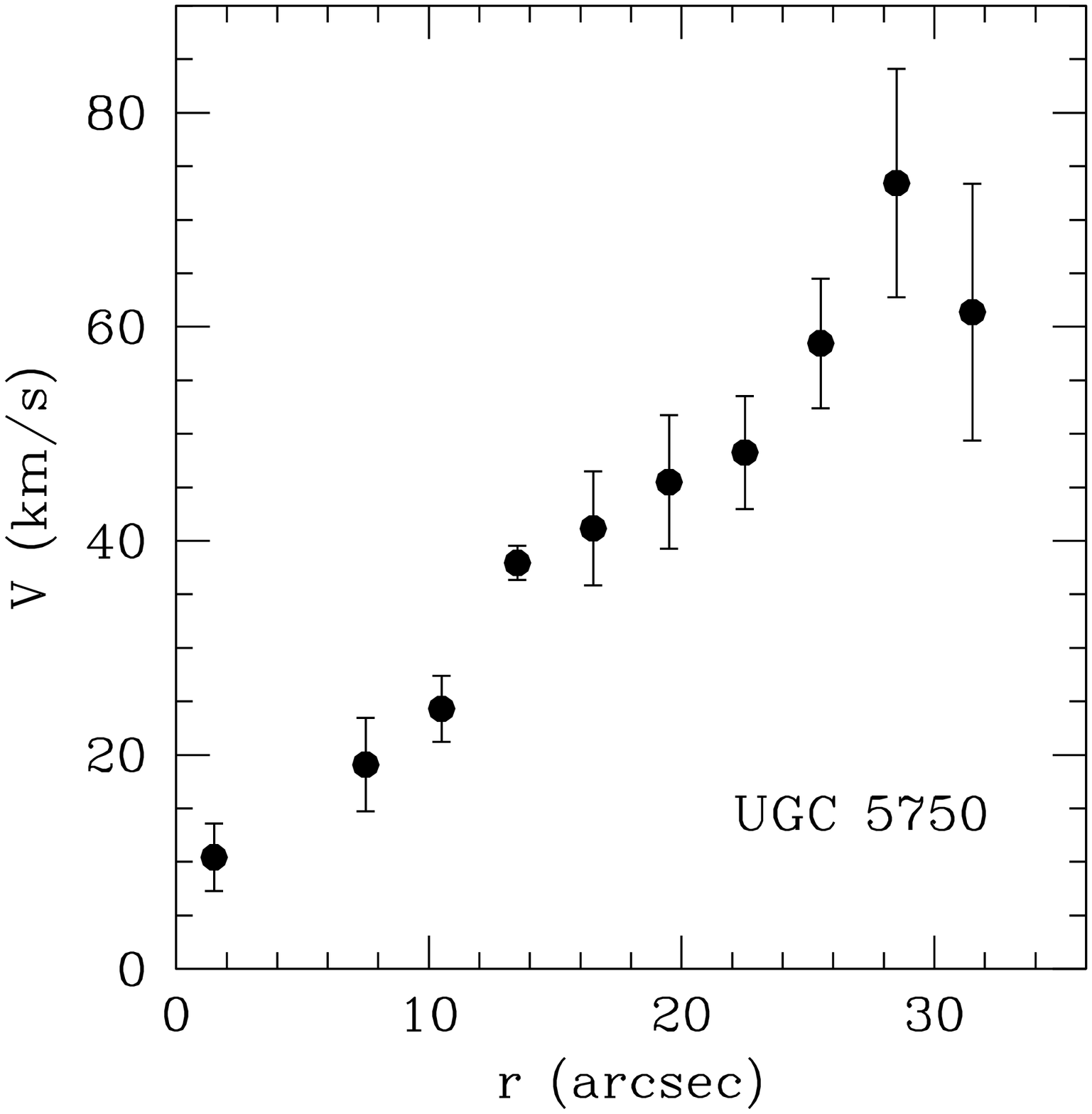}{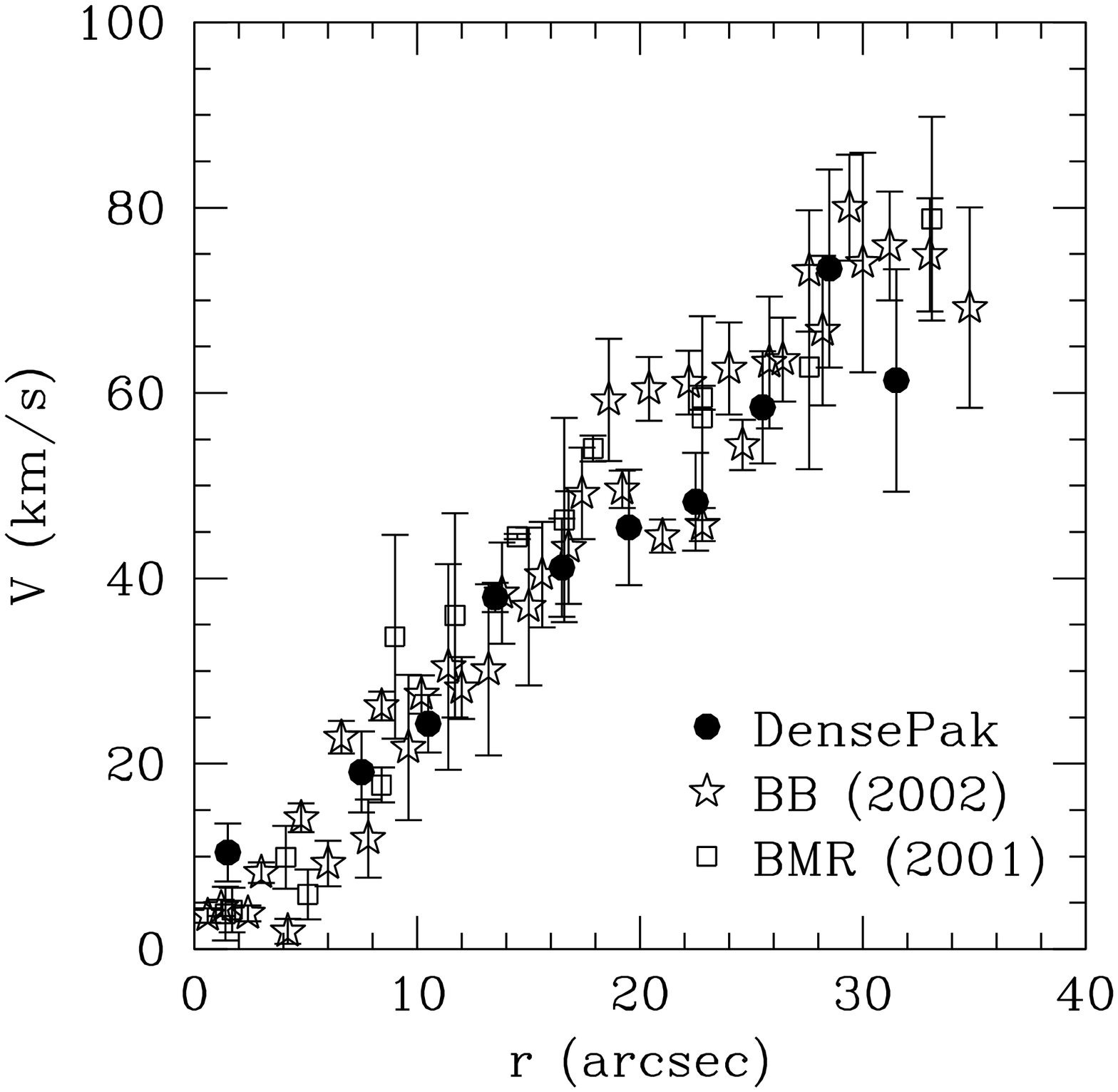}
\caption{Results for UGC 5750: {\it(Upper left)} Position of DensePak 
  array on an \Ha\ image of the galaxy. {\it(Upper right)} Observed 
  \Dpak\ velocity field.  Empty fibers are those without detections. 
  {\it(Lower left)} DensePak rotation curve. {\it(Lower right)}
  DensePak rotation curve plotted with the raw long-slit \Ha\ rotation 
  curves of \citet{dBB} and \citet{dBMR}.  Figure appears in color on-line.}
\end{figure*}
\begin{figure*}
\plottwo{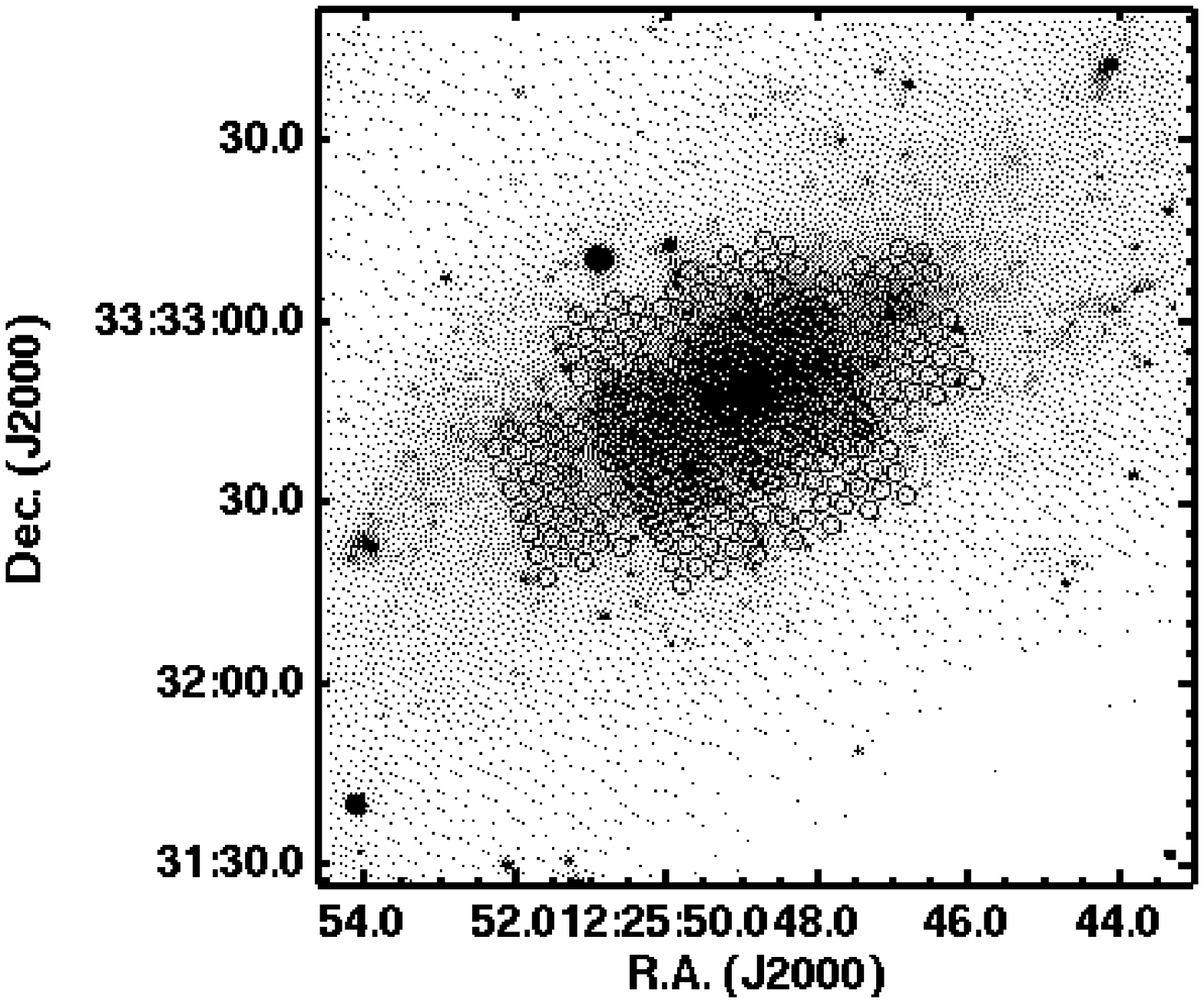}{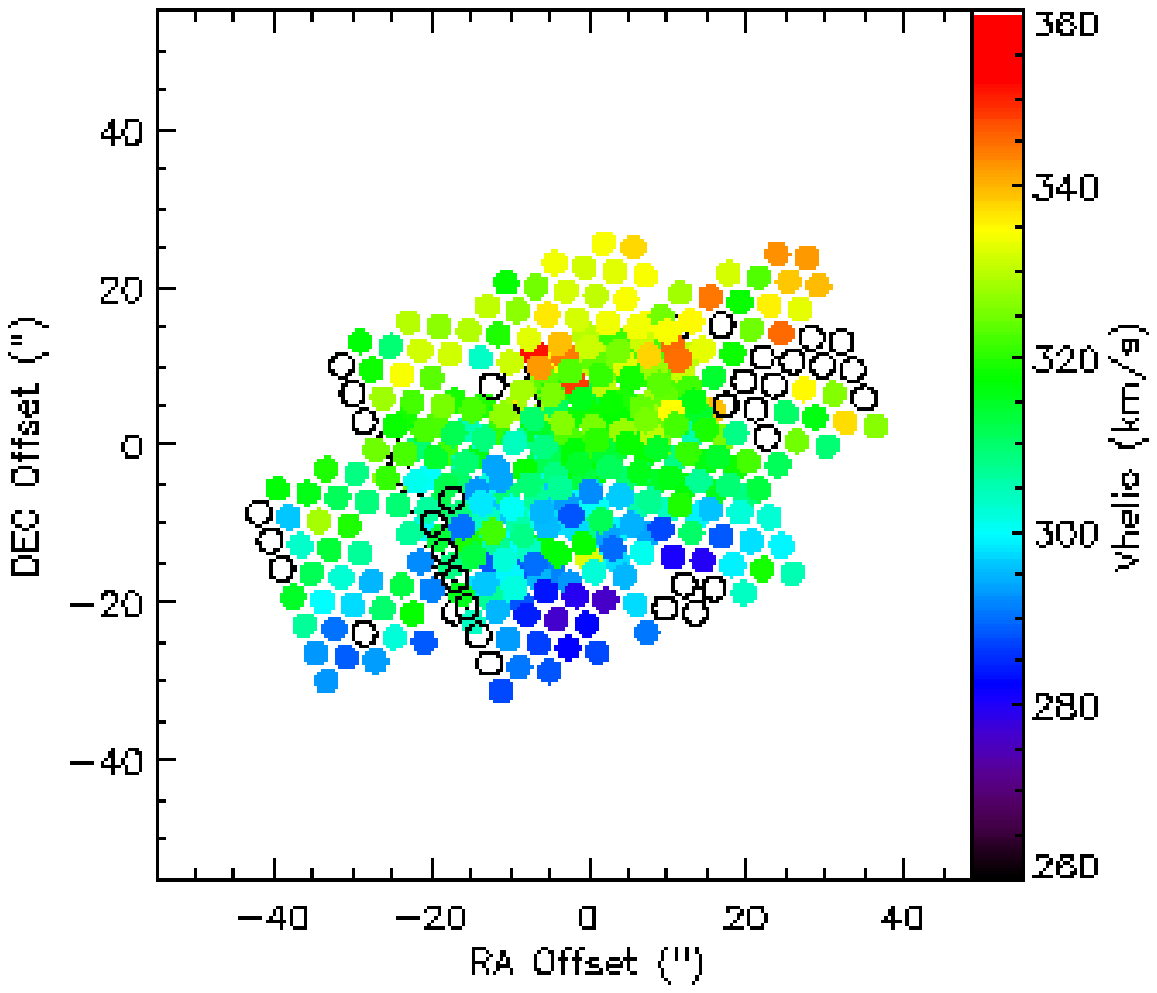}\\
\plottwo{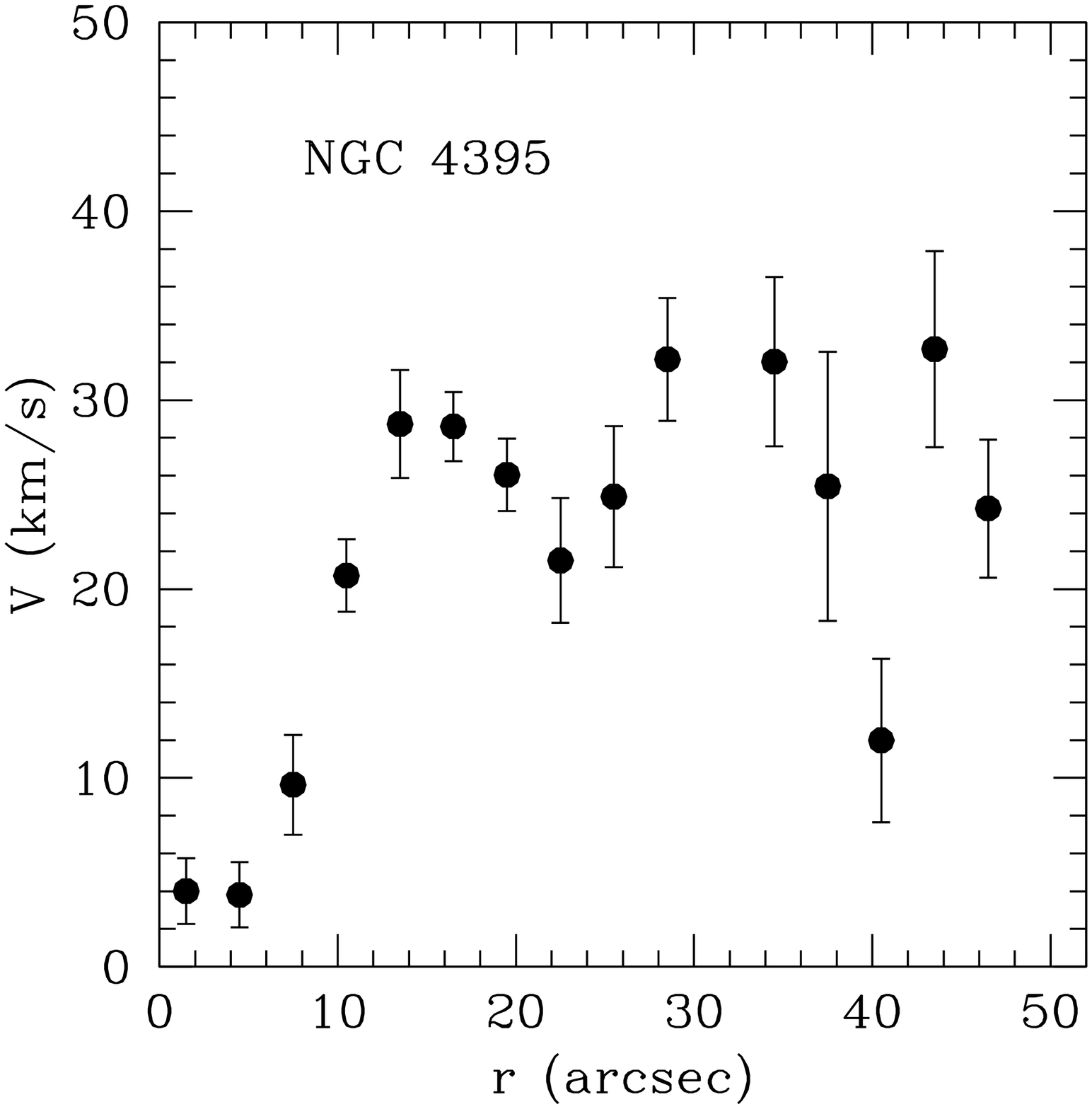}{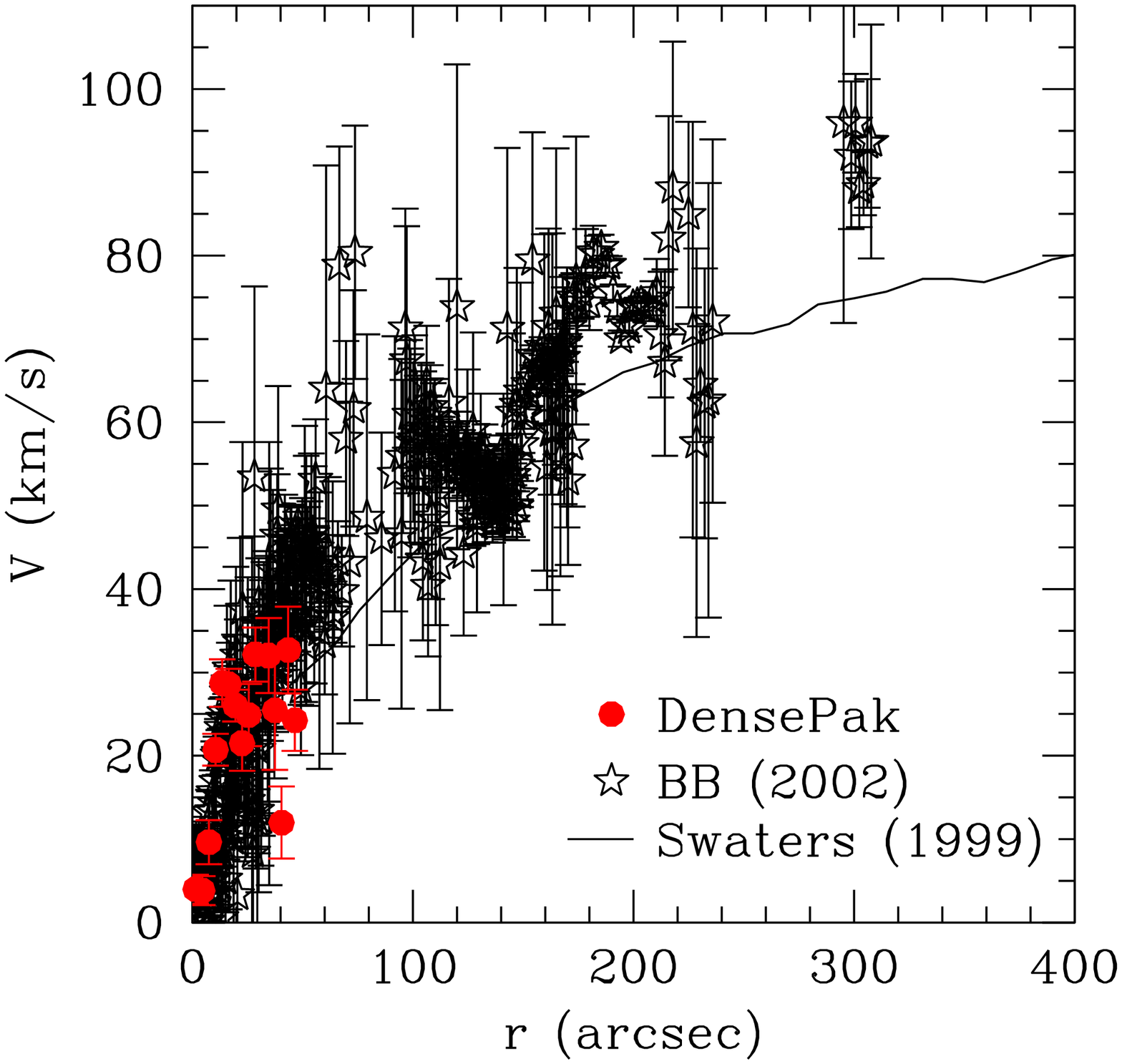}
\caption{Results for NGC 4395: {\it(Upper left)} Position of DensePak 
  array on an $R$-band image of the galaxy. {\it(Upper right)}
  Observed \Dpak\ velocity field.  Empty fibers are those without
  detections.
  {\it(Lower left)} DensePak rotation curve. {\it(Lower right)}
  DensePak rotation curve plotted with the raw long-slit \Ha\ rotation 
  curve of \citet{dBB} and the \HI\ rotation curve of 
  \citet{Swatersthesis}.  The \HI\ points used in the halo fits extend beyond
  the radial range of this plot.  Figure appears in color on-line.}
\end{figure*}
\begin{figure*}
\plottwo{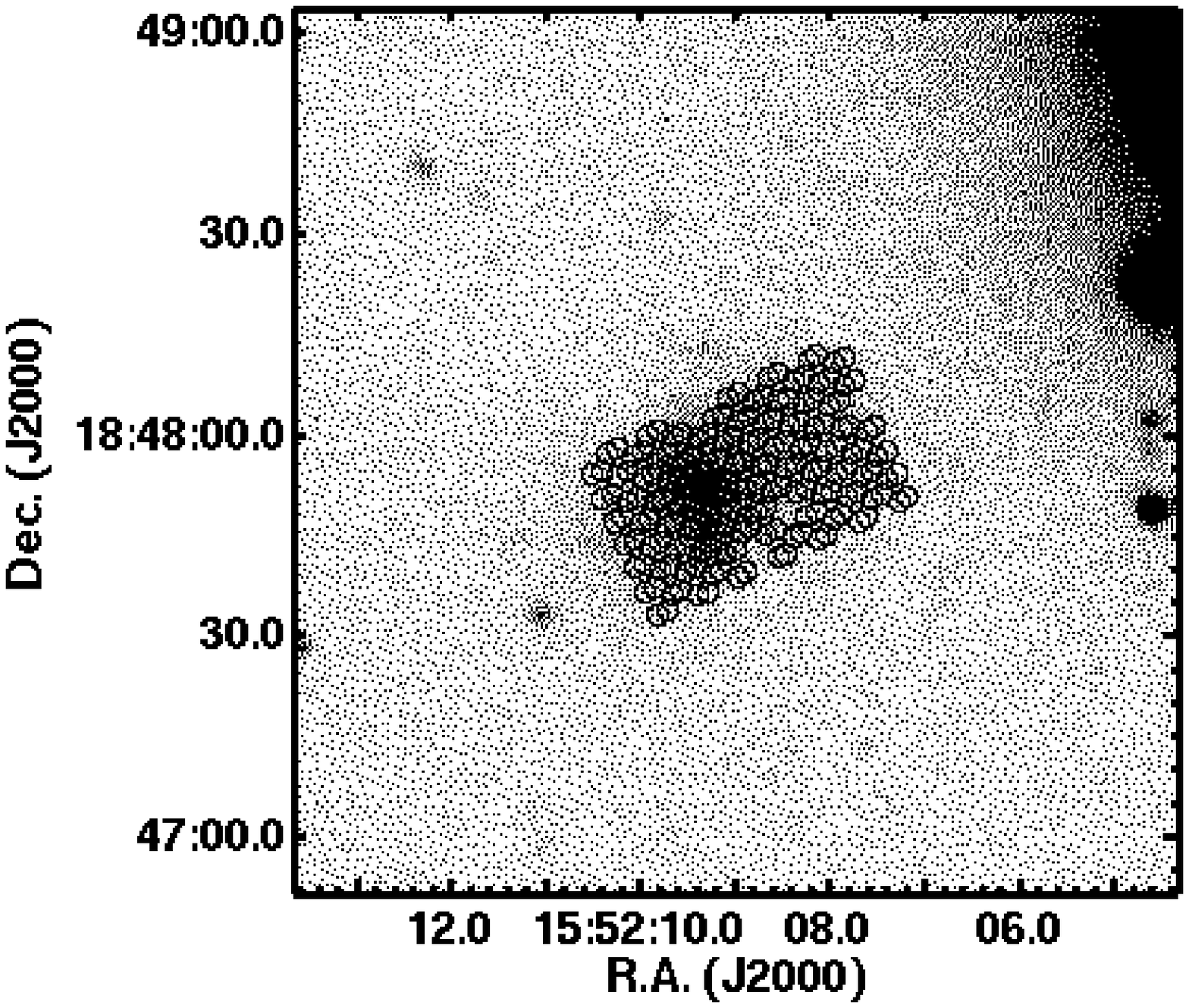}{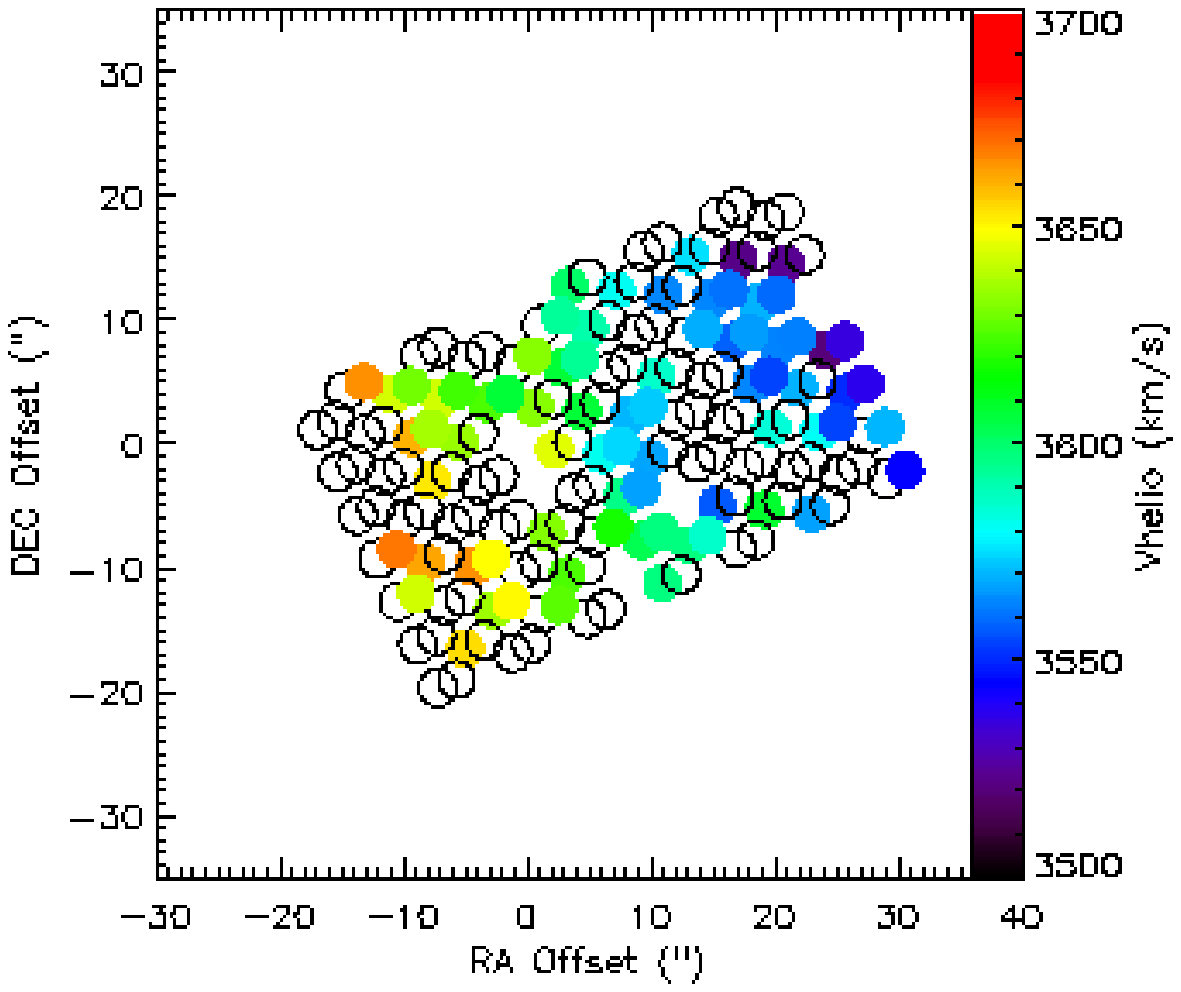}\\
\plottwo{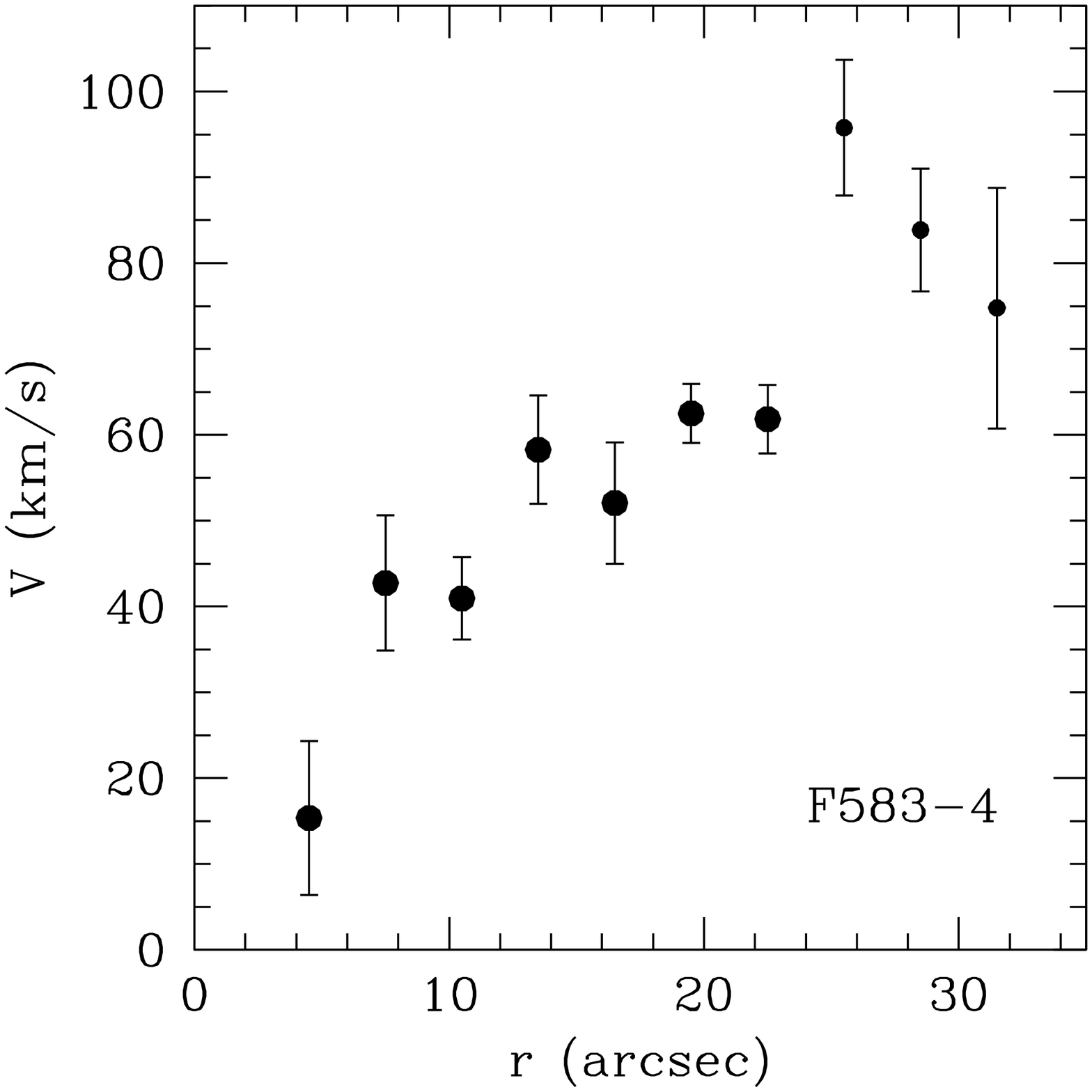}{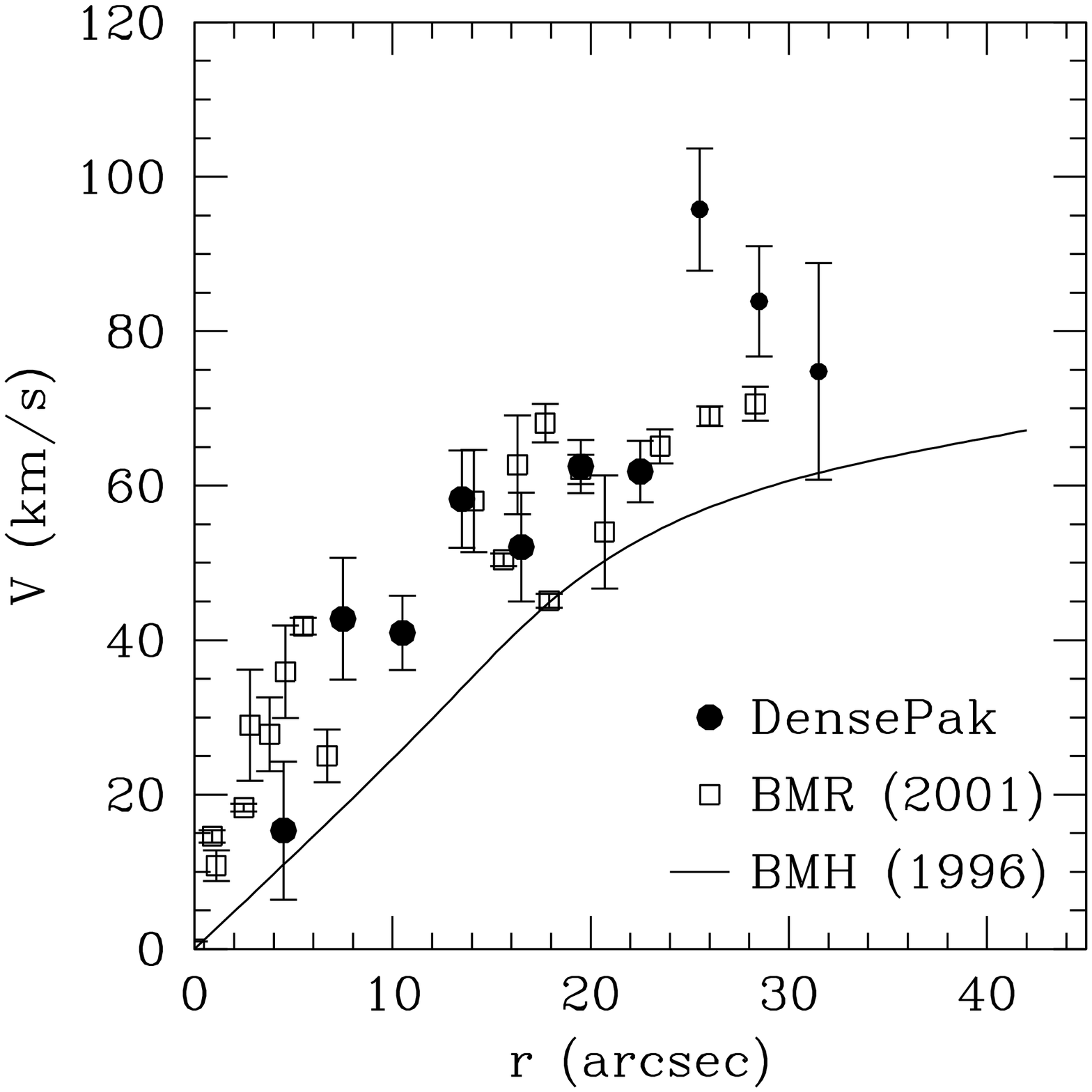}
\caption{Results for F583-4: {\it(Upper left)} Position of DensePak 
  array on an $R$-band image of the galaxy. {\it(Upper right)}
  Observed \Dpak\ velocity field.  Empty fibers are those without
  detections.
  {\it(Lower left)} DensePak rotation curve.  The last three points
  were excluded from the halo fits. {\it(Lower right)}
  DensePak rotation curve plotted with the raw long-slit \Ha\ rotation 
  curve of \citet{dBMR} and the \HI\ rotation curve of \citet{DMV}.  
  The \HI\ data was not included in the halo fits.  Figure appears in 
  color on-line.}
\end{figure*}
\begin{figure*}
\plottwo{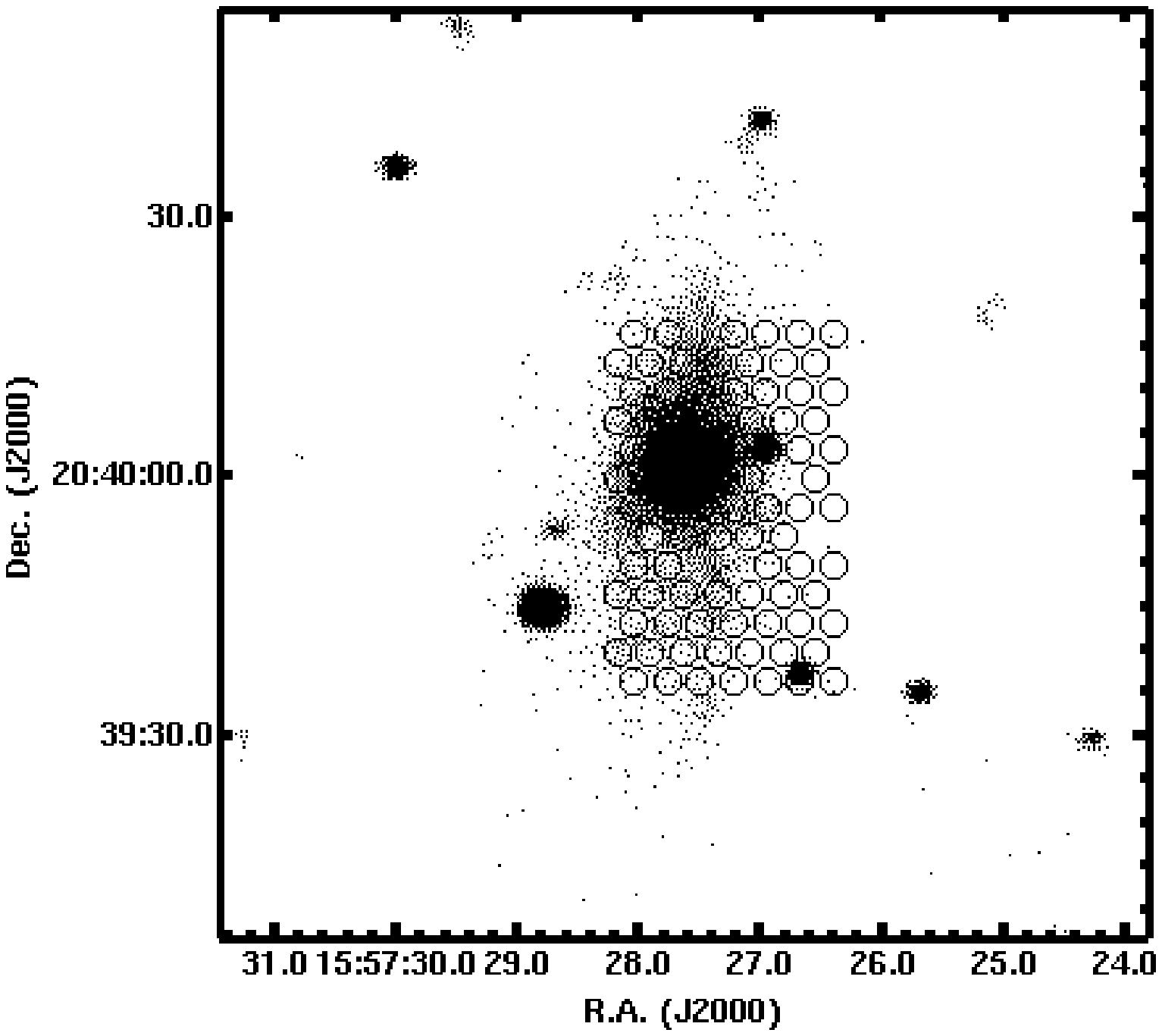}{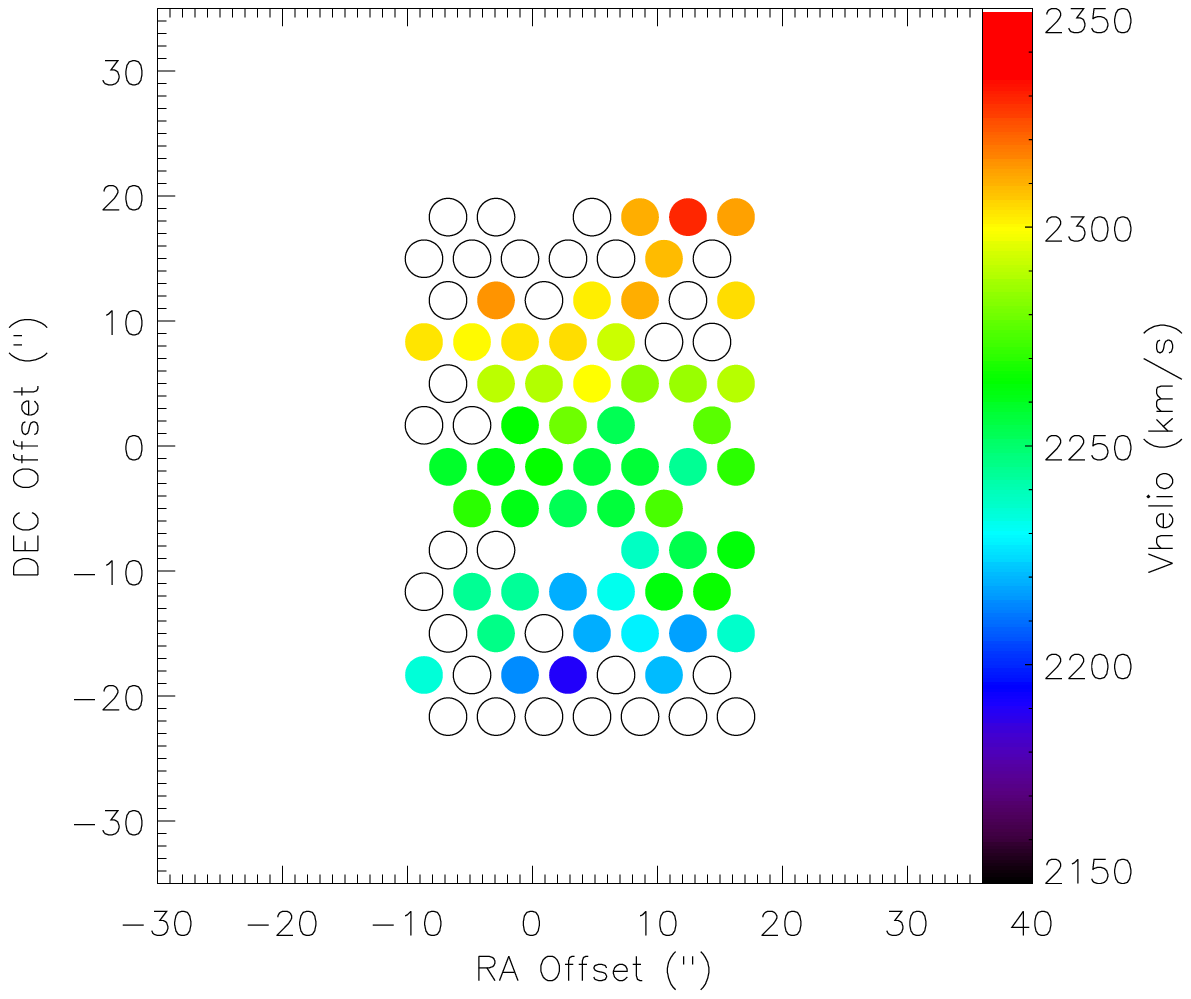}\\
\plottwo{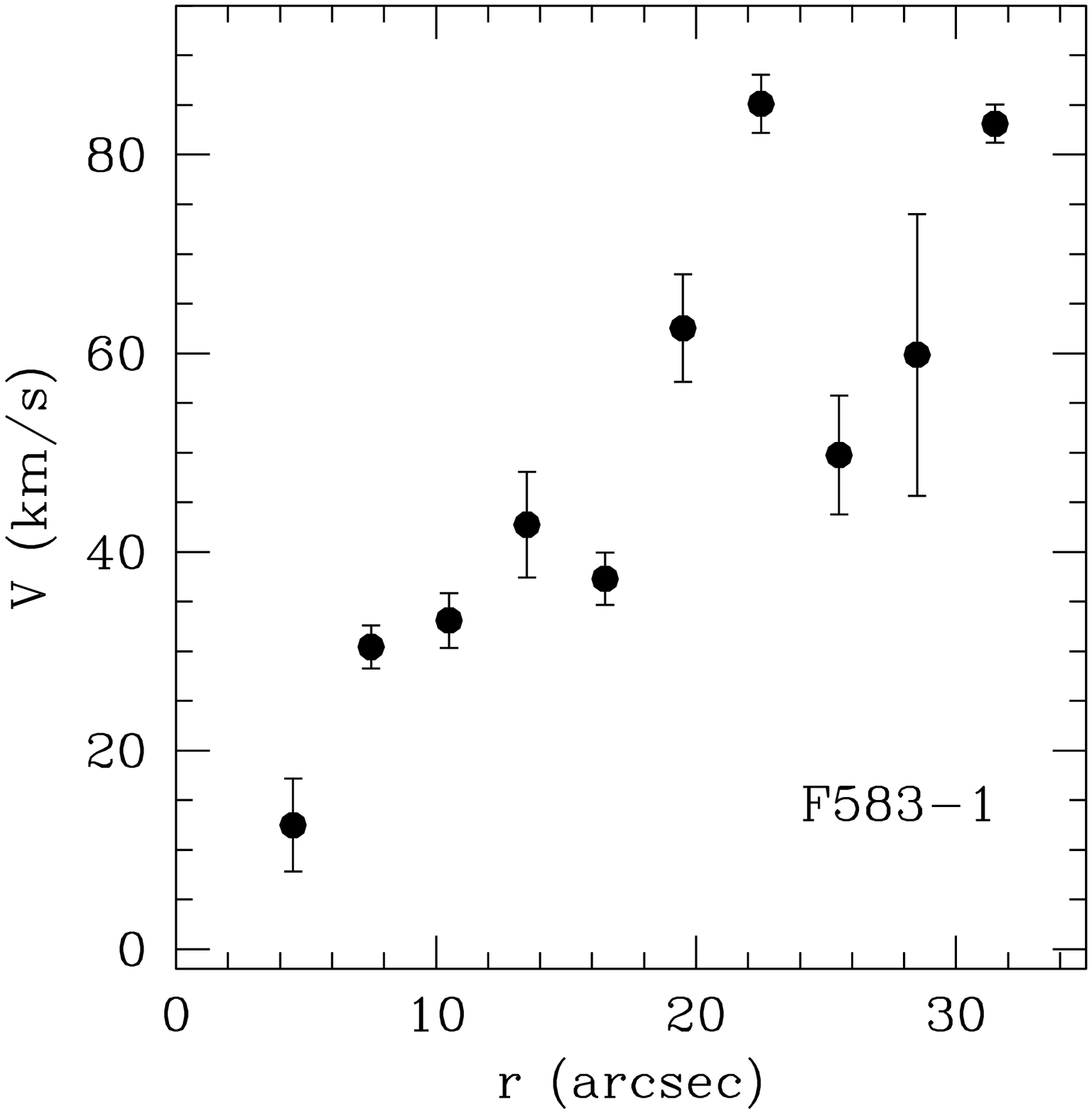}{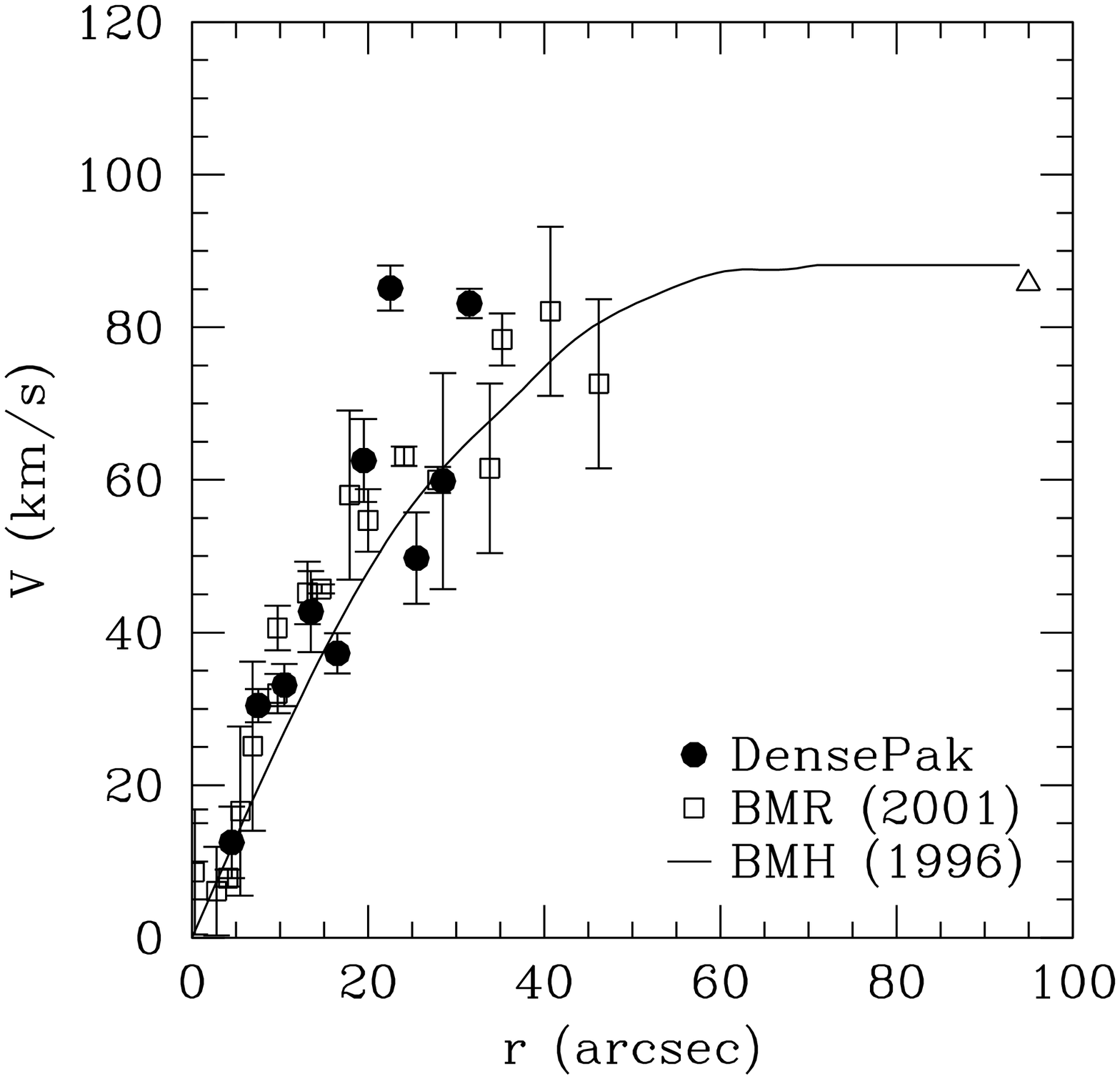}
\caption{Results for F583-1: {\it(Upper left)} Position of DensePak 
  array on a $R$-band image of the galaxy. {\it(Upper right)} 
  Observed \Dpak\ velocity field.  Empty fibers are those without
  detections.
  {\it(Lower left)} DensePak rotation curve. {\it(Lower right)} 
  DensePak rotation curve plotted with the raw long-slit \Ha\ rotation 
  curve of \citet{dBMR} and the \HI\ rotation curve of \citet{DMV}.  
  The triangle represents the \HI\ point used in the halo fits.  
  Figure appears in color on-line.}
\end{figure*}
\begin{figure*}
\plottwo{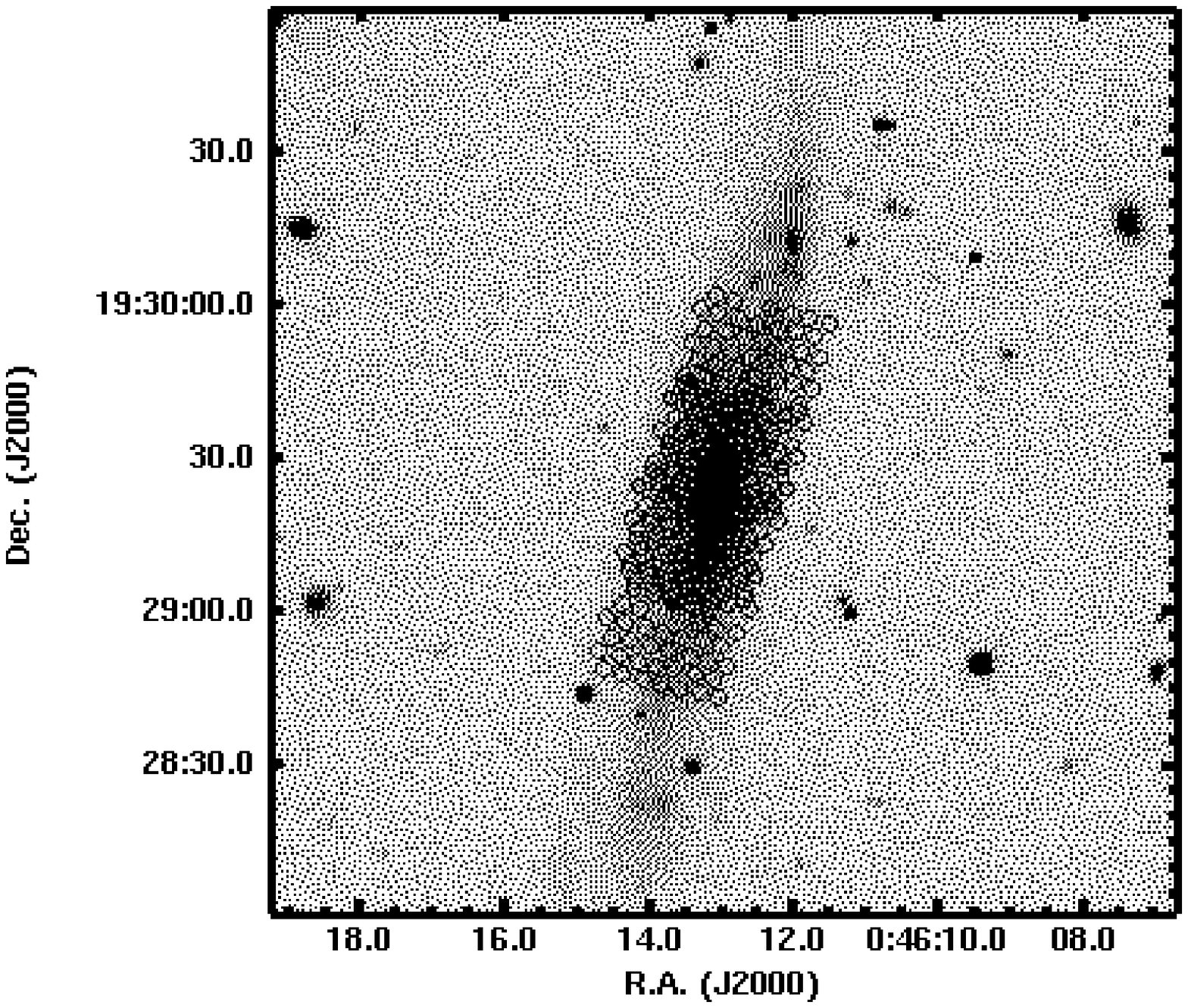}{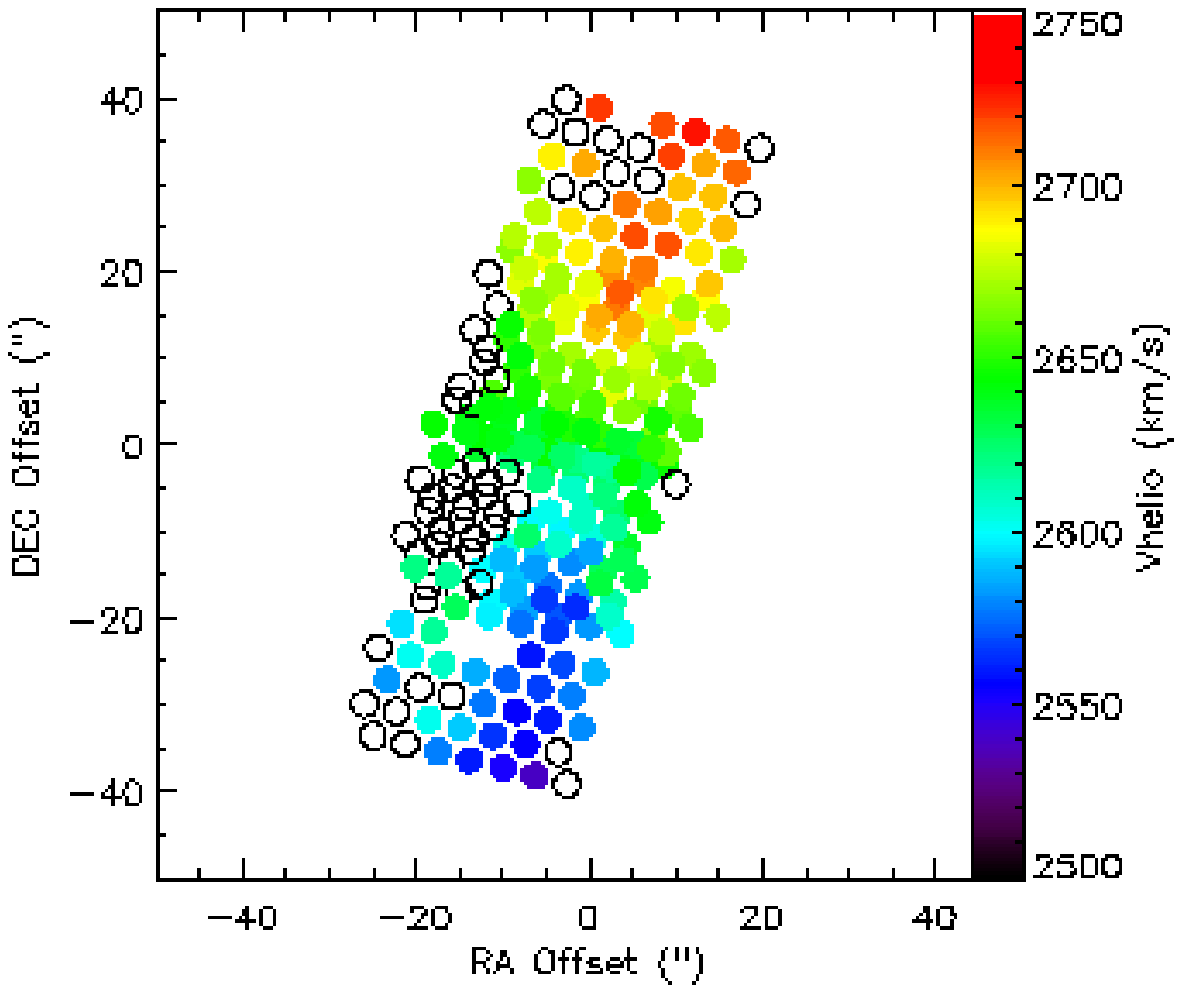}\\
\includegraphics[scale=0.4]{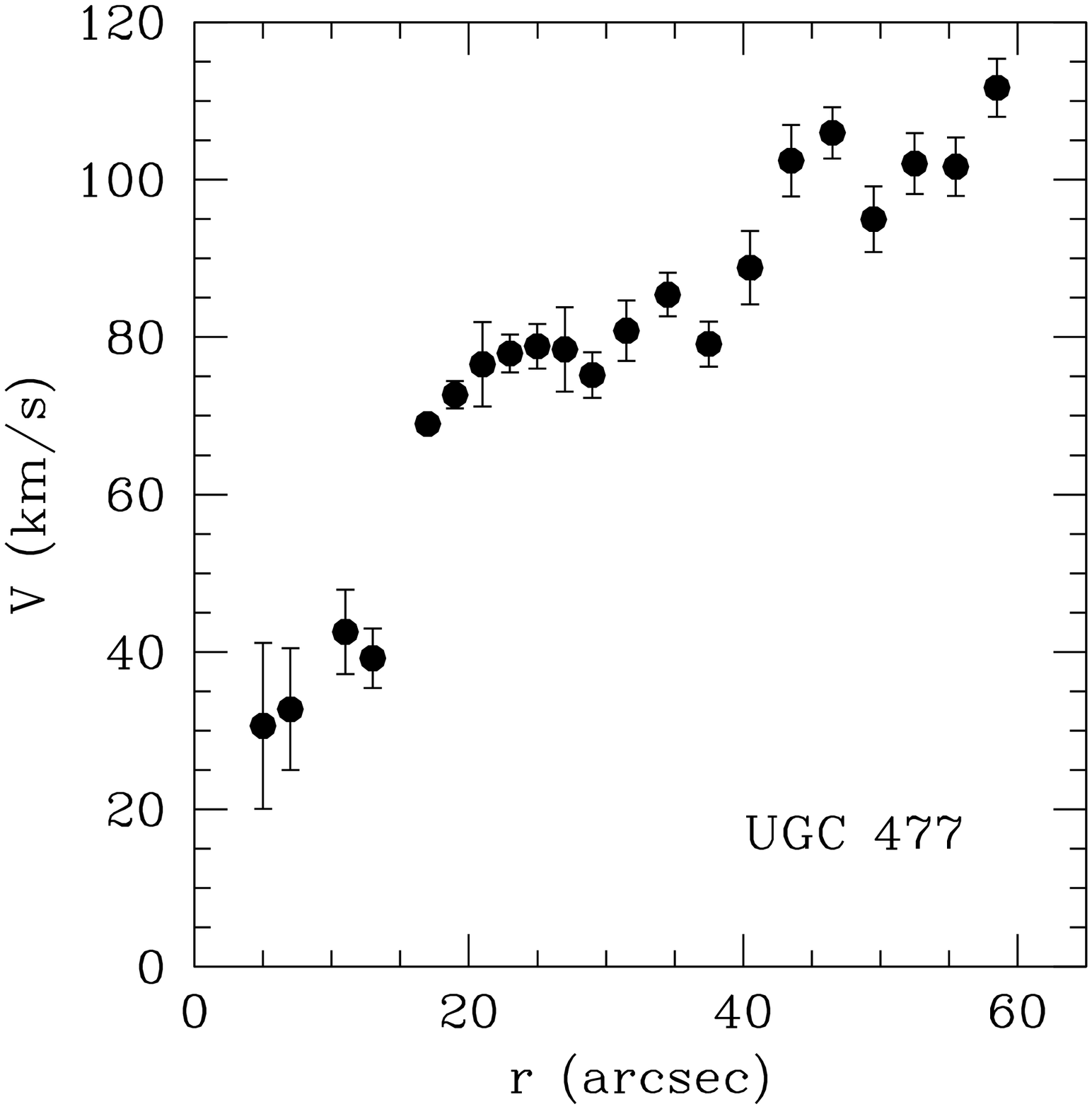}\\
\caption{Results for UGC 477: {\it(Upper left)} Position of DensePak 
  array on a $R$-band image of the galaxy. {\it(Upper right)} 
  Observed \Dpak\ velocity field.  Empty fibers are those without
  detections.
  {\it(Lower left)} DensePak rotation curve.  Figure appears in color 
  on-line. }
\end{figure*}
\begin{figure*}
\plottwo{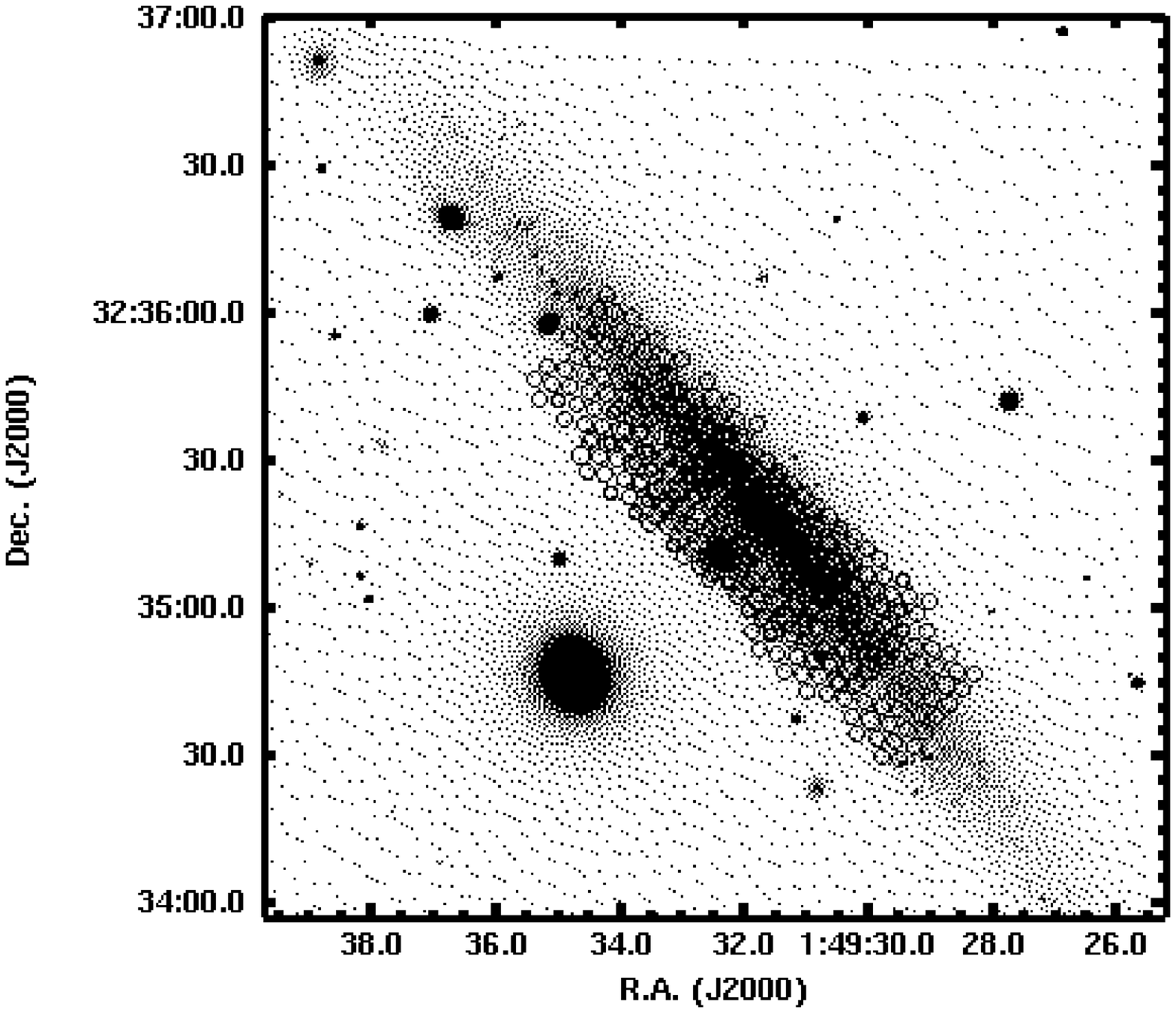}{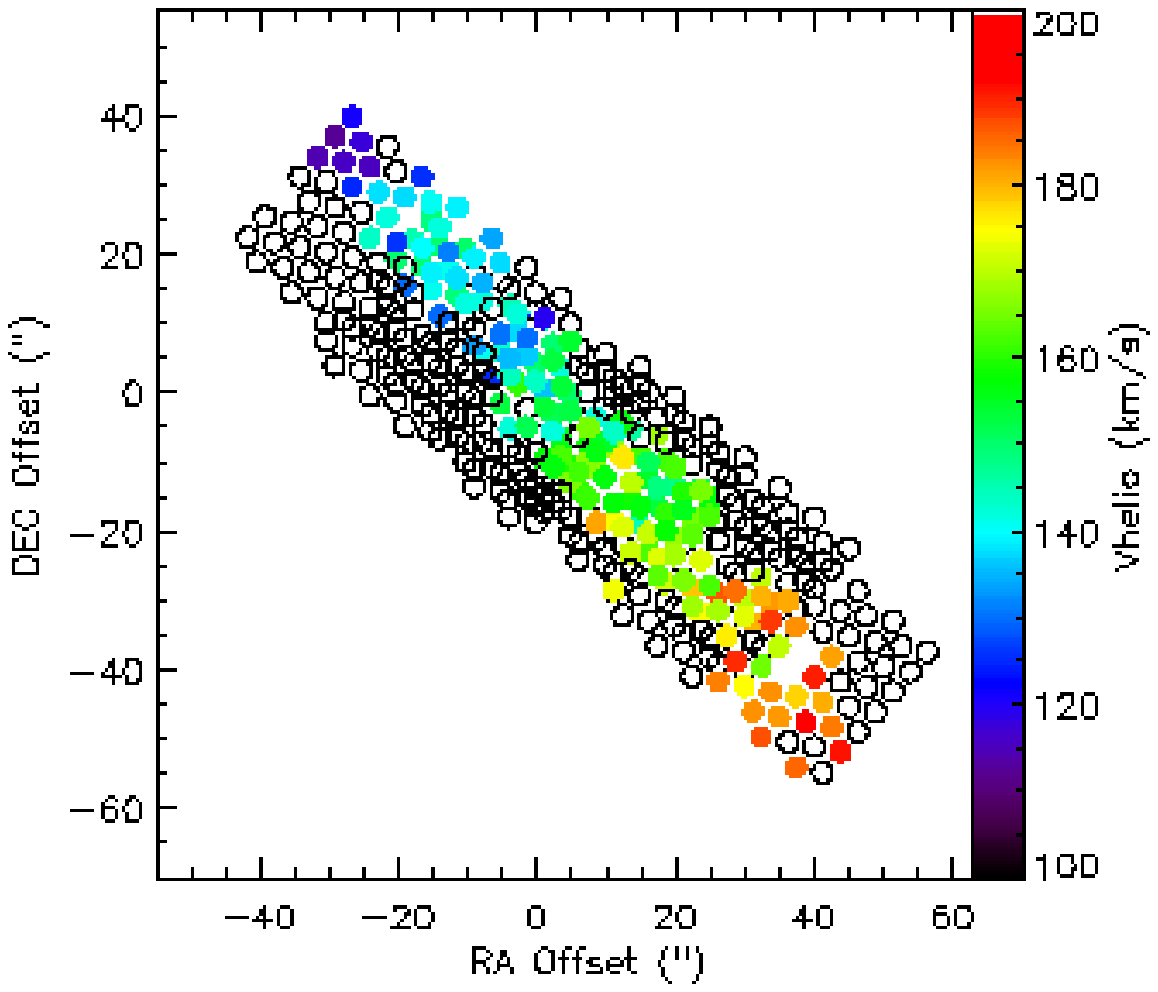}\\
\plottwo{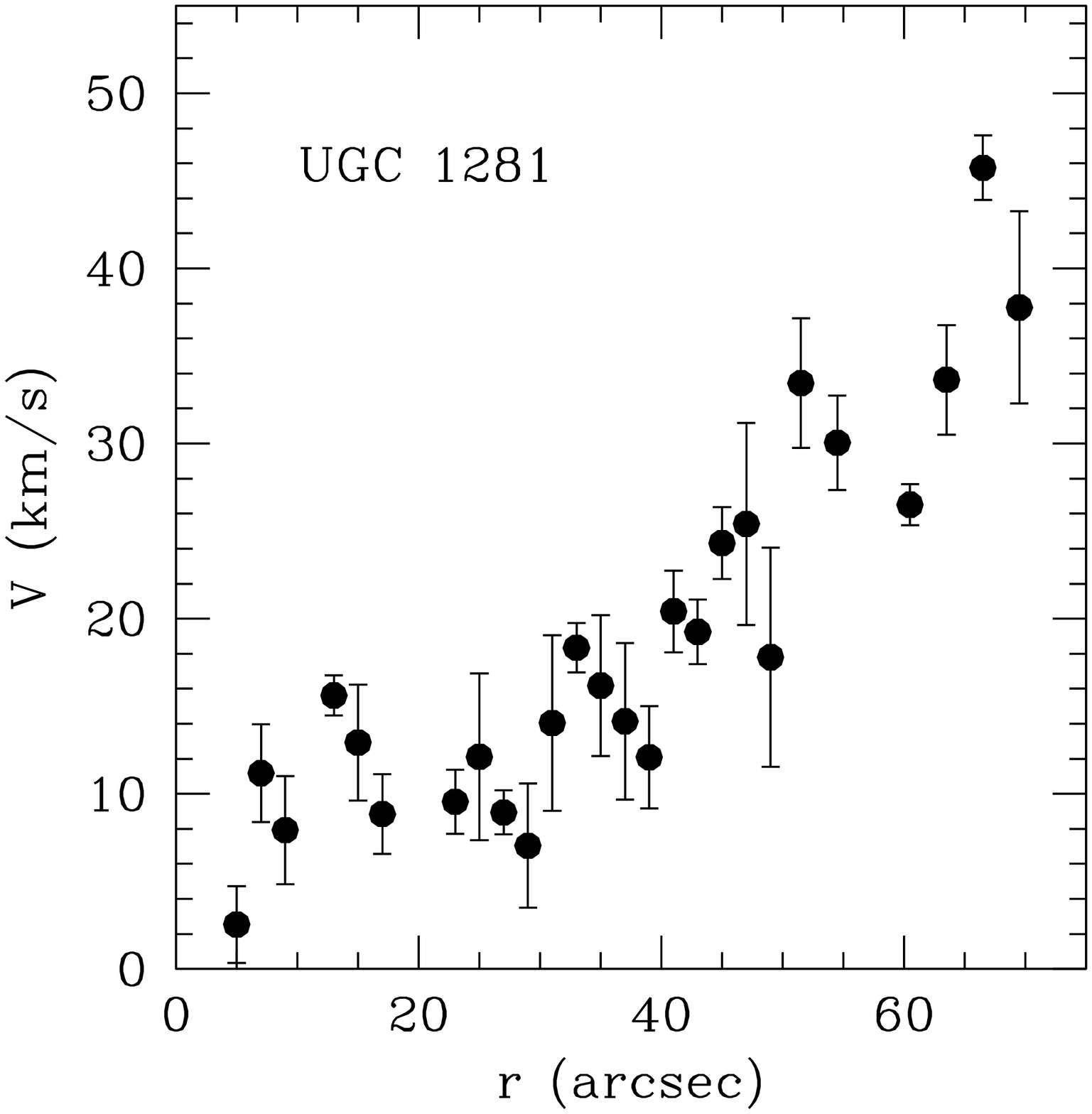}{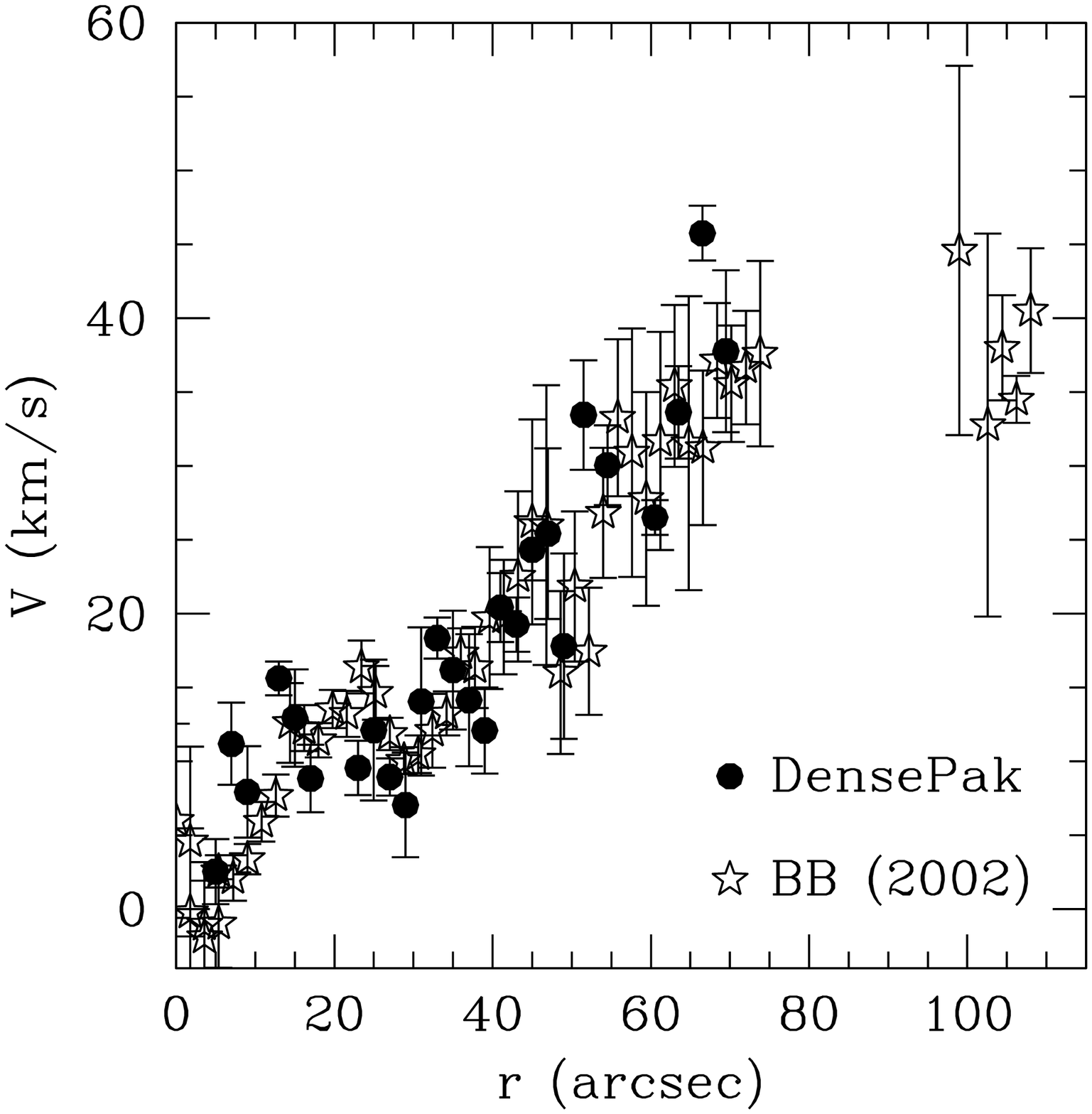}
\caption{Results for UGC 1281: {\it(Upper left)} Position of DensePak 
  array on a $R$-band image of the galaxy. {\it(Upper right)} 
  Observed \Dpak\ velocity field. Empty fibers are those without
  detections.
  {\it(Lower left)} DensePak rotation curve. {\it(Lower right)} 
  DensePak rotation curve plotted with the raw long-slit \Ha\ rotation 
  curve of \citet{dBB}.  Figure appears in color on-line.}
\end{figure*}

\section{Data Reduction}

The observations were reduced in IRAF \footnote{IRAF is distributed by 
the National Optical Astronomy Observatory, which is operated by the 
Association of Universities for Research in Astronomy (AURA), Inc., 
under agreement with the National Science Foundation.} using the HYDRA 
package.  The data were bias subtracted and flattened.  The IRAF task 
dohydra was used to extract the spectra.  
A wavelength solution created from the observations of the CuAr lamp 
was applied to the spectra.  The two exposures per pointing were 
combined to increase the signal-to-noise and to remove cosmic rays.   
Sky subtraction was not 
performed because the \Dpak\ sky fibers often fell inside the galaxies 
rather than on the sky and were therefore contaminated by galactic
emission.  We made use of the night sky emission lines by using them 
as the reference wavelengths \citep{Osterbrock} by which the
velocities of the galactic emission lines were measured.  We also
tried using the CuAr calibration to measure the velocities, but the 
night sky lines gave cleaner results.  Velocities were measured by 
fitting Gaussians to both the sky lines and the four galactic emission 
lines of interest: \Ha, \nii, \siis\ and \siio.  The average error on 
individual emission line velocities due to centroiding accuracy was 
roughly 1.5 \kms.  We used the arithmetic mean of the measured
emission line velocities in each fiber as the fiber velocity.  The
maximum difference between the measured velocities and the mean was
taken to be the error on the fiber velocity.  Many of these errors
were less than 5 \kms, though a few were as high as $\sim$ 20 \kms.  
If only \Ha\ was observed in a fiber, the observed \Ha\ velocity was 
taken as the fiber velocity and the error was set to 10 \kms.

The observed velocity fields were made by combining the individual 
\Dpak\ pointings using the input shifts at the telescope.  To confirm 
the accuracy of the offsets, an \Ha\ flux image of the galaxy
constructed from the \Dpak\ observations was compared to an actual 
\Ha\ image.  Relative positions between features in the galaxy (i.e., 
bright H{\sc ii} regions) in the two images were measured and
compared.  The fluxes in any overlapping fibers were also compared.  
The accuracy of the fiber positions is $\sim$ 0.6$\arcsec$ and we found 
the telescope to be capable of accurately shifting from the 
nearby bright offset stars (typically 1$\arcmin$ - 2$\arcmin$ shifts), 
as well as the smaller distances ($\sim$ 0.7$\arcsec$\ shifts)
required to observe the spaces between the \Dpak\ fibers.  The
telescope pointing is robust and repeatable.

We used the NEMO \citep{Peter} program ROTCUR \citep{Begeman} to 
derive rotation
curves from our two-dimensional data.  ROTCUR treats the observed 
velocity field as an ensemble of tilted rings and then fits for the 
center, systemic velocity, inclination, position angle and rotation 
velocity in each ring. Specifically, ROTCUR does a non-linear least
squares fit to the following equation:
\begin{equation}
V(x,y) = V_{sys} + V_{rot} cos(\theta) sin(i),
\end{equation}
where
\begin{equation}
cos(\theta) = \frac{-(x-x_{0}) sin(PA) + (y-y_{0}) cos(PA)}{r}
\end{equation}
and
\begin{equation}
r^{2} = (x-x_{0})^{2} + (y-y_{0})^{2}/cos^{2}(i) .
\end{equation}

The observed velocity at position $x,y$, $V(x,y)$, is a function of the 
systemic velocity ($V_{sys}$), the rotation velocity ($V_{rot}$), the 
inclination ($i$), the position of the center of rotation ($x_{0}$,$y_{0}$), 
and the position angle ($PA$) of the major axis.  The position angle is
defined as the angle between north and the receding side of the major 
axis, measured from north through east.  Each point is weighted 
by the inverse square of the error on the fiber velocity times 
cos($\theta$), the angle away from the major axis.

The \Dpak\ data cover the centers of the galaxies.  We find these to be
in the regime of solid-body rotation. Consequently, neither the center 
nor the inclination could be determined by the observations.  The center was 
therefore fixed to the position of the optical center and the inclination 
was fixed to published values  (BMR01; BB02; Tully 1988).  The systemic 
velocities were determined by ROTCUR.  We also used ROTCUR to determine the
position angle of the major axis, using published long-slit values as 
the initial guess.  If the position angle could not be
well-constrained by ROTCUR, then it was fixed to the long-slit value. 
The long-slit position angles are generally from either the \HI\
velocity field or the surface photometry, or in some cases, are the 
position angles indicated in a catalog such as the UGC \citep{Nilson}.  
Only two galaxies, UGC 477 and UGC 1281, had position angles 
well-constrained by ROTCUR.

Ring radii were set to the effective fiber resolution for each galaxy.  
The resulting rotation curve was inspected for rings with considerably 
higher or lower velocities than neighboring rings.  These highly
deviant points in the rotation curve were investigated and were
usually attributable to a single fiber in the ring having an extreme 
velocity. These extreme fibers were 
removed and the rotation curve was recalculated.  Lastly, we added in
quadrature to the errorbars on the final rotation curve the velocity error 
from centroiding accuracy corrected for inclination.  We do not impose
a minimum error on the rotation curve points, as was the case in some
previous works (eg. 5 \kms\ in \citet{Rob} and 4 \kms\ in 
\citetalias{dBMR}). The nature of the errors on the rotation curve points from
the velocity fields is different from the errors on long-slit rotation
curves.  In the long-slit case, the error is on a single velocity 
measurement and gives the accuracy with which a Gaussian could be fitted
to an emission line profile, or the error is given by the consistency of the
observed emission lines (\Ha, \nii, \siis\ and \siio).  The error does not 
contain information about the uncertainty in the rotation velocity.  Errors 
can become arbitrarily small in high signal-to-noise data and is the reason why
minimum errors were imposed.  In two-dimensional data, the rotation velocity 
is obtained from a tilted ring fit.  The error indicates something about the 
spread of the velocities in a ring.  In this case, a smaller error 
indicates that the gas in the ring has smaller non-circular motions.

\section{Results for Individual Galaxies}
In this section we present the \Dpak\ fiber positions, observed
velocity fields and rotation curves in Figures 1-11.  A description is 
given for each galaxy and we compare the rotation curves to previous 
raw long-slit \Ha\ rotation curves as well as \HI\ rotation curves when 
available.  Properties of the galaxies for which rotation curves are 
derived are listed in Table 1; Table 2 lists the observed galaxies for 
which meaningful velocity fields could not be constructed.  As a
result of remarkably poor weather, a number of galaxies have lopsided 
\Dpak\ coverage.

$\textbf{UGC 4325:}$  This galaxy is large on the sky and  four \Dpak\ 
pointings were made on the approaching  side.  The positions are shown 
on the \Ha\ image from \citet{vanzee}.  The fiber velocities were the 
average of the \Ha\ and \siis\ lines.  \Ha\ is prevalent and was
detected in almost every fiber. The position angle was fixed at the 
average of the position angles of previous long-slit observations 
(BB02; Swaters et al. 2003a).  There is excellent agreement between 
the \Dpak\ rotation curve and the long-slit \Ha\ rotation curve by 
\citetalias{dBB}.  Very similar results have also been obtained with 
SparsePak (Swaters 2005, private communication).  The \HI\ data 
\citep{Swatersthesis} lie above the optical data at the inner radii. This
 suggests that the \HI\ data have been over-corrected for beam smearing 
\citepalias{dBB}.  The outer \HI\ points are a bit lower than the optical 
data, but are within 1$\sigma$.  The decline in the rotation curve at 
the outer radii may be a real feature \citep{Bosma04}.

$\textbf{F563-V2:}$ There was one pointing for this galaxy.  The 
\Dpak\ fibers are shown on the \Ha\ image from \citet{ssmpix}.  The fiber 
velocities were the average of the \Ha, \siis\ and \siio\ lines.
There is little \Ha\ emission and only roughly half of the fibers have 
an \Ha\ detection.  The position angle was fixed to the value used in 
\citet{Rob}. The \Dpak\ rotation curve generally agrees with the \HI\ 
rotation curve of \citetalias{DMV}.

$\textbf{F563-1:}$ There was one pointing for this galaxy.  The fiber 
positions are shown on an \Ha\ image (W.J.G. de Blok 2005, private 
communication). The fiber velocities were the average of the \Ha,
\siis\ and \siio\ lines. The fibers were offset toward the receding 
side of the galaxy, and there was very little emission in the fibers 
on the approaching side.  The position angle was fixed to the value in 
\citetalias{MRdB} and \citetalias{dBB}.  When compared to the
long-slit \Ha\ rotation curves of \citetalias{dBMR} and
\citetalias{dBB}, there is good agreement up to  15\arcsec.  Beyond
that, however, the long-slit curves and the \HI\ curve
\citepalias{DMV} turn over and flatten out, while the \Dpak\ curve 
continues to rise.  Additional \Dpak\ coverage would be useful in 
determining where the \Dpak\ curve turns over.

$\textbf{DDO 64:}$ There were three pointings of the center and 
approaching side of this galaxy.  The \Dpak\ fiber positions are 
shown on a Digitized Sky Survey \footnote{The Digitized Sky Surveys 
were produced at the Space Telescope Science Institute under U.S. 
Government grant NAG W-2166. The images of these surveys are based on 
photographic data obtained using the Oschin Schmidt Telescope on 
Palomar Mountain and the UK Schmidt Telescope. The plates were
processed into the present compressed digital form with the 
permission of these institutions.} image.  Fiber velocities were the 
average of the \Ha\ and \siis\ lines.  The amount of emission in this 
galaxy is very high and nearly all fibers detected emission.  The 
position angle was fixed to the \citetalias{dBB} value. 
The \Dpak\ points  rise and fall in a pattern similar
to the long-slit data \citepalias{dBB}.  The structure in the rotation
curve appears to be real.  That the inner two points fall
below the long-slit data may suggest the presence of non-circular motions
in the inner regions.  The  \HI\ data \citep{Stil} lies slightly 
below, but within the errors of, the optical data.  It should be noted, 
however, that over the radial range plotted in Figure 4 there are only 2 
\HI\ points. 

$\textbf{F568-3:}$ There was one pointing for this galaxy and the 
fiber positions are shown on an \Ha\ image (W.J.G. de Blok 2005, 
private communication).   The fiber velocities were the average of 
the \Ha\ and the \nii\ lines.  \Ha\ emission was detected in roughly 
60\%\ of the fibers.  The position angle was fixed to the 
\citetalias{MRdB} and \citet{Rob} long-slit value.   The \Dpak\ 
rotation curve is  consistent with the long-slit \Ha\ curve of 
\citetalias{dBMR} and the \HI\ curve of \citetalias{DMV}.

$\textbf{UGC 5750:}$ There was one pointing for this galaxy.  The 
fiber positions are shown on an \Ha\ image (W.J.G. de Blok 2005, 
private communication).  The average of the \Ha, \nii, \siis\ and
\siio\ lines was taken as the fiber velocity.  \Ha\ emission was
sparse in this galaxy and only roughly half of the fibers had a 
detection.  The position angle was fixed to the value listed in 
\citetalias{MRdB} and \citetalias{dBB}.  There is good agreement 
with the long-slit \Ha\ rotation curves of both \citetalias{dBB} 
and \citetalias{dBMR}.

$\textbf{NGC 4395:}$ There were five pointings for this galaxy and 
the positions of the \Dpak\ fibers are shown on an $R$-band image 
(W.J.G. de Blok 2005, private communication).  The \Ha\ line 
overlapped slightly with a sky line.  This was not a serious problem 
for this galaxy because the emission lines (particularly the \Ha\
line) were very strong.  The \Ha\ line was measured in a fiber if the 
\nii, \siis\ and \siio\ lines were visible and strong and if the \Ha\ 
line was stronger than the neighboring sky lines.  With these
criteria, \Ha\ was detected in nearly all the fibers.  The fiber 
velocities were the average of the \Ha\ and \siis\ lines.  The 
position angle of \citetalias{dBB} was used as the \Dpak\ position
angle.  The \Dpak\ rotation curve is consistent with the long-slit
\Ha\ curve of \citetalias{dBB} as well as the \HI\ curve 
\citep{Swatersthesis} at the innermost radii. At the outer radii the 
\HI\ curve falls below the optical data at about the 2$\sigma$ level.

$\textbf{F583-4:}$ There were two pointings for this galaxy: a central 
pointing and an interstitial pointing; the fiber positions are shown
on the $R$-band image from \citetalias{DMV}.  The fiber velocities are 
the averages of the \Ha, \siis\ and \siio\ lines, and  the approaching 
side of the galaxy has slightly better coverage  than the receding
side.  The position angle was fixed to the value in \citetalias{MRdB}.  
The \Dpak\ rotation curve is generally consistent with  the long-slit 
\Ha\ rotation curve of \citetalias{dBMR}.  Both optical rotation
curves are offset to noticeably higher velocities than the \HI\ rotation 
curve of \citetalias{DMV}. This is one of the few galaxies where 
beam smearing in the \HI\ data turned out to be important 
\citepalias{dBMR}. 

$\textbf{F583-1:}$ There is one pointing for this galaxy.  The \Dpak\ 
fibers are shown on the $R$-band image of the galaxy from
\citetalias{DMV}.  The fiber velocities are the averages of the \Ha, 
\siis\ and \siio\ lines, and more of the approaching side of the
galaxy is seen than the receding side.  The \citetalias{MRdB} position 
angle was taken as the \Dpak\ position angle.  The \Dpak\ rotation
curve is consistent with both the long-slit \Ha\ rotation curve of 
\citetalias{dBMR} and the \HI\ rotation curve of \citetalias{DMV}.

$\textbf{UGC 477:}$ There were three interstitial pointings along the 
length of this galaxy and spatial coverage of the galaxy was optimal.  
The \Dpak\ fibers are shown on the $R$-band image obtained at the KPNO 
2m telescope.  Fiber velocities were the average of the H$\alpha$,
\nii, \siis\ and \siio\ lines.  The amount of emission was very high
in this galaxy and almost every fiber had a detection.  The position 
angle was well-constrained by the data.

$\textbf{UGC 1281:}$ There were five interstitial pointings for this
galaxy.  The amount of emission was sparse as less than 50\%\ of the
fibers had a detection.  The \Dpak\ fibers are shown on the $R$-band
image obtained at the KPNO 2m telescope.  Fiber velocities were the
average of the H$\alpha$, \siis\ and \siio\ lines.  The position angle
was well-constrained by the data and the \Dpak\ rotation curve is
consistent with the long-slit H$\alpha$ rotation curve of
\citetalias{dBB}.  UGC 1281 is a nearly edge-on galaxy.  At high
inclination, line-of-sight integration effects become important and
may cause an intrinsically steeply rising rotation curve to appear
slowly rising.  The shape of the emission line profiles can be used to
constrain the shape of the rotation curve.  If instrumental resolution
is high enough and there is no line-of-sight obscuration, then
symmetric line profiles indicate a solid-body rotation curve and
skewed line profiles indicate a curved, NFW-like rotation curve 
\citep{Kregel}.  The line profiles in the \Dpak\ fibers are symmetric.
UGC 1281 is unlikely to be so optically thick that the observations are
hampered by obscuration \citep{deNaray}.  Our instrumental resolution, 
however, is probably not high enough to resolve any skewing which may 
be present.  We will present the minimum disk case analysis for this 
galaxy in $\S$ 5.2 but exclude it from further modeling.

Obtaining high-quality velocity fields is not trivial.  LSB galaxies are
difficult to observe.  The \Ha\ emission is faint.  Additionally, the 
emission may not be spread out enough to be detected across the entire
\Dpak\ fiber array.  Though we tried to select galaxies with promising
\Ha\ emission, the remaining 17 galaxies of our sample (listed in Table 2)
were observed, but unfortunately the detections were not good enough to 
construct meaningful velocity fields.

For the eleven galaxies for which we have constructed velocity fields, there 
is good overall agreement between the new \Dpak\ rotation curves
and the previous data sets.  In all cases, the data are broadly consistent
with previous long-slit rotation curves.  In only two cases, F563-1 and 
F583-4, do these independent data seem to differ, and then only over a 
very limited range in radius.  In cases where \HI\ data are available, 
five are in good overall agreement.  In two cases, DDO 64 and F583-4, the
\HI\ data are somewhat lower than the optical data.  In one case, UGC 4325, 
the \HI\ data are too high.  It is tempting to blame these cases on 
beam smearing, or in the case of UGC 4325, over-correction for beam 
smearing, though other factors such as the intrinsic \HI\ distribution are
also relevant.  On the whole, we are encouraged by the extent to which
various independent data sets agree given the difficult observational 
challenge posed by LSB galaxies.

\section{Preliminary Analysis and Discussion}
In this section we wish to give an impression of how well the data 
are described by various halo models.  We limit this discussion to 
the minimum disk case in which we ignore the contribution of the stars 
and gas and attribute all rotation to dark matter.  This puts an upper 
limit on the slope and/or concentration of the halo density profile.  
More detailed analysis including mass modeling will be presented in a 
future paper.

\subsection{Halo Models}
Two of the most prominent competing  dark matter halo density profiles 
are the pseudo-isothermal halo and the NFW profile.  We provide a
brief description of each below.   

\subsubsection{Pseudo-Isothermal Halo}
The density profile of the pseudo-isothermal halo is
\begin{equation}
\rho_{iso}(R) = \rho_{0}[1 + (R/R_{C})^{2}]^{-1} ,
\end{equation}
with $\rho_{0}$ being the central density of the halo and $R_{C}$ 
representing the core radius of the halo.  The rotation curve 
corresponding to this density profile is
\begin{equation}
V(R) = \sqrt{4\pi G\rho_{0} R_{C}^{2}\left[1 - \frac{R_{C}}{R}\arctan\left(\frac{R}{R_{C}}\right)\right]} .
\end{equation}
This form has traditionally been used in rotation curve fitting
because it works well.  By construction it produces flat rotation 
curves at large radii.  This halo form is empirically motivated, 
predating those stemming from simulations.

\subsubsection{NFW Profile}
Numerical simulations produce the NFW profile and its variants. It 
predicts the same functional behavior for CDM halos of galaxy clusters 
as for the CDM halo of a single galaxy.  The NFW mass-density 
distribution is described as
\begin{equation}
\rho_{NFW}(R) = \frac{\rho_{i}}{(R/R_{s})(1 + R/R_{s})^{2}} ,
\end{equation}
in which $\rho_{i}$ is related to the density of the universe at the 
time of halo collapse, and $R_{s}$ is the characteristic radius of 
the halo.  The NFW rotation curve is given by
\begin{equation}
V(R) = V_{200}\sqrt{ \frac{\ln(1+cx) - cx/(1 + cx)}{x[\ln(1 + c) - c/(1+c)]}},
\end{equation}
with $x$ = $R$/$R_{200}$.  The rotation curve is parameterized by 
a radius $R_{200}$ and a concentration parameter $c$ = $R_{200}$/$R_{s}
$, both of which are directly related to $R_{s}$ and $\rho_{i}$.  
$R_{200}$ is the radius at which the density contrast exceeds 200,  
roughly the virial radius. $V_{200}$ is the circular velocity
at $R_{200}$ \citep{NFW96}.  As previously mentioned, there are other 
cuspy halo models with slopes steeper than the NFW profile \citep[e.g.][]
{Moore, Reed, Navarro2004, Diemand}.
The NFW profile thus serves as a lower limit to the slope of cuspy 
density profiles and as such, gives the cuspy halo the best possible 
chance to fit the data.  From an observational perspective, there is 
very little to distinguish the various flavors of cuspy halos.

\subsection{Halo Fits to Combined Data}
In this section we combine the \Dpak\ data with the previous smoothed 
long-slit \Ha\ and \HI\ rotation curves when available.  For the ten 
galaxies with long-slit and/or \HI\ data, we supplement 
the \Dpak\ data with the entire long-slit rotation curve, and include 
only those \HI\ points which extend beyond the radial range of both the 
\Dpak\ and long-slit data.  Using only the outer \HI\ points where the
rotation curves have usually begun to flatten lessens possible 
resolution effects.  As discussed below, there are three galaxies,
F563-1, F583-4 and UGC 4325, for which we exclude some 
of the \Dpak\ and/or \HI\ data.

We find the best-fit isothermal halo and NFW halo.  This NFW halo fit 
is referred to as $NFW_{free}$.  The $NFW_{free}$ halo fits do not 
necessarily have parameters that are realistic or consistent with 
$\Lambda$CDM.  There is a tendency for the fits to drive towards very 
low $c$ and very high $V_{200}$.  There is also a $c$-$V_{200}$ 
degeneracy which allows halos of different $c$,$V_{200}$ to look the 
same over a finite range of radius.  It is common for the $NFW_{free}$ 
halos to overshoot the data at small radii, then undershoot the data 
and then overshoot the data again and provide a poor description of the 
data at large radii.  To address this, we also make a fit which we refer 
to as $NFW_{constrained}$.  This halo was required to match the velocities 
at the outer radii of each galaxy while constraining the concentration to 
agree with cosmology.  We chose a value of $V_{200}$ which forced the NFW
velocities to match as many of the data points in the turn-over region 
as closely as possible, with a minimum requirement of falling within 
the errorbars.  This is a reasonable constraint because dark matter 
must explain the high velocities at large radii where the contribution
of the baryons has fallen off.    Equation 7 of \citet{dBBM}, which gives the
concentration as a function of $V_{200}$ \citep{NFW97}, was then used to 
calculate the concentration.  We adjusted this concentration to agree with the 
cosmology of \citet{Tegmark} by subtracting 0.011 dex \citep{McGaugh03}.

We did not include the last four \Dpak\ points of F563-1 nor the last
three \Dpak\ points of F583-4 in the halo fits.  The F563-1 points
were excluded because they are based on few fibers and because of the
significant inconsistency with the long-slit and \HI\ rotation curves
as discussed in $\S$ 4.1.  The outer \Dpak\ points of F583-4 spike to
higher velocities than the long-slit data at the equivalent radii.
Neither the isothermal nor NFW halo model will
be able to fit this feature in the \Dpak\ rotation curve.  These
\Dpak\ points also  have little influence on the halo fits; there is 
essentially no change in the values of the isothermal or NFW halo
parameters, only much improved $\chi^{2}$, when the three \Dpak\
points are removed.  There are 2 and 3 \HI\ points beyond the optical
data for UGC 4325 and F583-4, respectively, which were also not used.  
Though the rise then sudden decline suggested by the combined optical
and \HI\ data for UGC 4325 may well be real, no simple, smooth model can
fit it. Since we are interested in the inner halo structure as probed
by the new data, we exclude the 2 HI points.  For a thorough
comparison of many more independent data for this galaxy see
\citet{Bosma04}.  The situation is similar for the \HI\ points of F583-4.

In Figure 12 we plot the halo fits over the data and list the
parameters in Table 3.  We plot $\log$$(V)$ against $\log$$(r)$ to
emphasize the fits to the data at small radii.  The
$NFW_{constrained}$ halos have already been required to match the data
at large radii, and although the best-fit $NFW_{free}$ halos do not
necessarily match the velocities at large radii, the magnitude of the 
discrepancy is generally smaller than it is at small radii.

No $NFW_{free}$ fit could be made to four of the galaxies: UGC 4325, 
DDO 64, F568-3, UGC 1281.  Three galaxies, UGC 5750, F583-4 and
F583-1, have concentrations too low to be consistent with 
$\Lambda$CDM.  Specifically, the concentrations are farther from the 
concentrations of the $NFW_{constrained}$ halos than the expected log 
scatter in $c$ of 0.18 \citep{Bullock01}.  Only the concentrations
from the $NFW_{free}$ fits of the remaining four galaxies, NGC 4395, 
UGC 477, F563-V2 and F563-1, are consistent with $\Lambda$CDM.

\begin{deluxetable*}{lccccccccccc}
\tabletypesize{\scriptsize}
\tablecaption{Best-Fit Halo Parameters}
\tablewidth{0pt}
\tablehead{
\colhead{Galaxy}  &\multicolumn{3}{c}{ISO} &\colhead{} &\multicolumn{3}{c}{$NFW_{free}$} &\colhead{} &\multicolumn{3}{c}{$NFW_{constrained}$}\\
\cline{2-4} \cline{6-8} \cline{10-12} 
&\colhead{$R_{c}$}  &\colhead{$\rho_{0}$}  &\colhead{$\chi^{2}_{r}$} &\colhead{}  &\colhead{c} &\colhead{$V_{200}$}  &\colhead{$\chi^{2}_{r}$} &\colhead{}  &\colhead{c} &\colhead{$V_{200}$} &\colhead{$\chi^{2}_{r}$}  
}
\startdata
UGC 4325 &3.3$\pm$0.2  &91$\pm$4  &3.8 & &\nodata &\nodata &\nodata  & &6.9 &249 &40 \\
F563-V2 &1.5$\pm$0.1  &119$\pm$6 &0.71 & &7.7$\pm$2.0 &128$\pm$32 &0.40 & &7.9 &130 &0.58\\
F563-1 &2.1$\pm$0.1 &67$\pm$2 &0.43 & &7.8$\pm$1.3 &106$\pm$10 &0.88 & &8.4 &101 &0.95\\
DDO 64 &4.4$\pm$0.9 &38$\pm$3 &5.5 & &\nodata &\nodata &\nodata  & &9.2 &62 &20\\
F568-3 &3.8$\pm$0.2  &27$\pm$1 &1.2 & &\nodata &\nodata &\nodata  & &8.2 &110 &12\\
UGC 5750 &5.7$\pm$0.4 &7.1$\pm$0.3 &0.83 & &0.5$\pm$0.1 &320$\pm$43 &1.7 & &9.1 &67 &25\\
NGC 4395 &0.7$\pm$0.1 &258$\pm$9 &2.9 & &10.1$\pm$0.6 &77$\pm$4 &2.1 & &8.6 &87 &2.2\\
F583-4 &1.3$\pm$0.1 &67$\pm$2 &0.67 & &5.5$\pm$2.2 &92$\pm$32 &0.41 & &9.1 &67 &1.1\\
F583-1 &2.7$\pm$0.1 &35$\pm$2 &5.4 & &4.7$\pm$0.7 &133$\pm$21 &8.7 & &8.7 &83 &11  \\
UGC 477 &2.2$\pm$0.1 &57$\pm$2 &4.6 & &6.9$\pm$0.6 &120$\pm$9 &4.6 & &8.3 &105 &5.4 \\ 
UGC 1281 &2.6$\pm$0.1 &23$\pm$1 &3.8 & &\nodata &\nodata &\nodata & &9.3 &58 &25 \\
\enddata
\tablecomments{Best-fit halo parameters for the combined \Dpak, long-slit and \HI\ rotation curves.  $R_{c}$ is the core radius in kpc, $\rho_{0}$ is the central density in $10^{-3}$ $\textrm{M}_{\sun}$ $\textrm{pc}^{-3}$, and $V_{200}$ is in \kms\ .  }
\end{deluxetable*}

One should bear in mind that these fits are taken in the limit of minimum 
disk.  Baryons do matter some in LSB galaxies.  This will drive the 
concentrations even lower in proper mass models.

The four galaxies for which no $NFW_{free}$ fits could be made are fit
significantly better by isothermal halos than $NFW_{constrained}$
halos.   The case of F568-3 is typical: the $NFW_{constrained}$ halos
overpredict the velocity at all radii interior to where they were
forced to match the observed velocity.  Of the galaxies with $NFW_{free}$ 
fits, UGC 5750 is a good example of the over-under-over fitting trend of 
of the NFW halo. This fitting trend has 
been observed before (e.g. BMR01; Gentile et al. 2004) and is what is 
being referred to when it is said that the NFW rotation curve has the 
wrong shape.   Of the galaxies with $NFW_{free}$ fits, UGC 5750, F583-1
and F563-1 are best fit by isothermal halos in terms of $\chi^{2}_{r}$.
UGC 5750 is the strongest case, as the value of its best-fit concentration,
0.5, is far too low according to current ideas from cosmological 
simulations.  NGC 4395 has a reasonable NFW concentration and a 
$\chi^{2}_{r}$ favoring the NFW halo.  There are signs of a possible bar
or oval structure at the center of this galaxy.  Mass modeling beyond
the minimum disk case may help to determine the central structure of the
galaxy and whether or not the NFW halo remains a good description of the
data.  F583-4 and F563-V2 also have
$\chi^{2}_{r}$ in favor of the NFW halo; however, the best-fit
concentration for F583-4 is bordering on the low side of expected 
values, and F563-V2 has too few data points to really distinguish between 
halo types.  It is worth noting that when the small uncertainty in the
first \Dpak\ point of F563-V2 is increased to 5 \kms, neither the 
isothermal nor NFW fit is significantly changed.  Finally, UGC 477 
is equally fit by both isothermal and NFW halos.

In total, seven galaxies are well-described by the isothermal halo,
one is consistent with NFW, and three are indistinguishable. For the 
majority of the galaxies,  NFW halos could either
not be fit to the data, or the halos had concentrations too low to be
consistent with $\Lambda$CDM.  As shown by NGC 4395, velocities
consistent with cuspy halos can be detected in the two-dimensional data.

\begin{figure*}
 \includegraphics[scale=0.23]{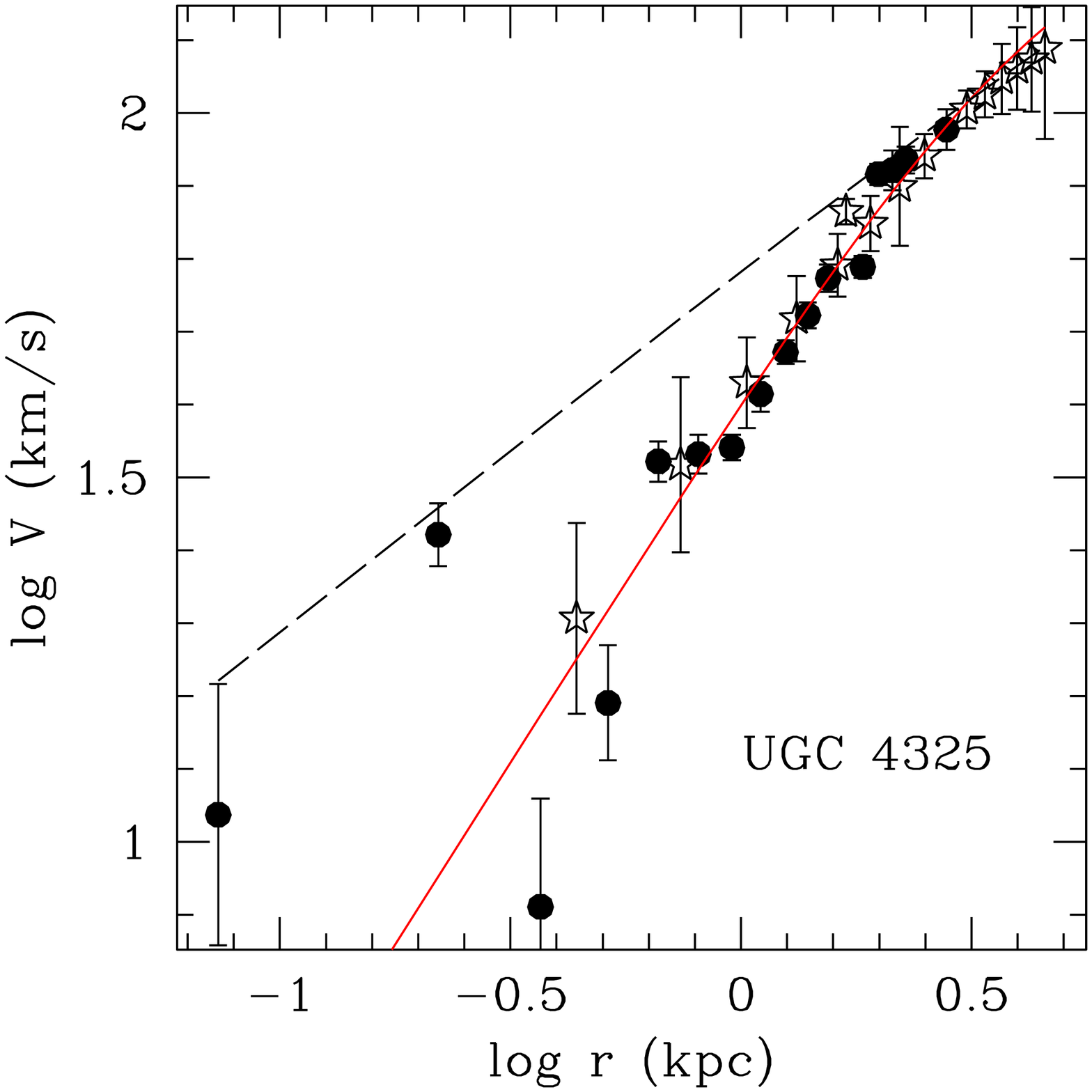}
 \hfill
 \includegraphics[scale=0.23]{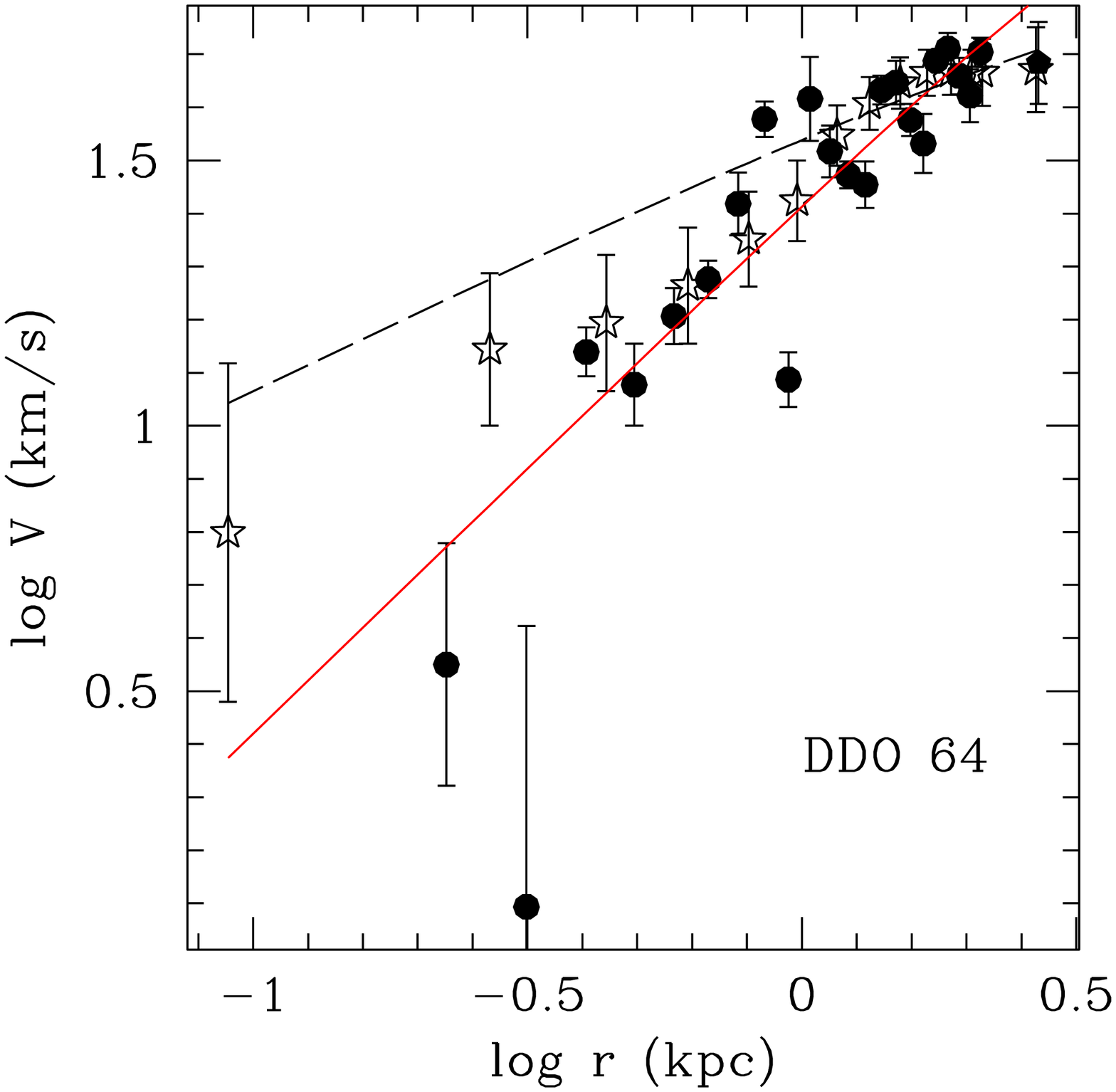}
 \hfill
 \includegraphics[scale=0.23]{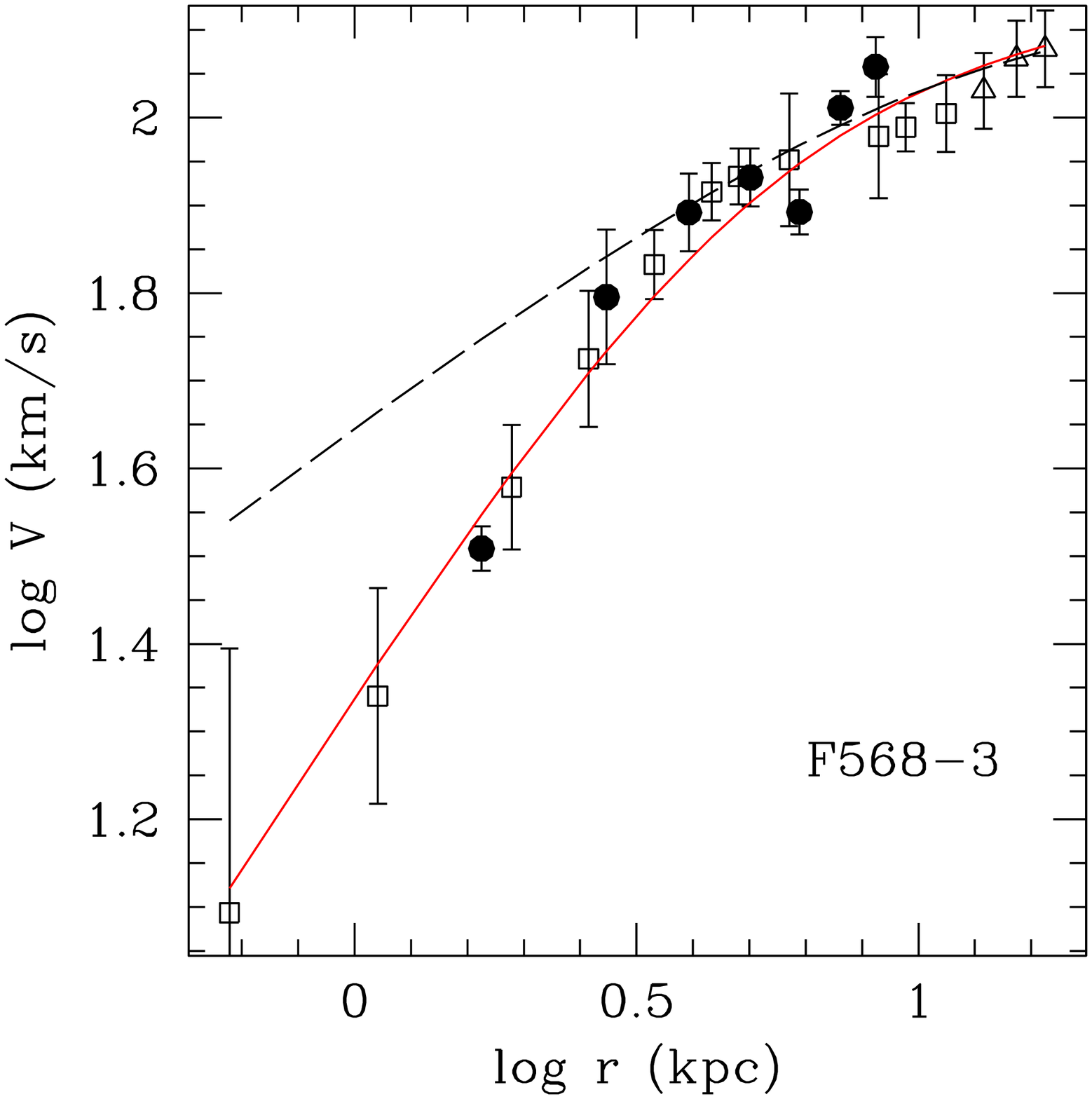}\\
 \includegraphics[scale=0.23]{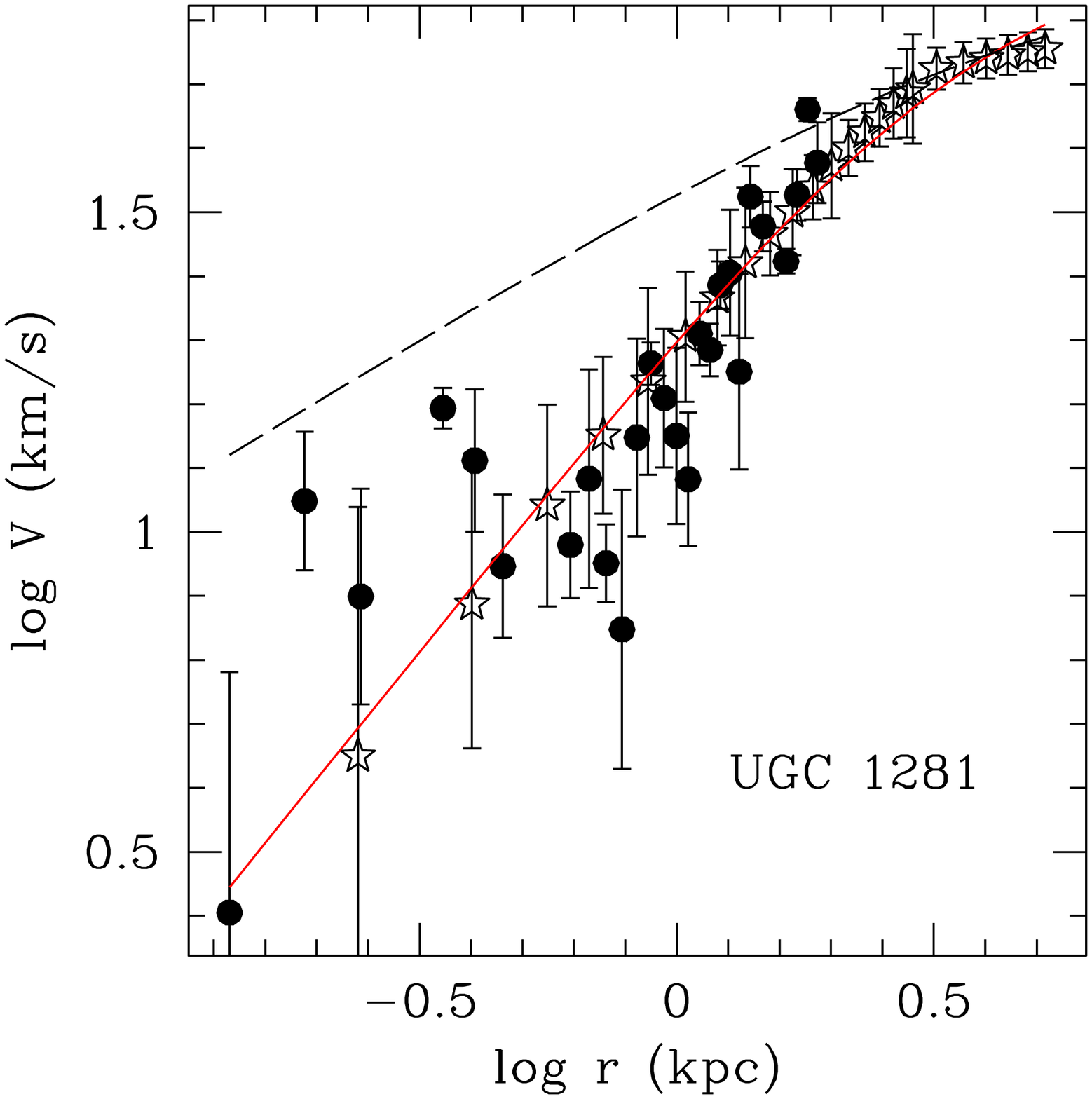}
 \hfill
 \includegraphics[scale=0.23]{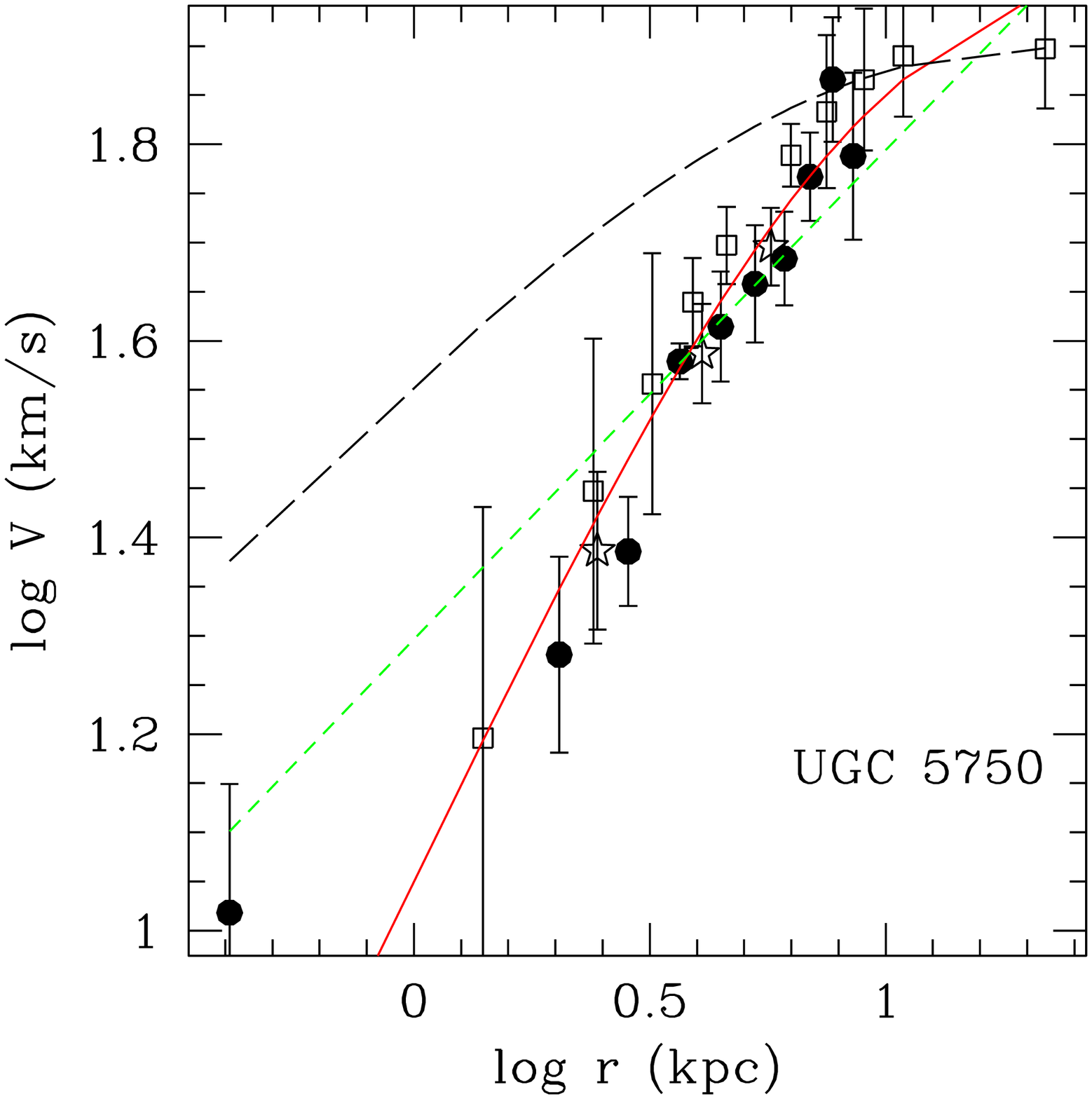}
 \hfill
 \includegraphics[scale=0.23]{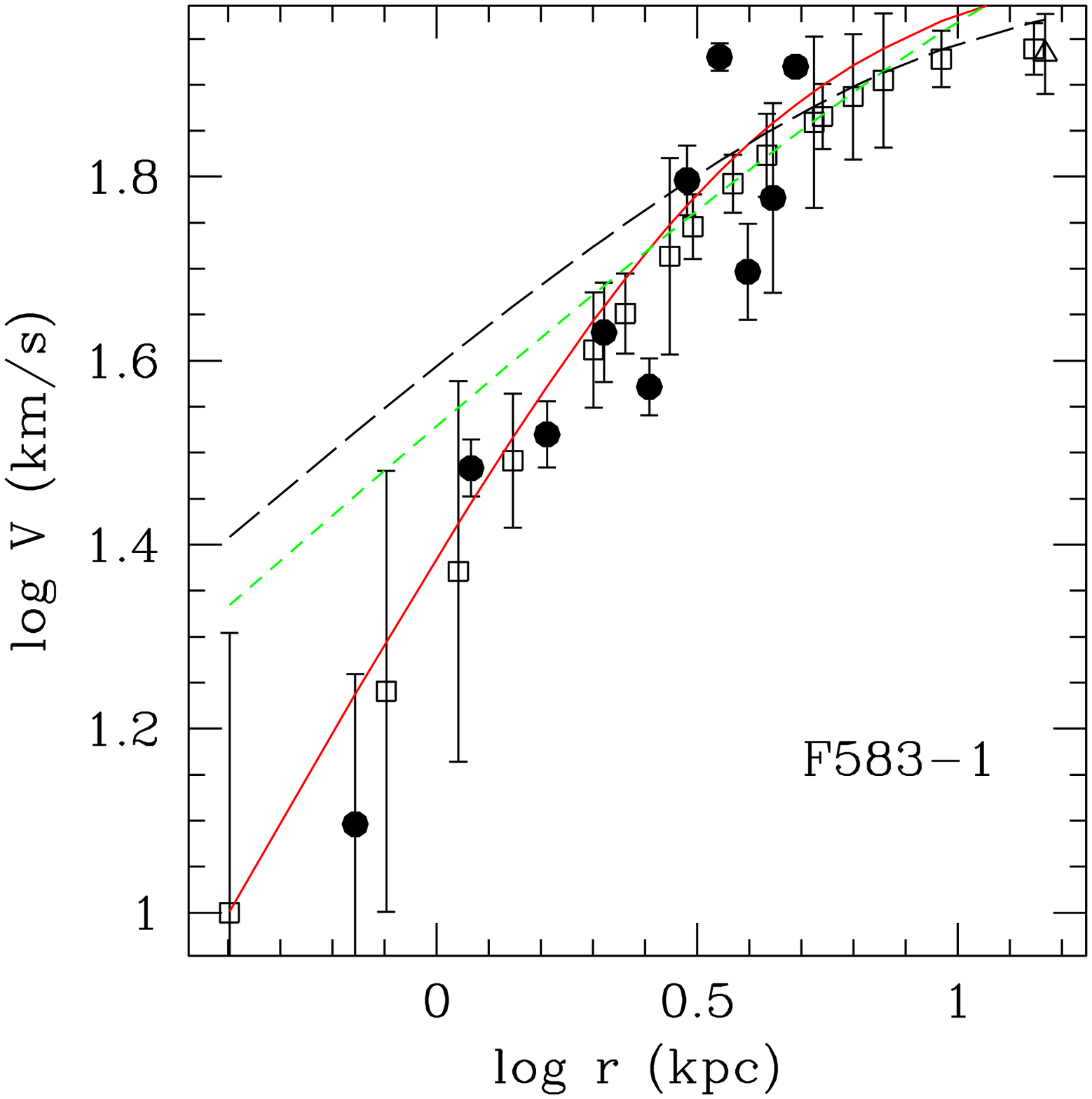}\\
 \includegraphics[scale=0.23]{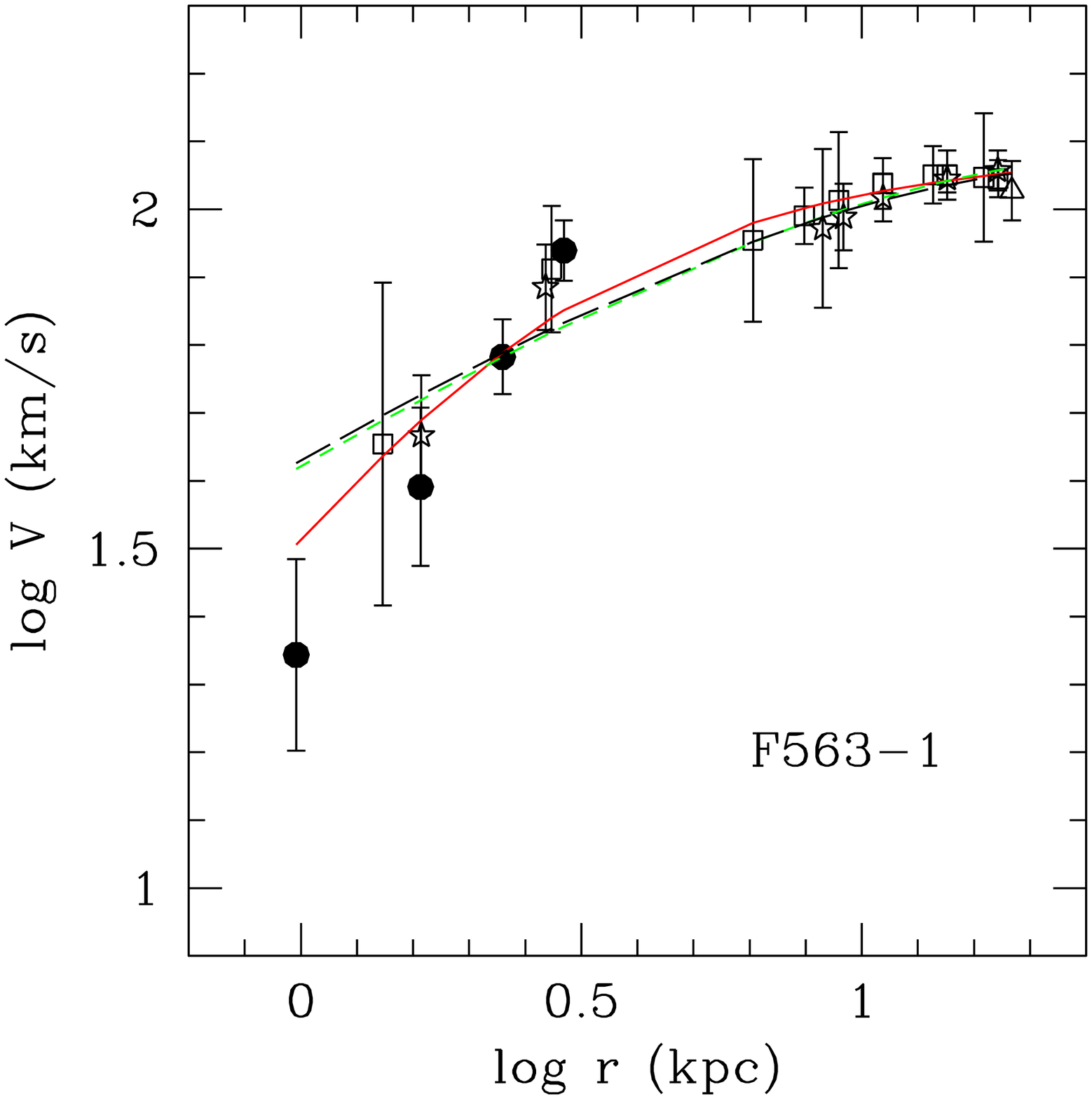}
 \hfill
 \includegraphics[scale=0.23]{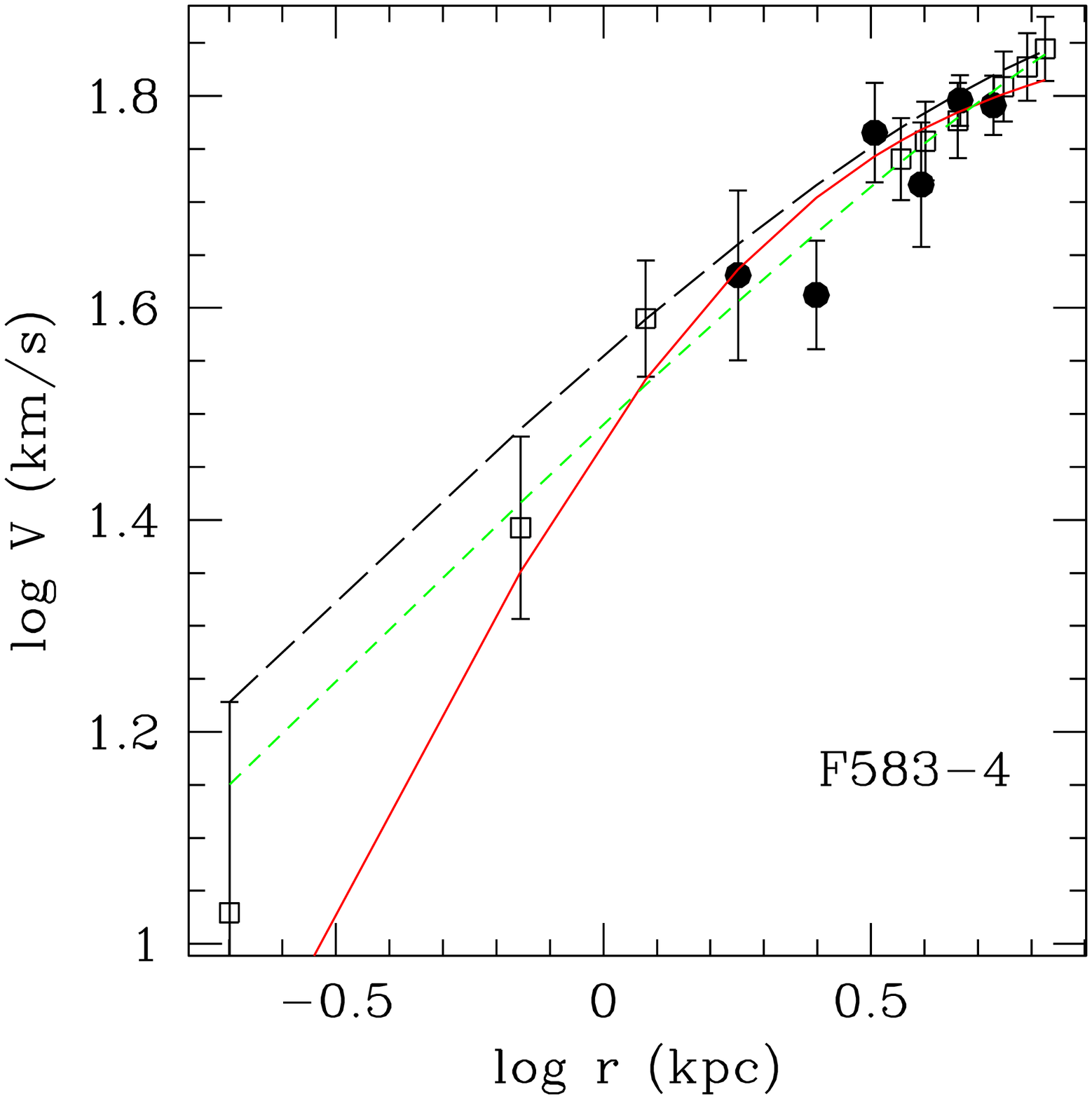}
 \hfill
 \includegraphics[scale=0.23]{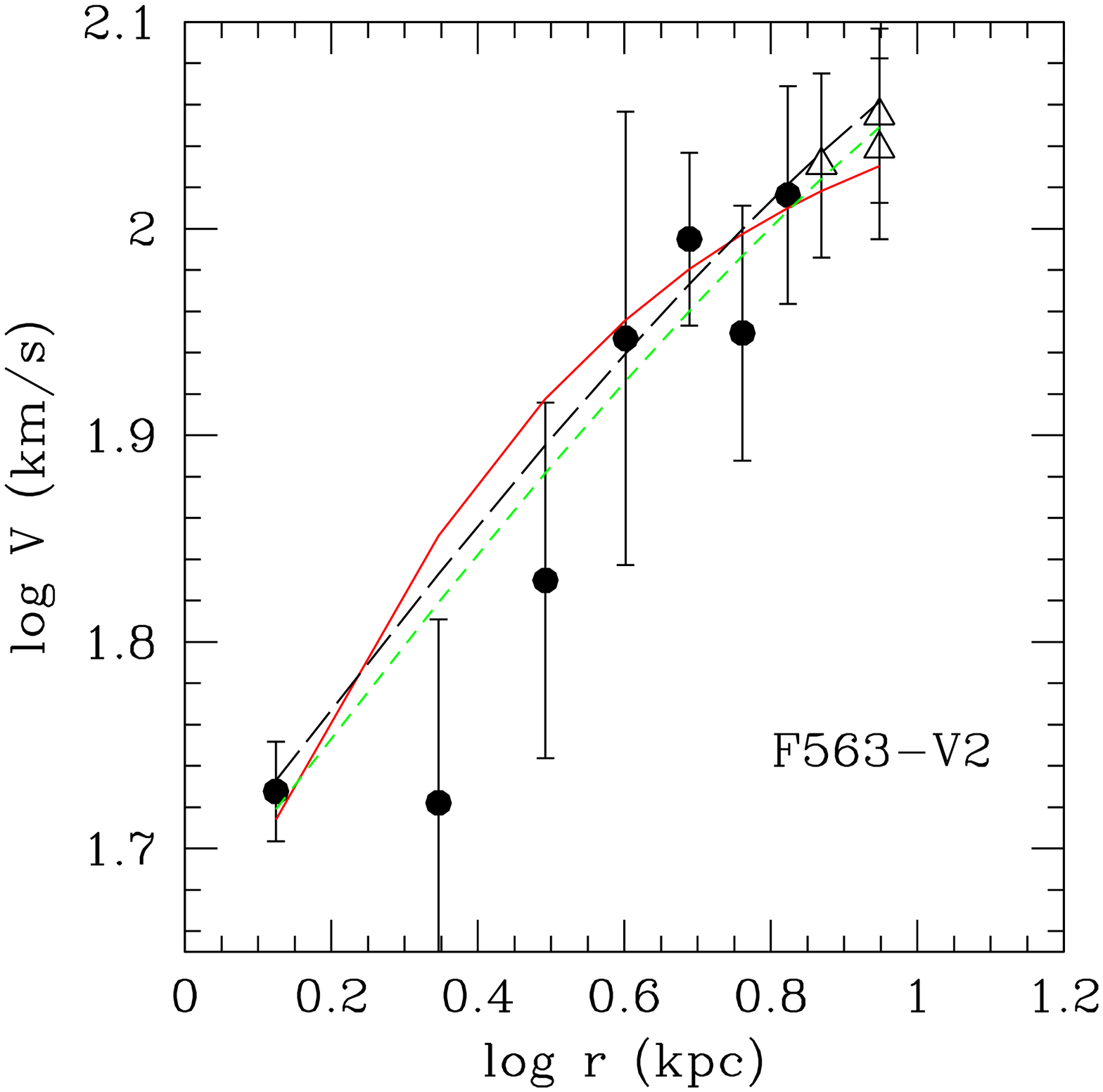}\\
 \includegraphics[scale=0.23]{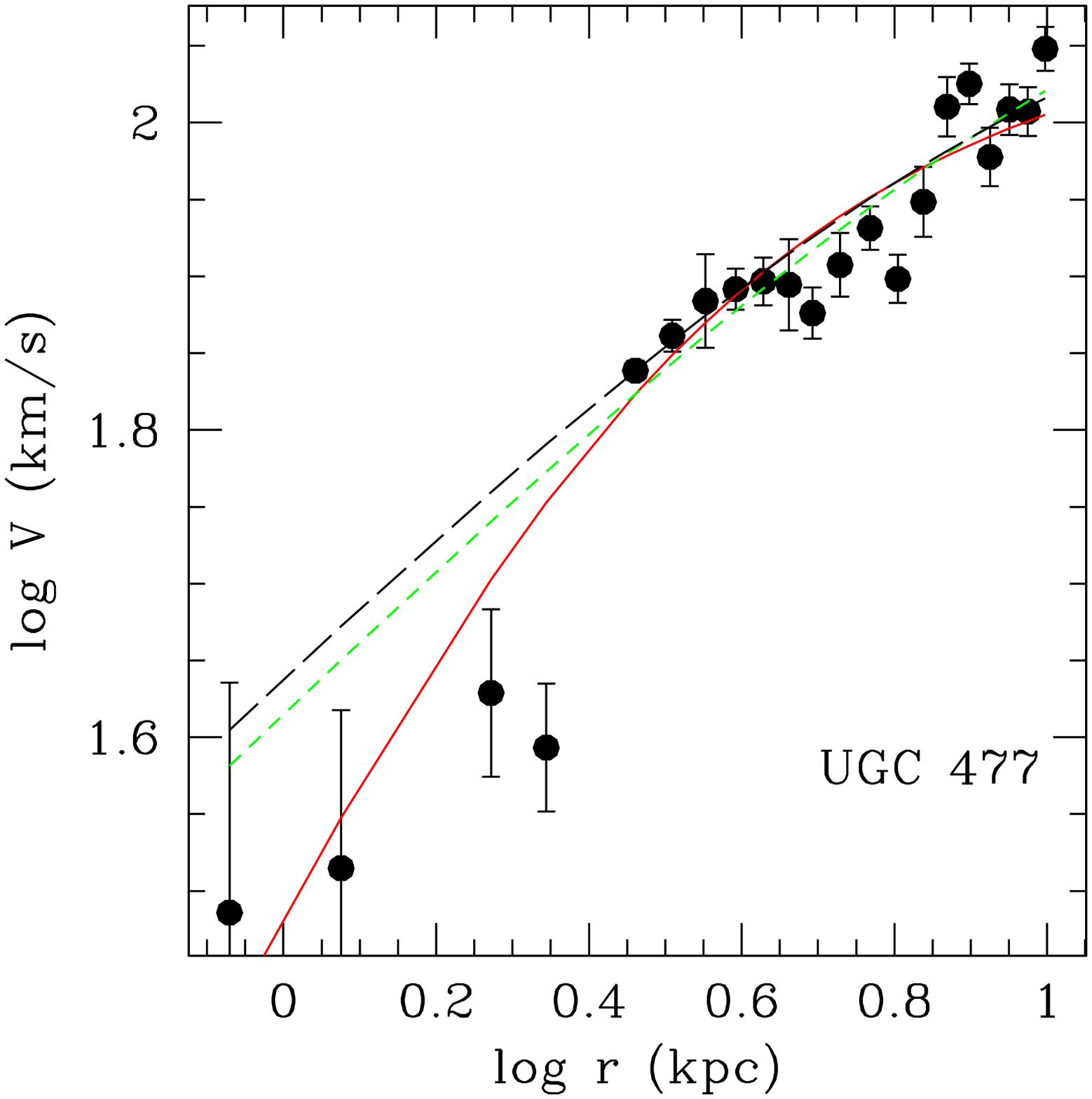}
 \hfill
 \includegraphics[scale=0.23]{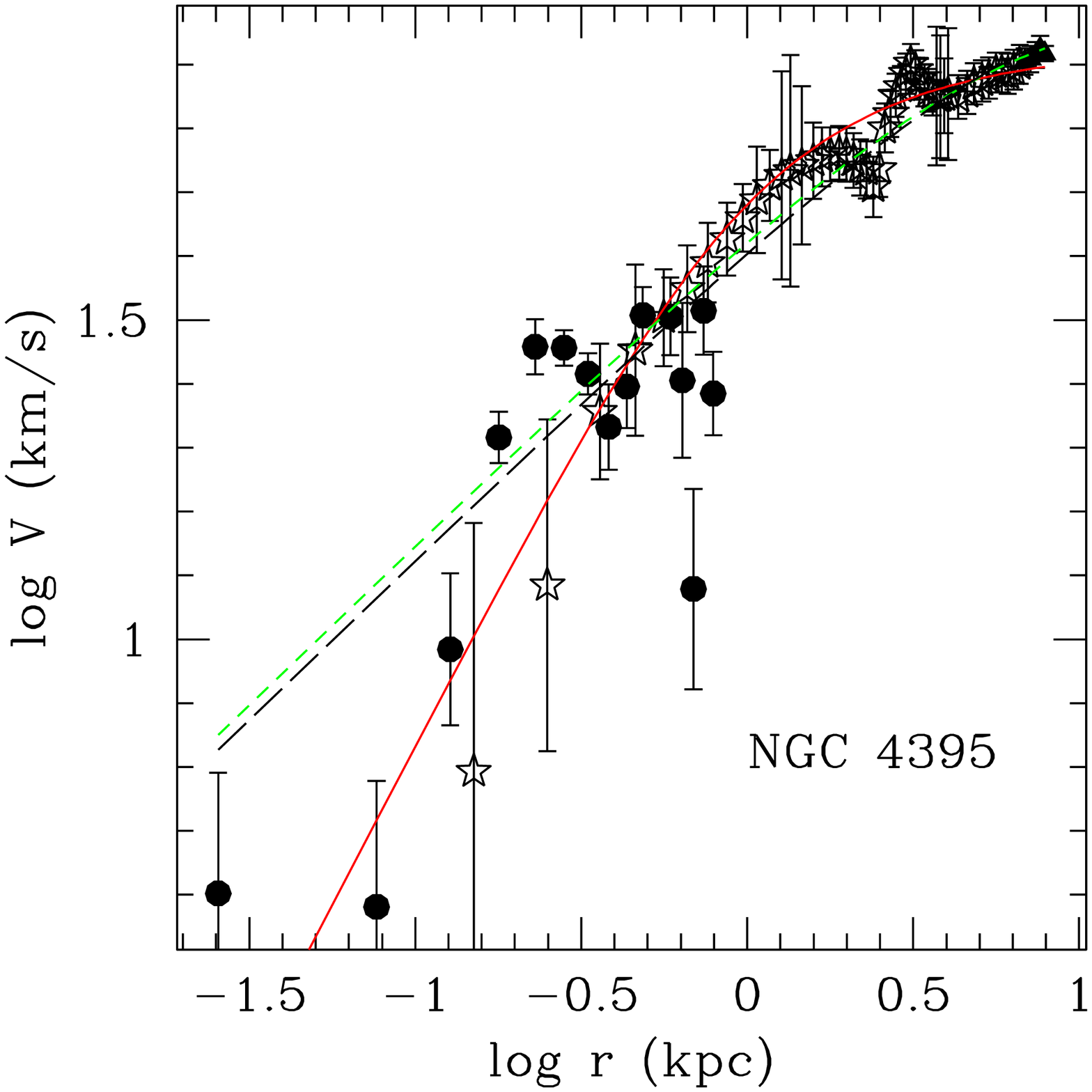}
 \hfill
 \includegraphics[scale=0.23]{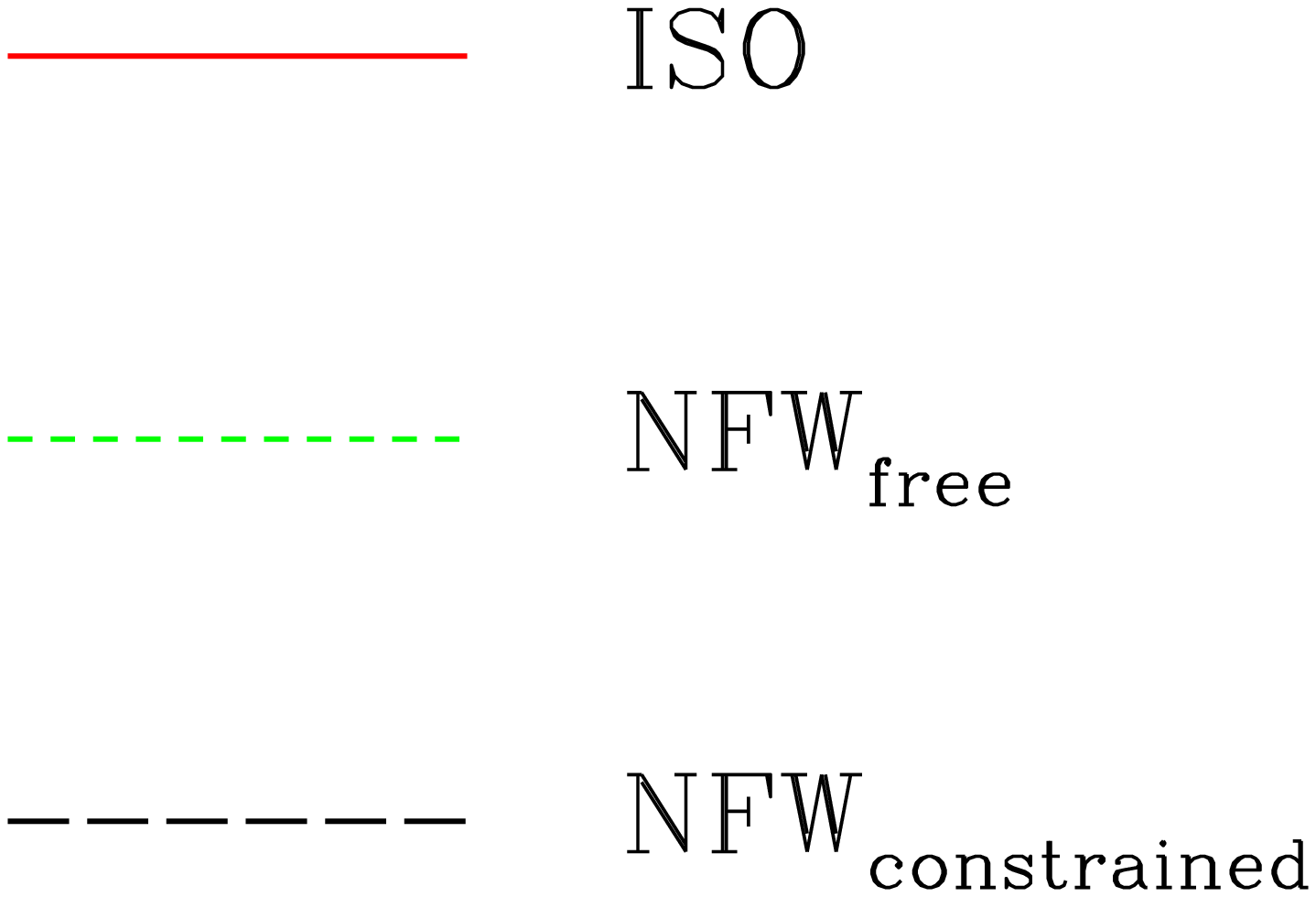}
   \caption{Halo fits to the combined \Dpak\ and smoothed long-slit and \HI\ rotation curves.  Symbols remain the same as in Figures 1-11.  The order of the galaxies has been changed from the order of Figures 1-11 in order to show a progression in the halo fits.  The fits change from isothermal to indistinguishable to NFW from top to bottom, left to right.  Figure appears in color on-line.}
\end{figure*}
\begin{figure*}
\plottwo{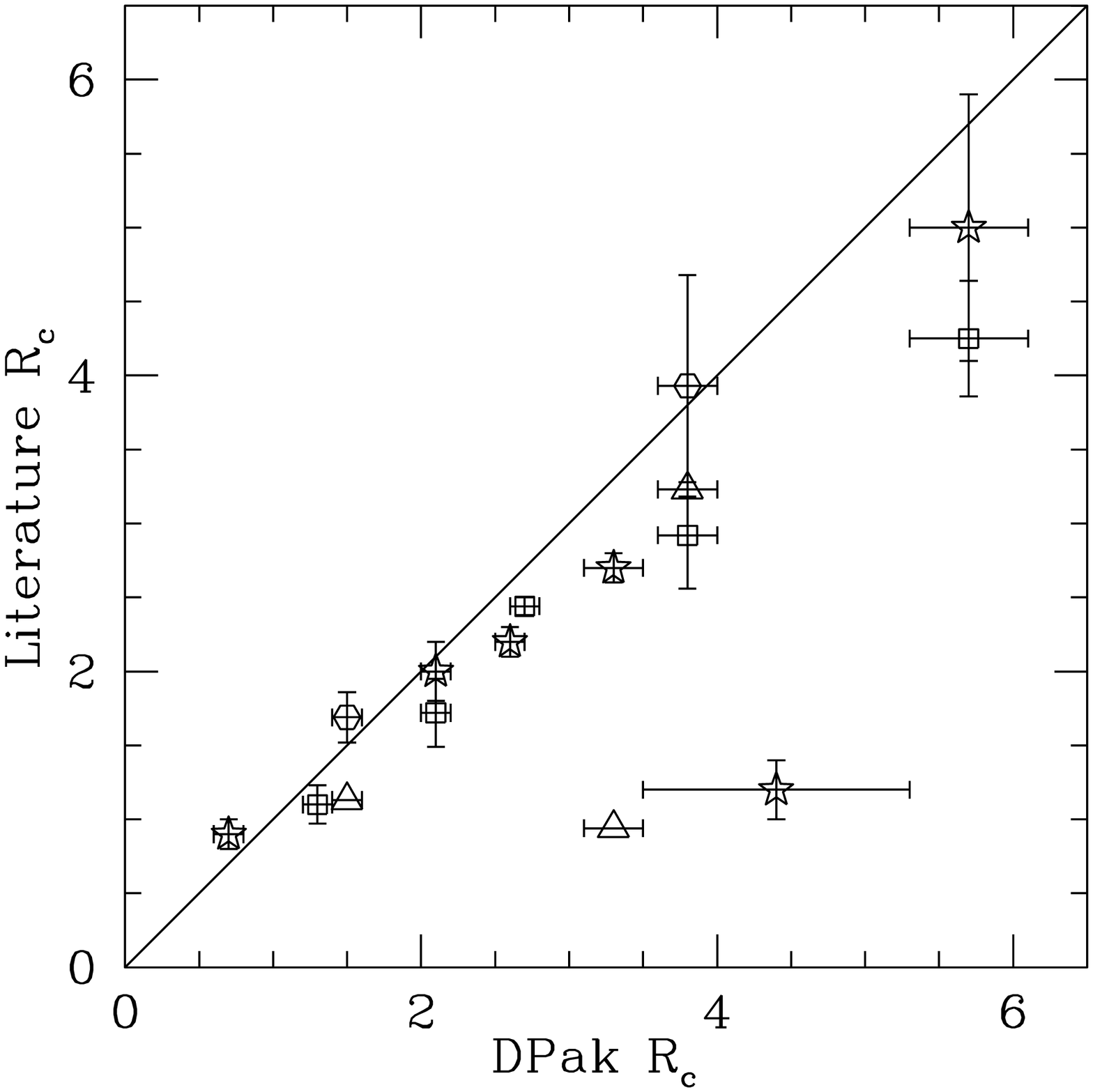}{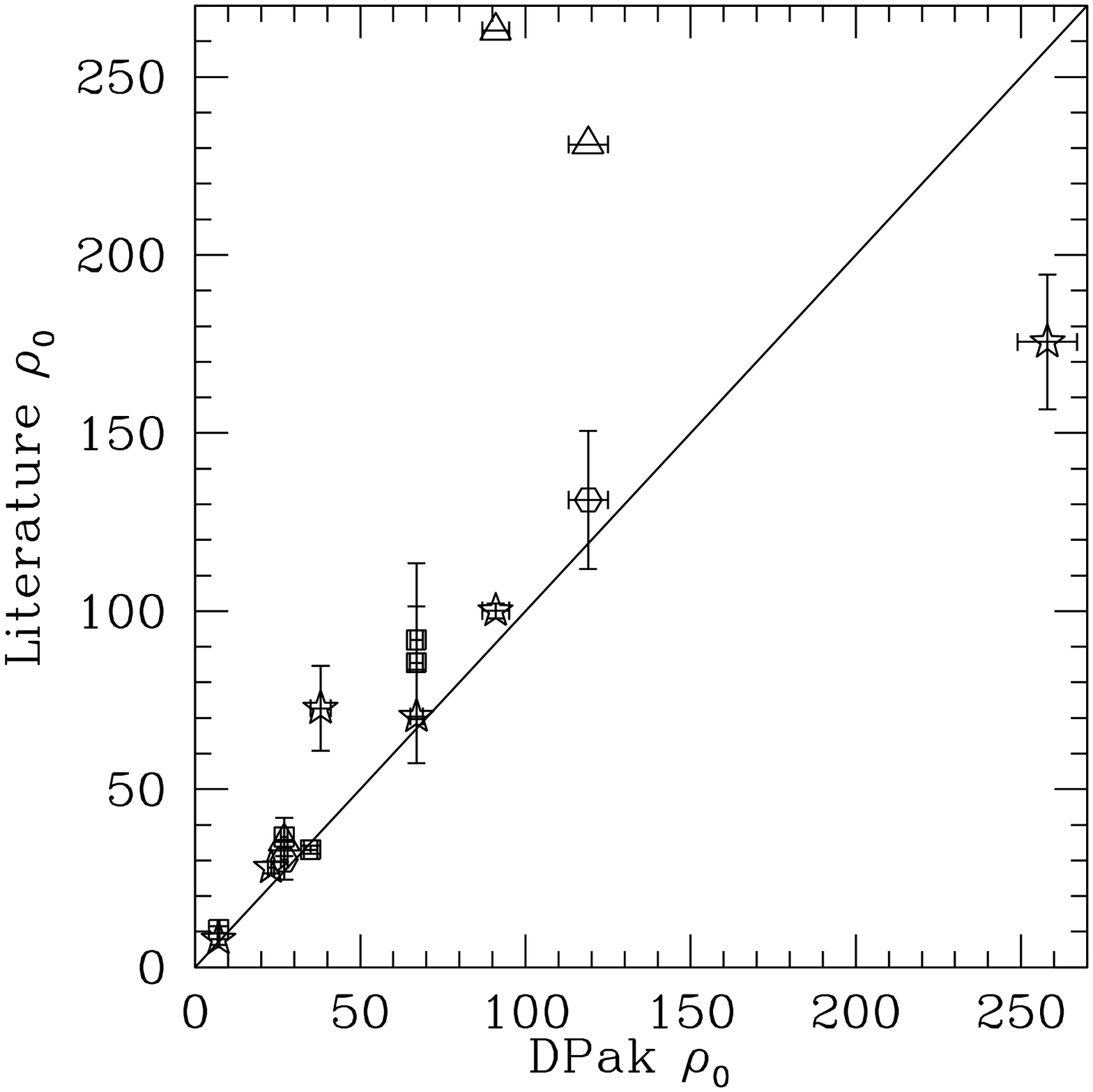}\\
\plottwo{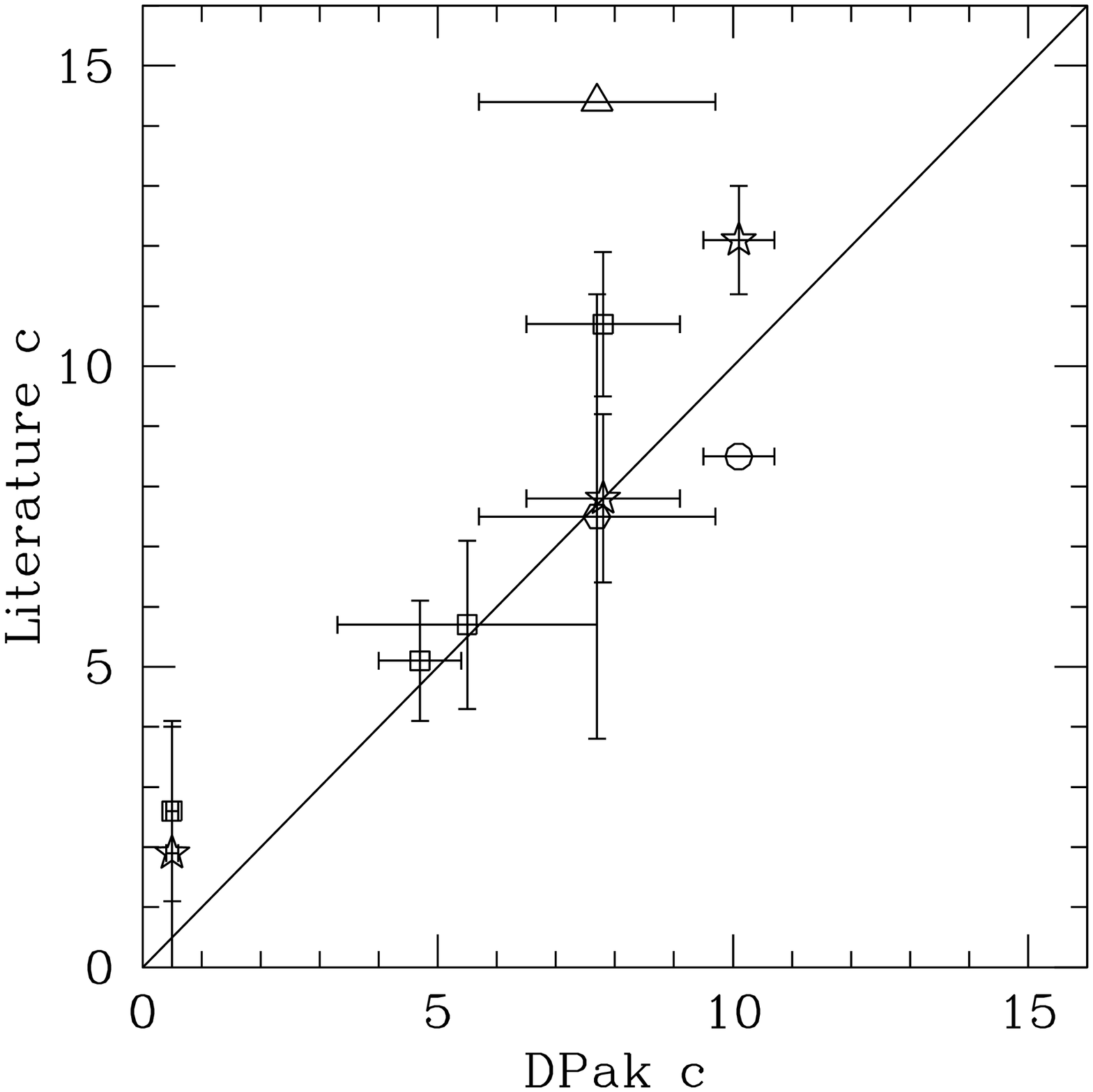}{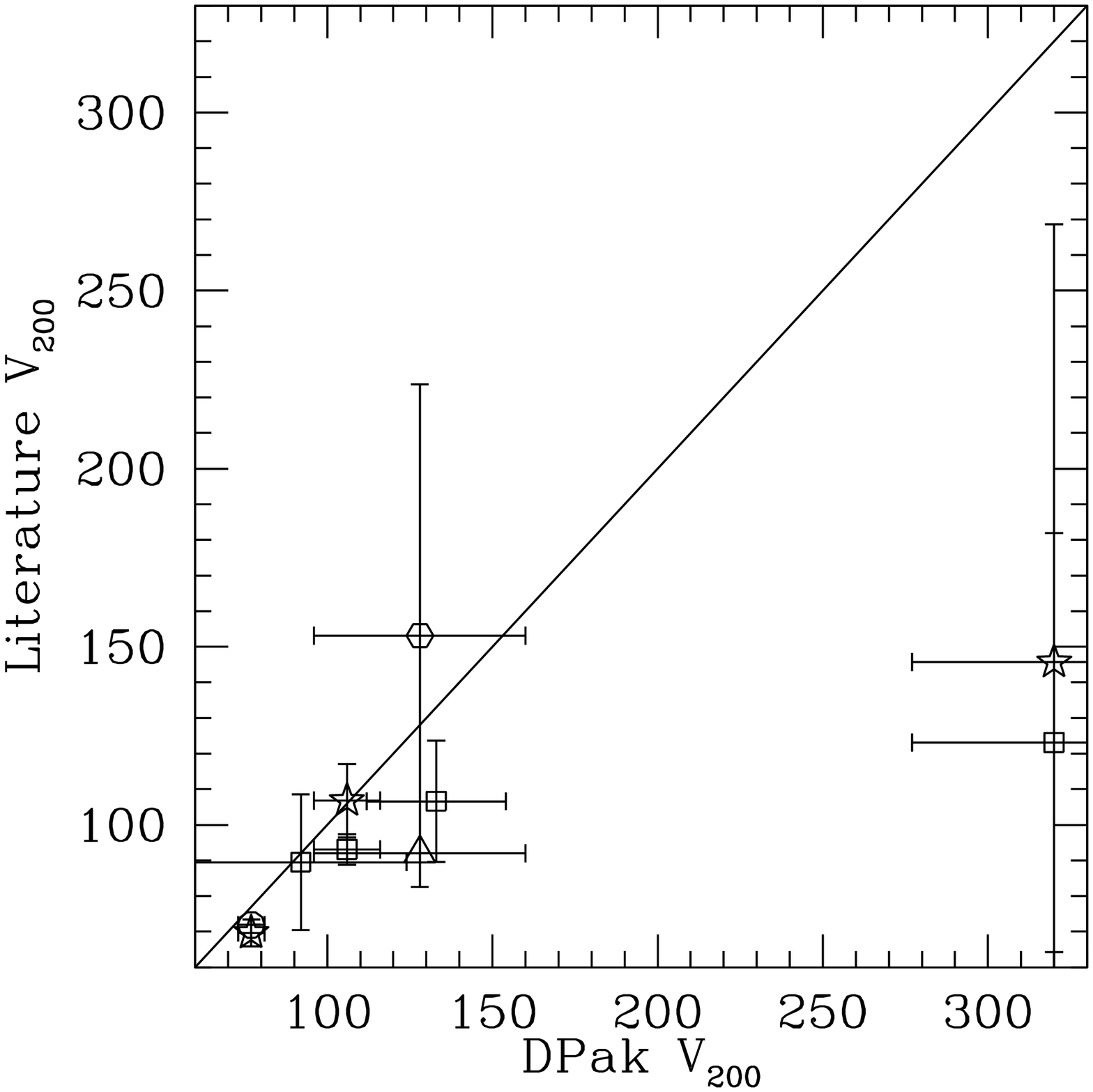}
\caption{Comparison of \Dpak\ halo parameters to previously published values.
 $R_{c}$ and $\rho_{0}$ are displayed for the isothermal halo and $c$ and 
 $V_{200}$ are shown for the NFW halo.  Stars represent the data from 
 \citetalias{dBB}, squares the data from \citetalias{dBMR}, triangles the
 data from \citet{Rob}, circles the data from \citet{vandenB01} and
 hexagons the \citetalias{dBMR} analysis of the \citet{Swaters00} data.  
 The addition of the \Dpak\ does not significantly alter the previous halo
 fits, but does reduce the errors on the halo parameters. }
\end{figure*}

\subsection{Comparison to Literature}
Ten of the eleven galaxies have published pseudo-isothermal and NFW halo 
fits to
the long-slit and/or \HI\ data.  In Figure 13 we plot the \Dpak\ halo
parameters for each galaxy ($R_{c}$ and $\rho_{0}$ for isothermal, $c$ and 
$V_{200}$ for NFW) against the minimum disk literature values.  There is
gross agreement between the \Dpak\ and literature halo parameters, showing
that the addition of the new two-dimensional optical data has not 
substantially altered the fits.  The addition of the \Dpak\ data does, 
however, bring down the errors on the halo parameters by roughly a factor 
of 2.  The \Dpak\ 
parameters are listed in Table 3; the numbers mentioned in the text below 
are from the cited references.

$\textbf{UGC 4325:}$  The \Dpak\ isothermal halo parameters agree well with 
the results of \citetalias{dBB} ($R_{c}$ = 2.7$\pm$0.1; $\rho_{0}$ = 
100.1$\pm$2.1), and the agreement between those two datasets is better than
the agreement of either set with the parameters of \citet{Rob} 
($R_{c}$ = 0.94; $\rho_{0}$ = 263).  No $NFW_{free}$ fit could be made to
the \Dpak\ data, and the \citetalias{dBB} data preferred a concentration
less than 0.1 (c = 0.1, $V_{200}$ = 3331.6).  Many of the \citetalias{dBB} 
galaxies have NFW fits with unphysical (very small or negative) values of the 
concentration.  In these cases, the concentration was set to 0.1.  The NFW 
fits by both \citet{Rob} and \citet{vandenB01} required very high concentration
values (c = 14.8, $V_{200}$ = 83; c = 30.9, $V_{200}$ = 53.5, respectively).
 The \citet{Rob} NFW fit shows the over-under-over fitting trend and is 
not a good representation of the data.
 For this galaxy, all of the listed concentrations
are far beyond the reasonable range of expected values.  The \Dpak, 
\citetalias{dBB} and \citet{Rob} results all favor the isothermal
halo as the best fit. 

$\textbf{F563-V2:}$  For both the isothermal and NFW halo parameters, 
there is excellent agreement between the \Dpak\ values and the results 
of the \citetalias{dBMR} analysis of the \citet{Swaters00} data ($R_{c}$ = 
1.69$\pm$0.17, $\rho_{0}$ = 131.2$\pm$19.4).  The error on each 
parameter has also been reduced by the \Dpak\ data.   The
agreement is not as good with the isothermal parameters of \citet{Rob} 
($R_{c}$ = 1.13; $\rho_{0}$ = 231). Their NFW concentration (c = 14.4)
is again much higher than the \Dpak\ value, and $V_{200}$ much lower 
($V_{200}$ = 92).  As previously discussed, there are few points in the 
\Dpak\ data, making a clear distinction between halo fits difficult.  
Certainly there is nothing to contradict the conclusions of both previous 
studies that found the data to prefer the isothermal halo.

$\textbf{F563-1:}$  The \Dpak\ results agree 
extremely well with both the isothermal ($R_{c}$ = 2.0$\pm$0.2, 
$\rho_{0}$ = 70.4$\pm$13.1) and NFW (c = 7.8$\pm$1.4, 
$V_{200}$ = 106.8$\pm$10.3) fits of \citetalias{dBB}.  The new data 
shrink the formal uncertainties on the isothermal halo parameters by a 
considerable amount.  The level of
agreement with the \citetalias{dBMR} results ($R_{c}$ = 1.72$\pm$0.23,
$\rho_{0}$ = 91.9$\pm$21.6; c = 10.7$\pm$1.2, $V_{200}$ = 93.1$\pm$4.3)
is only slightly less.  The isothermal halo is preferred by all three
studies.

$\textbf{DDO 64:}$  There is a difference between the isothermal halo
parameters determined by the \Dpak\ data and the results of 
\citetalias{dBB} ($R_{c}$ = 1.2$\pm$0.2, $\rho_{0}$ = 72.7$\pm$11.9), 
but both studies prefer the isothermal halo to the NFW halo.
The NFW results, however, are similar: no $NFW_{free}$ fit could be 
made to the \Dpak\ data, and the \citetalias{dBB} data favored a 
concentration less than 0.1 (c = 0.1, $V_{200}$ = 1182.3).  While the
two isothermal fits are distinguishable, they are not too different.  
This galaxy seems to have real structure which is reflected in the 
rotation curve.  The independent long-slit and \Dpak\ data both 
show non-monotonic features (``bumps and wiggles") that cannot be 
fit by any simple, smooth halo models of the type considered here.  
There is more information in the data than a simple model can represent.

$\textbf{F568-3:}$  The isothermal halo parameters of the \citetalias{dBMR} 
analysis 
of the \citet{Swaters00} data ($R_{c}$ = 3.93$\pm$0.75, $\rho_{0}$ = 
30.2$\pm$5.6) are in very good agreement with the \Dpak\ results.  Again,
the \Dpak\ parameters have lower errors.  There is also moderate agreement
between the \Dpak\ parameters and isothermal halo parameters of 
\citetalias{dBMR} ($R_{c}$ = 2.92$\pm$0.36, $\rho_{0}$ = 36.6$\pm$5.4)
and \citet{Rob} ($R_{c}$ = 3.23, $\rho_{0}$ = 35.3).  No 
$NFW_{free}$ fit could be made to the \Dpak\ data, and all three other
datasets require halos with low concentrations (\citetalias{dBMR} analysis
of \citet{Swaters00}: c = 1.2, $V_{200}$ = 591.1; 
\citetalias{dBMR}: c = 3.2$\pm$3.7, $V_{200}$ = 214.6$\pm$233.9;
\citet{Rob}: c = 1.0, $V_{200}$ = 637).  In all datasets, the
isothermal halo is a better fit.

$\textbf{UGC 5750:}$  There is decent agreement with the \Dpak\ isothermal halo 
parameters and the results of both \citetalias{dBB} ($R_{c}$ = 
5.0$\pm$0.9, $\rho_{0}$ = 7.9$\pm$1.6) and \citetalias{dBMR} 
($R_{c}$ = 4.25$\pm$0.39, $\rho_{0}$ = 10.6$\pm$1.0).  The errors are 
lower on the \Dpak-derived parameters.  All three datasets require a
very low NFW concentration (\citetalias{dBB}: c = 1.9$\pm$2.1, $V_{200}$ 
= 145.7$\pm$122.9; \citetalias{dBMR}: c = 2.6$\pm$1.5, $V_{200}$ = 
123.1$\pm$58.8).  The NFW halo is not a good description of the 
data.  Though the formal fit parameters differ for the NFW halo, the 
degeneracy between halo parameters is such that there is little to 
distinguish the resulting halo rotation curve.

\begin{deluxetable}{lc}
\tabletypesize{\footnotesize}
\tablecaption{Galaxy Velocity Dispersions}
\tablewidth{0pt}
\tablehead{
\colhead{Galaxy}    &\colhead{Vel.Disp.}\\
  &\kms 
}
\startdata
UGC 4325 &7.8\\
F563-V2  &8.2\\
F563-1  &8.1\\
DDO 64  &6.2\\
F568-3  &8.7\\
UGC 5750  &8.7\\
NGC 4395  &9.7\\
F583-4  &9.5\\
F583-1  &7.3\\
UGC 477  &8.1\\
UGC 1281  &6.9\\
\enddata
\tablecomments{Velocity dispersion in the \Dpak\ data for each galaxy.  Values are between 6 and 10 \kms\ and are consistent with the typical dispersions for the gas components of galaxies.}
\end{deluxetable}

$\textbf{NGC 4395:}$   There is moderate agreement between the \Dpak\ 
isothermal halo parameters and the values of the parameters found by 
\citetalias{dBB} ($R_{c}$ = 0.9$\pm$0.1, $\rho_{0}$ = 175.6$\pm$18.9).  
The $NFW_{free}$ concentration determined by the \Dpak\ data is between 
the values listed in \citetalias{dBB} (c = 12.1$\pm$0.9, $V_{200}$ = 
69.7$\pm$3.8) and \citet{vandenB01} (c = 8.5, $V_{200}$ = 71.9).  
As with UGC 5750, though the formal NFW fit parameters differ, the 
degeneracy between halo parameters is such that there is little to 
distinguish the resulting halo rotation curve.
\citetalias{dBB} find the isothermal halo to be a slightly better 
fit to the data than the NFW halo, whereas the \Dpak\ data have a 
slight preference for the NFW halo.  The NFW halo certainly cannot be 
excluded as it has a  reasonable concentration in all three fits. The 
misalignment of the minor axis in the \Dpak\ velocity field 
\citep[see also][]{Garrido,Noord01} and effects from star formation
\citepalias{dBB} need to be considered.  The presence of a bar may create 
strong enough non-circular motions that the true potential is underestimated. 
Correcting for this would cause the halo profile to become more NFW-like. 
However, bars are disk dynamical features and imply that the disk has mass 
(which has so far been ignored in the minimum disk case) and would cause the 
halo profile to become more core-like.

$\textbf{F583-4:}$  The agreement between the \Dpak\ halo parameters and the 
halo parameters of \citetalias{dBMR} is slightly better for the NFW 
halo (c = 5.7$\pm$1.4, $V_{200}$ = 89.5$\pm$19.0) than the isothermal halo
($R_{c}$ = 1.10$\pm$0.13, $\rho_{0}$ = 85.5$\pm$15.8).  Both datasets find
the NFW halo to be a slightly better fit to the data.

$\textbf{F583-1:}$  There is good agreement of the isothermal ($R_{c}$ = 
2.44$\pm$0.06, $\rho_{0}$ = 33.0 $\pm$1.1) and NFW (c = 5.1 $\pm$1.0,
$V_{200}$ = 106.6$\pm$17.0) halo parameters of \citetalias{dBMR} with
the \Dpak\ parameters.  Both datasets find the isothermal halo to be
the better description of the data.

$\textbf{UGC 1281:}$  There is moderate agreement between the isothermal halo 
parameters of the \Dpak\ data and the results of \citetalias{dBB} 
($R_{c}$ = 2.2$\pm$0.1, $\rho_{0}$ = 28.0$\pm$1.7).  No $NFW_{free}$
halo could be fit to the \Dpak\ data.  The \citetalias{dBB} data 
required a very low concentration (c = 0.1, $V_{200}$ = 785).

\subsection{A Word about Non-Circular Motions}
The presence of non-circular motions may cause the circular velocity
to be underestimated or sometimes overestimated \citep{Rob}.  This effect has 
been suggested as a reason why 
cored halos appear to be preferred to those with cusps \citep[e.g.,][]
{Rob, vandenB01}.  
The presence of non-circular motions and 
the magnitude of the effect on the system can be qualitatively
ascertained by looking at the velocity field.  For instance, the 
alignment of the major and minor axes will begin to deviate from 
perpendicular (a mild example being NGC 4395).  Or, the isovelocity 
contours will become noticeably more kinked, wiggly or twisted as 
non-circular motions increase.  There are
indications of non-circular motions in some of our sample of 
galaxies, but how significant are they?

One way of measuring this is to look at the velocity dispersion about
the mean difference of the individual fiber velocities from the circular 
velocity.  The velocity dispersions measured in this fashion for each 
galaxy are listed in Table 4.  The velocity dispersions are all in the 
range of 6 - 10 \kms.  These values are totally consistent with the typical
dispersions for the gas component of galaxies, and suggest that we are
not seeing signs of extreme non-circular motions.  If added in
quadrature to the rotation velocity, dispersions of this magnitude
affect only the innermost points where the rotation velocity and
velocity dispersion are comparable.  We tried this exercise assuming
an isotropic dispersion ($v_{circ}^{2}$ = $v_{rot}^{2}$ + 3$\sigma^{2}$) for 
the UGC 5750 \Dpak\ data.  We chose UGC 5750 because,
of the galaxies with NFW fits, it is the least consistent with the NFW
halo and because it has the largest differences between the halo fits 
at low radii.  With the exception of only the
first data point, all corrected velocities remain within the errors of
the uncorrected velocities.  The first data point increases by $\sim$ 8
\kms, but this does not improve the $NFW_{free}$ fit: the 
concentration remains virtually unchanged (c = 0.4$\pm$0.1).    
A more in-depth analysis of non-circular motions will be discussed 
in a future paper, but this simple analysis already suggests that the 
magnitude of realistic non-circular motions is not likely to be 
sufficient to recover the high concentration cuspy halos expected 
from $\Lambda$CDM structure formation simulations.

\section{Conclusions and Future Work}
We have presented the two-dimensional velocity fields and rotation
curves of a sample of LSB galaxies that have been observed with \Dpak.  
The majority of these new data have been shown to be consistent with
the rotation curves of previous long-slit \Ha\ and \HI\ observations.
In a preliminary analysis, we have combined these data and have fit 
the minimum disk case for three halo models:   the best-fit 
isothermal halo, the $NFW_{free}$ halo with no constraints on the
parameters, and the $NFW_{constrained}$ halo which was constructed 
to agree with $\Lambda$CDM cosmology.  We found seven galaxies to prefer the 
isothermal halo, one to prefer the NFW halo and three to show no 
clear preference.  When $NFW_{free}$ fits could be made, the 
concentrations  were often too low compared to the expected values for 
$\Lambda$CDM.

We have compared our \Dpak\ halo fits to the results of previous studies. 
The \Dpak\ halo parameters change little, but do have improved uncertainties. 
Future work will include a detailed assessment of non-circular motions, slit 
placement and mass-modeling to determine the distribution of dark matter 
in more realistic cases than the minimum disk scenario.

\section{Acknowledgments}
We would like to thank the referee for helpful comments.  The work of 
RKD and SSM was supported by NSF grant AST0505956.  This research
has made use of the NASA/IPAC Extragalactic Database (NED) which is 
operated by the Jet Propulsion Laboratory, California Institute of 
Technology, under contract with the National Aeronautics and Space 
Administration.  Our velocity field plots were made using a modified
version of the program found at 
\url{http://www.astro.wisc.edu/$\sim$mab/research/densepak/ DP/dpidl.html}

\end{document}